\documentstyle[12pt]{article}

\renewcommand{\theequation}{\arabic{section}.\arabic{equation}}
\newcommand{\lbl}[1]{\label{eq:#1}}
\newcommand{ \rf}[1]{(\ref{eq:#1})}
\newcommand{\vs}[1]{\rule[- #1 mm]{0mm}{#1 mm}}
\newskip\humongous \humongous=0pt plus 1000pt minus 1000pt

\newif\ifdtup

\newcommand{\eq}{\vs{2}\begin{equation}}
\newcommand{\en}{\\[2mm]\end{equation}}
\newcommand{\bea}{\begin{eqnarray}}
\newcommand{\ena}{\end{eqnarray}}
\newcommand{\lapprox}{%
\mathrel{%
\setbox0=\hbox{$<$}
\raise0.6ex\copy0\kern-\wd0
\lower0.65ex\hbox{$\sim$}
}}
\newcommand{\gapprox}{%
\mathrel{%
\setbox0=\hbox{$>$}
\raise0.6ex\copy0\kern-\wd0
\lower0.65ex\hbox{$\sim$}
}}

\def\mpi{M_\pi}
\def\fpi{F_\pi}

\newcommand{\NP}[1]{Nucl.\ Phys.\ {\bf #1}}

\newcommand{\PL}[1]{Phys.\ Lett.\ {\bf #1}}

\newcommand{\AN}[1]{Ann. Phys. {\bf #1}}

\newcommand{\PR}[1]{Phys.\ Rep.\ {\bf #1}}
\newcommand{\PRev}[1]{Phys.\ Rev.\ {\bf #1}}
\newcommand{\PRL}[1]{Phys.\ Rev.\ Lett.\ {\bf #1}}
\newcommand{\MPL}[1]{Mod.\ Phys.\ Lett.\ {\bf #1}}

\begin{document}

\begin{titlepage}

\renewcommand{\thefootnote}{\fnsymbol{footnote}}

\rightline{IPNO/TH 95-45}
\rightline{PURD-TH-95-05}
\vspace {1cm}

\begin{center}

\indent

{\Large{\bf{ The Low Energy $\pi\,\pi$ Amplitude}}}

\indent

{\Large{\bf{ To One and Two Loops}}}

\vspace{2cm}

{\bf{M. Knecht, B. Moussallam, J. Stern}}

\indent

{\sl Division de Physique Th\'eorique
\footnote{Unit\'e de Recherche des Universit\'es Paris XI et Paris VI
associ\'ee au CNRS.}, Institut de Physique Nucl\'eaire\\
F-91406 Orsay Cedex, France} \\

\indent

and

\indent

{\bf{N. H. Fuchs}}

\indent

{\sl Department of Physics, Purdue University\\
 West Lafayette, IN 47907, USA}

\vspace{2cm}

\end{center}

\begin{abstract}

The low-energy $\pi\pi$ amplitude is computed explicitly  to two-loop
accuracy in the chiral expansion. It depends only on six independent
(combinations of) low-energy constants which are not fixed by chiral
symmetry. Four of these constants are determined {\it via} sum rules
which are evaluated using $\pi\pi$ scattering data at higher energies.
Dependence of the low-energy phase shifts and of the threshold
parameters on the remaining two constants (called $\alpha$ and $\beta$)
are discussed and compared to the existing data from $K_{l4}$
experiments.  Using generalised $\chi$PT, the constants $\alpha$ and
$\beta$ are related to fundamental QCD parameters such as the quark
condensate $\langle 0|\bar{q}q|0\rangle$ and the quark mass ratio
$m_s/\widehat{m}$.  It is shown that forthcoming accurate low-energy
$\pi\pi$ data can be used to provide, for the first time, experimental
evidence in favour of or against the existence of a large quark-antiquark
condensate in the QCD vacuum.

\end{abstract}

%\vspace{5cm}

\end{titlepage}

\setcounter{footnote}{0}
\renewcommand{\thefootnote}{\arabic{footnote}}

\indent

\pagestyle{plain}
\setcounter{equation}{0}
\setcounter{subsection}{0}
\setcounter{table}{0}
\setcounter{figure}{0}

\section{\bf Introduction}

\indent
\setcounter{equation}{0}

Current understanding of the mechanism of spontaneous breakdown of
chiral symmetry (SB$\chi$S) in QCD is still lacking clear experimental
support. The basic fact that in the limit $m_u=m_d=m_s=0$ the chiral
symmetry of the lagrangian is spontaneously broken down to
$SU(3)_{\rm V}$ is
known to be a mathematical consequence of anomalous Ward identities
\cite{tHooft}, quark confinement and the vector-like character of the
theory \cite{VW}. It just means that  there are eight massless
Goldstone bosons coupled to eight conserved axial-vector currents. In
the real world where $m_u,\ m_d,\ m_s$ are nonzero but much smaller than
the typical mass scale $\Lambda_H\sim 1$ GeV of the first massive
hadrons, the theoretical fact of  SB$\chi$S and the Goldstone character
of the eight lightest  pseudoscalars are experimentally well founded.
The chiral world, i.e. the first step in the expansion in powers of
$m_{\rm quark}/\Lambda_H$, indeed strongly resembles the observed world.

That much is known and  is sufficient to establish a low-energy
effective theory in which the role of effective degrees of freedom
is played by weakly interacting Goldstone bosons \cite{W2}.
Equivalence of this effective theory with QCD is ensured step by step
in a systematic expansion in powers of external momenta and quark
masses, referred to as chiral perturbation theory \cite{GL1,GL2}
($\chi$PT). The standard wisdom, however, involves much more than this
established general framework of SB$\chi$S and of the corresponding
low-energy effective theory since, in addition, it assumes a particular
mechanism of symmetry breaking. What is commonly assumed is that in
QCD, as in the Nambu-Jona-Lasinio model \cite{NJL},
SB$\chi$S is triggered by strong condensation of quark-antiquark
pairs in the vacuum. The quark condensate parameter
\eq
B_0=-\lim_{m_u,m_d,m_s\to 0} {1\over F^2_\pi}
\langle\Omega\vert \bar q q\vert\Omega\rangle
\en
is then assumed to play a dominant role \cite{GOR} in the description
of symmetry-breaking effects induced by quark masses. The assumption of
a strong quark condensation and its consequences have not so far
been tested experimentally. Moreover, in a vector-like theory like
QCD, a natural alternative scenario is conceivable. Goldstone
bosons with a non-vanishing coupling $F_0=\lim_{m_{\rm quark}\rightarrow0}
F_\pi$
to conserved axial currents can be formed even if quark condensation
is marginal or even absent. For all these reasons an experimental probe
of the mechanism of SB$\chi$S in QCD becomes of fundamental importance.
It has been pointed out that the most promising
framework for testing the
strength of quark condensation is low-energy $\pi\pi$
scattering \cite{FSS2,SSF}. In the present work the theoretical basis
for such a test is elaborated in detail, in view of two experimental
projects to improve considerably the accuracy of existing
low-energy $\pi\pi$ scattering data \cite{dirac,BF}. The $\pi\pi$
scattering amplitude is worked out explicitly to two-loop
accuracy in the general low-energy expansion, independently of any
prejudice concerning the size of the quark condensate.

Before describing the content of this article in detail, let us
briefly comment on the theoretical possibility that in QCD,
SB$\chi$S need not necessarily imply a strong quark condensation.
This is illustrated by an analogy with antiferromagnetic
spin systems \cite{And}. In this case, the spontaneous breakdown
of rotation symmetry does not necessarily yield a large spontaneous
magnetisation (as it would happen for a ferromagnet). The latter can
even vanish to the extent that the structure of the ground state
reaches the N\'eel-type magnetic order. From renormalised perturbation
theory, we are used to consider as ``unnatural" a situation where a
quantity would vanish (or be kept unusually small) if it is not forced
to do so for reasons of symmetry. The above example of
antiferromagnetism suggests that this reasoning need not necessarily
apply to quantities of genuine nonperturbative origin such as
spontaneous magnetisation or the quark condensate. Indeed, in QCD
(as in any local relativistic field theory) the necessary and
sufficient criterion for SB$\chi$S is the non-vanishing of the
coupling $F_0$ of the Goldstone bosons to the Noether currents
rather than a non-zero value of the quark condensate. It is instructive
to formally express $F^2_0$ and the quark condensate by means of the
euclidian functional integral of QCD in a finite box of size $L$.
Integrating over the fermions first (giving them a mass $m$), the
double limit $\lim_{m\to0}\lim_{L\to\infty}$ of order parameters
of SB$\chi$S becomes sensitive to the small eigenvalues $\lambda_n
(A)$ of the Dirac operator $iD\!\!\!\!\slash(A)$, averaged over all gluonic
configurations. In particular, the crucial question is how dense
does the spectrum become as $L\to\infty$.\cite{dense,VW,LSmil} ~If the
average level spacing
$\overline{\Delta\lambda}$ behaves as $1/L$, then chiral symmetry remains
unbroken. Both $F_0$ and the quark condensate vanish. In order to
have $F_0\neq0$, it is necessary and sufficient that for $L\to\infty$,
$~~\overline{\Delta\lambda}$ behaves {\sl at least} as $1/L^2F_0$
\cite{JSinprep}. It is well
known that quarks start to condense only provided \cite{LSmil}
$~~\overline{\Delta\lambda}\sim1/L^4\vert\langle\bar qq\rangle\vert$.
The above discussion does not tell us what actually does happen
in the QCD ground state. However, it demonstrates that the
theoretical possibility of SB$\chi$S (i.e., $F_0\ne0$) without
quark condensation might naturally arise within QCD. In fact, it is
conceivable that small eigenvalues with
$\overline{\Delta\lambda}\sim L^{-2}$ and
$\overline{\Delta\lambda}\sim L^{-4}$ coexist and compete, giving
rise to a marginal quark condensate, which then would be difficult to
estimate in advance.

At this point, one should recall that the alternative scenario
of a small or vanishing condensate, though not contradicted by a
single experimental fact, does not seem to be favoured by existing
lattice simulations when they are extrapolated to the chiral limit.
However, lattice regularisation is known to mistreat chiral symmetry,
in one way or another, especially in the quenched approximation, in
which the most significant results are obtained.  (For a recent review,
see Ref.  \cite{Gup}.) Lattice results
constitute another reason to attempt a direct experimental test
of theoretical prejudices regarding the value of
$\langle\bar qq\rangle$.

The quark condensate manifests itself exclusively through
symmetry-breaking effects proportional to quark masses.  The order
parameter $B_0$ (1.1) is one of the constants of  the low-energy
effective lagrangian, but it enters into physical observables multiplied
by a quark mass. An example of a meaningful question to ask concerns,
for instance, the magnitude of the  renormalisation group invariant
product $(m_u+m_d)B_0$ measured in  units of the pion mass. The
smallness of most symmetry-breaking effects is the main reason why
experimental tests of quark condensation are difficult and have not been
performed so far. Actually, the only symmetry-breaking effects which are
easily accessible experimentally are the Goldstone boson masses. Their
expansion in powers of quark masses can be schematically written as (for
more details, see \cite{HBK1,GL2})
\bea
&&M^2_{\pi^+}=(m_u+m_d) B_0 +  (m_u+m_d)^2 A_0 +...\\ \nonumber
&&M^2_{K^+}=(m_u+m_s) B_0 + (m_u+m_s)^2 A_0 +... \ ,
\lbl{M2}
\ena
where $F_0^2A_0$ is a (suitably subtracted) massless QCD two-point
function of scalar and pseudoscalar quark densities. $A_0$ is
of the order of 1, typically $A_0=1\div5$. (A similar expression
holds for the $\eta$ mass, except that at the quadratic level, a
new constant appears, reflecting the axial $U(1)$ anomaly.) While
the overall convergence of the expansion (1.2) is controlled by the
small parameter $m_{\rm quark}/\Lambda_H$, the relative importance of the
first two terms is determined by the relative sizes of the quark
mass and of the mass scale
\eq
m_0={B_0\over2A_0}
\en
set by the quark condensate. If $m_{\rm quark}\ll m_0$, the condensate term
dominates the expansion of Goldstone boson masses. On the contrary, if
$m_{\rm quark}\sim m_0$, the first and second order terms in Eq. (1.2) are of
comparable size. This is what distinguishes the standard large-condensate
wisdom from the alternative, low-condensate, scenario.
The standard case $(m_0\sim\Lambda_H)$ is the basis of the standard
$\chi$PT, i.e. the expansion in powers of quark masses and external
momenta in which one power of a quark mass is counted as two powers of
the momenta:
\eq
m_{\rm quark}=O({\rm p}^2)\ .
\en
The consequences of this scheme are well known. At leading order,
pseudoscalar masses satisfy the Gell-Mann--Okubo (GMO) formula, and
the ratio of strange to non-strange quark masses gets determined
to be \cite{Rabi}
\eq
r={m_s\over \widehat{m}}=2{M^2_K\over M^2_\pi}-1\simeq26,~~~~~~
\widehat{m}={1\over2}(m_u + m_d) \ .
\en
Historically, the emergence of the GMO formula at leading order was
at the origin of the large $B_0$ scenario \cite{GOR}. Today, the GMO
formula can still be considered as an argument of plausibility,
but not as an experimental test of the large-condensate hypothesis,
for the following two reasons: i) The validity of the GMO relation does
not imply the dominance of the condensate term in Eq. \rf{M2} and it does
not rule out the small condensate alternative, and ii) the standard
$\chi$PT is not yet able to predict the size of corrections to the leading
order GMO formula; this would require an independent determination of the
constant $L_8$ \cite{GL2} from other symmetry-breaking effects.

On the other hand, if the quark condensate turned out to vanish, then
the quark mass ratio (1.5) would become
\eq
r=2{M_K\over M_\pi}-1\simeq6.3 \ .
\en
Notice that even in this extreme case, the $\eta$ mass would remain
unrelated to $M_K$ and $M_\pi$, because of the anomaly contribution.
However, the small-condensate alternative is more general than this: it
covers the whole range $m_0 \ll \Lambda_H$. For instance, if $B_0\sim F_0
\sim 90$ MeV (i.e. $-\langle\bar q q\rangle\sim F_0^3$), then $m_0$ could
easily be as small
as $\sim 20$ MeV. In this case, $m_{\rm quark}\sim m_0$ could happen even
for $u$
and $d$ quarks, implying a value of the ratio $(m_u+m_d)B_0/M^2_\pi$
significantly below one. In the generic case
$m_0 \ll \Lambda_H$, the quark mass ratio
$r$ interpolates between the two extreme values (1.5) and (1.6).
The standard $\chi$PT treats the terms $m_{\rm quark}/\Lambda_H$ and
$m_{\rm quark}/m_0$ in a similar way:  they are both considered small and
of order
$O({\rm p}^2)$. In the small-condensate alternative, the
standard expansion of the (same) effective lagrangian has to be
modified since, now, $m_{\rm quark}/m_0$ is of order 1. This requires a
modification of the chiral counting rule (1.4). $B_0$ now becomes an
expansion parameter that counts as a small quantity of order $M_\pi$
\cite{SSF}.
The new counting rule reads
\eq
m_{\rm quark}=O({\rm p})\ ,\qquad B_0=O({\rm p}) \ .
\en
The corresponding expansion of ${\cal L}^{\rm eff}$ defines the so-called
``generalized chiral perturbation theory" (G$\chi$PT)
\cite{FSS2,SSF,ggpp,HBK1}.
Order by order,
the G$\chi$PT contains the standard expansion as a special case, because
at each order it includes additional terms, which the standard $\chi$PT
relegates to higher orders.  Notice that in both schemes the symbol
$O({\rm p}^n)$ has the same meaning: it represents a quantity with an
order of magnitude $(M_\pi/\Lambda_H)^n$. The symbol $O(m^n_{\rm quark})$,
in contrast, has to be interpreted according to (1.4) or (1.7),
depending on the scheme. More details on the structure of the
G$\chi$PT may be found in Refs. \cite{SSF,ggpp,HBK1}.

In order to test the strength of quark condensation, the constraints
arising from the discussion of Goldstone boson masses have to be
combined with experimental information on different symmetry-breaking
effects. There are two particularly significant examples
of such effects:
\begin{itemize}
\item[i)] Comparing the observed deviation from the
Goldberger-Treiman relation in $\pi N$, $K\Lambda$ and $K\Sigma$
channels one obtains a sum rule for the ratio $r=m_s/\hat m$ valid up
to higher order corrections \cite{FSS1}. The output of the sum rule is
particularly sensitive to the precise value of the charged
pion-nucleon coupling constant $g_{\pi N}$, which is nowadays
rather controversial. For the old value of Koch and
Pietarinen \cite{KP}, recently confirmed by the analysis of
a new $n-p$ charge-exchange scattering experiment \cite{npchgex}, the
value of $r$
turns out to be significantly smaller than the standard value $r\simeq26$,
by at least a factor of 2 \cite{FSS1}. For lower values of
$g_{\pi N}$ the outcome is less conclusive. In any case, more
accurate data as well as an estimate of higher order corrections
are needed to transform this indication into a real measurement
of $m_s/\hat m$.
\item[ii)] The second relevant symmetry-breaking
effect concerns the I=0 S-wave $\pi\pi$ scattering
length. The standard $\chi$PT, based
on the large-condensate hypothesis, predicts (at one-loop accuracy)
$a_0^0=0.20\pm0.01$ \cite{GLlett,GL1},
while the current experimental value is \cite{nagels} $a_0^0=
0.26\pm0.05$. One of the purposes of this work is to show
that this one standard deviation effect finds a natural explanation
within the low-condensate alternative. It has been known for some
time \cite{FSS2} that already at tree level Weinberg's
prediction \cite{W3} $a_0^0=0.16$ (which implicitly uses the
large-condensate hypothesis) can gradually increase towards a
value $a_0^0=0.27$ provided the ratio $(m_u+m_d)B_0/M^2_\pi$ is
allowed to decrease from 1 to 0. Here, this statement will be made
more quantitative including the one-loop \cite{HBK2} and the
two-loop G$\chi$PT corrections. In view of forthcoming precise
low energy $\pi\pi$ scattering data \cite{dirac,BF}, our analysis
might imply a realistic possibility to disentangle the large- and
low-condensate alternatives experimentally.
\end{itemize}

The paper is organized as follows. We first describe the relevant
features of the effective action at order $O({\rm p}^4)$ in the generalized
case and obtain the corresponding expression for the $\pi\pi$ amplitude in
terms of four independent combinations of the low-energy constants (Section 2).
A preliminary version of this calculation has already been
published \cite{HBK2}.
In Section 3, we show how this result can be rederived and easily extended
to two-loop accuracy using a slightly different approach.
At two-loop order, the $\pi\pi$ amplitude involves
two additional constants which are not fixed by chiral symmetry. We determine
four out of this total of six constants from sum rules and $\pi\pi$
production data at
higher energies in Section 4, where the dependence of low-energy
phase shifts and threshold parameters on the remaining two parameters is also
discussed. The possibilities of exploiting future precise data on low-energy
$\pi\pi$ scattering in order to distinguish between the small and large
condensate alternatives are investigated in Section 5. A summary and concluding
remarks are presented in Section 6, and miscellaneous formulae and auxiliary
results have been collected in four Appendices.

Finally, it should be stressed that although the paper is motivated mainly
by questions raised by the chiral structure of the QCD vacuum,
this work contains new material concerning the phenomenology of
low-energy $\pi\pi$ scattering which might be of interest for its own
sake. The reader mostly interested in these aspects can directly read
Sections 3 and 4 as well as appendices B, C and D, which form a
self-contained core.

\section {\bf The $\pi\pi$ Amplitude to Order $O({\rm p}^4)$ in G$\chi$PT}

\indent
\setcounter{equation}{0}

In this section, we briefly discuss the structure of the generating
functional for the effective theory in the generalized framework up to
order $O({\rm p}^4)$. Then we derive the low-energy expansion, to this
order, of the amplitude $A(s\vert t,\, u)$ which describes $\pi\pi$
scattering in the absence of isospin breaking ({\it i.e.} we shall take
$m_u = m_d = {\widehat m}$ and we shall neglect electromagnetic
corrections).  We work within $SU(3)_{\rm L}\times SU(3)_{\rm R}$,
although $\pi\pi$
scattering is merely an $SU(2)_{\rm L}\times SU(2)_{\rm R}$ problem.
The reason for
this choice should be obvious from the preceding discussion on Goldstone
boson masses.  Working within the $SU(3)_{\rm L}\times SU(3)_{\rm R}$
chiral expansion
offers the opportunity to make use of relationships between constants which
characterise low-energy $\pi\pi$ scattering and QCD parameters (such
as $m_s/\widehat{m}$) or observables (such as $F_K/F_\pi, ~M_K/M_\pi$).

The strength of the quark-antiquark condensate in the QCD
vacuum does not affect the structure of the effective lagrangian ${\cal L}^{\rm
eff}$. The latter is merely determined by chiral symmetry  \cite{Leut} and the
transformation properties of the chiral symmetry breaking mass term of QCD.
Thus, ${\cal L}^{\rm eff}$ consists of an infinite tower of invariants
\eq
{\cal L}^{\rm eff} = \sum_{(k,l)} {\cal L}_{(k,l)}\ ,\lbl{Leff}
\en
where ${\cal L}_{(k,l)}$ contains $k$ powers of covariant derivatives of the
Goldstone boson fields, and $l$ powers of scalar or pseudoscalar sources. In
the chiral limit, ${\cal L}_{(k,l)}$ vanishes like the $k$-th power of
external momenta ${\rm p}$ and the $l$-th power of the quark masses,
\eq
{\cal L}_{(k,l)} \sim {\rm p}^km_{\rm quark}^l\ .
\en
For sufficiently small quark masses, such that {\it both} $m_{\rm quark}\ll
\Lambda_H$ and $m_{\rm quark}\ll m_0 = B_0/2A_0$, one has ${\cal L} = O({\rm
p}^{k+2l})$. In this case, the double expansion  \rf{Leff} can be reorganized
as
\eq
{\cal L}^{\rm eff} = {\cal L}^{(2)} + {\cal L}^{(4)} + {\cal L}^{(6)} + \cdots
\ ,\lbl{Lstd}
\en
where  \cite{GL1,GL2}
\eq
{\cal L}^{(d)} = \sum_{k+2l=d}{\cal L}_{(k,l)}\ .
\en
This expansion defines the standard $\chi$PT. On the other hand, if for the
actual values of the quark masses one has $m_{\rm quark}\sim m_0\ll \Lambda_H$,
then ${\cal L}_{(k,l)}= O({\rm p}^{k+l})$, and this new counting yields a
{\it different expansion} of the {\it same} effective lagrangian  \rf{Leff}:
\eq
{\cal L}^{\rm eff} = {\tilde{\cal L}}^{(2)} + {\tilde{\cal L}}^{(3)} +
{\tilde{\cal L}}^{(4)} + {\tilde{\cal L}}^{(5)} + {\tilde{\cal L}}^{(6)}
+ \cdots\ ,\lbl{Lgen}
\en
with now \cite{SSF}
\eq
{\tilde{\cal L}}^{(d)} = \sum_{k+l+n=d} B_0^n{\cal L}_{(k,l)}\ .
\en
Although Eqs.  \rf{Lstd} and  \rf{Lgen} sum up the same effective lagrangian
${\cal L}^{\rm eff}$ to all orders, their truncations at any finite order may
differ.

The structure of the effective action ${\cal Z}^{\rm eff}$ in the generalized
case has been discussed in  \cite{K} up to order $O({\rm p}^4)$ in the
framework
of the $SU(3)_{\rm L}\times SU(3)_{\rm R}$ chiral expansion. It is given by
\eq
{\cal Z}^{\rm eff} = \int d^4 x \left\{ {\tilde{\cal L}}^{(2)} +
                     {\tilde{\cal L}}^{(3)} + {\tilde{\cal L}}^{(4)}\right\}
                  + {\tilde{\cal Z}}^{(4)}_{\rm 1\, loop}\ .\lbl{Zeff}
\en
Whereas ${\tilde{\cal L}}^{(n)}$, $n = 2,3,4$, gives the tree level
contributions at order $O({\rm p}^n)$, ${\tilde{\cal Z}}^{(4)}_{\rm 1\, loop}$
contains the contributions from one-loop graphs with an arbitrary number of
vertices from ${\tilde{\cal L}}^{(2)}$ only. Thus, the leading order of the
generalized expansion is described by ${\tilde{\cal L}}^{(2)}$, which was first
given in  \cite{FSS2}:
\bea
{\tilde{\cal L}}^{(2)}&=&{1\over 4}F_0^2
\left\{\langle D_\mu U^+D^\mu U\rangle +2B_0\langle U^+\chi+\chi^+U\rangle+
\right.\nonumber\\
&&\qquad + A_0\langle (U^+\chi)^2+(\chi^+U)^2\rangle
+Z_0^S\langle U^+\chi+\chi^+U\rangle ^2\lbl{glead} \lbl{L2}\\
&&\qquad + Z_0^P\langle U^+\chi-\chi^+U\rangle ^2
+\left. H_0\langle \chi^+\chi\rangle\right\}\ .\nonumber
\ena
The notation is as in Refs.  \cite{GL1,GL2}, except for the consistent
removal of the factor $2B_0$ from $\chi$, the parameter that collects
the scalar and pseudoscalar sources,
\eq
\chi={s}+i{ p}={\cal M}+\cdots \ ,\, {\cal M}={\rm diag}(m_u,m_d,m_s)
\ .\lbl{chi}
\en
In G$\chi$PT, the next-to-leading-order corrections are of order
$O({\rm p}^3)$, and still
occur at tree level only. They are
embodied in ${\tilde{\cal L}}^{(3)}={\cal L}_{(2,1)}+{\cal L}_{(0,3)}$,
which reads  \cite{SSF,HBK1}
\bea
\tilde{\cal L}^{(3)}&=&{1\over 4}F_0^2
\left\{\xi\langle D_\mu U^+D^\mu U(\chi^+U+U^+\chi)\rangle
+\tilde\xi\langle D_\mu U^+D^\mu U\rangle
\langle\chi^+U+U^+\chi\rangle\right.\nonumber\\
&&\qquad + \rho_1\langle (\chi^+U)^3+(U^+\chi)^3\rangle
+\rho_2\langle (\chi^+U+U^+\chi)\chi^+\chi\rangle\nonumber\\
&&\qquad + \rho_3\langle\chi^+U-U^+\chi\rangle
\langle(\chi^+U)^2-(U^+\chi)^2\rangle\lbl{L3} \\
&&\qquad + \rho_4\langle(\chi^+U)^2
+(U^+\chi)^2\rangle
\langle\chi^+U+U^+\chi\rangle\nonumber\\
&&\qquad + \rho_5\langle\chi^+\chi\rangle
\langle\chi^+U+U^+\chi\rangle\nonumber\\
&&\qquad + \left.\rho_6\langle\chi^+U-U^+\chi\rangle^2
\langle\chi^+U+U^+\chi\rangle+\rho_7\langle\chi^+U+U^+\chi\rangle^3
\right\}\ .\nonumber
\ena
The tree-level contributions at order $O({\rm p}^4)$ are contained in
\footnote{At order $O({\rm p}^4)$,
terms with odd intrinsic parity coming from the Wess-Zumino term are
also present in the effective lagrangian, but since they start to contribute
to $A(s\vert t,\, u)$ only at order $O({\rm p}^8)$,
we do not mention them further here.}
\eq
\tilde{\cal L}^{(4)}={\cal L}_{(4,0)}+{\cal L}_{(2,2)}+{\cal L}_{(0,4)}+
B_0^2{\cal L}'_{(0,2)} + B_0{\cal L}'_{(2,1)} + B_0{\cal L}'_{(0,3)}\ .
\lbl{L4}
\en
The part without explicit chiral symmetry breaking, ${\cal L}_{(4,0)}$,
is described by the same low-energy constants
$L_1$, $L_2$, $L_3$, $L_9$ and $L_{10}$ as
in the standard case  \cite{GL2}. The part with two powers of momenta and two
powers of quark masses is given by:

\bea
{\cal L}_{(2,2)}&=& {1\over{4}}F_0^2\bigg\{
A_1 \langle D_{\mu}U^+D^{\mu}U (\chi^+\chi + U^+\chi\chi^+U)\rangle
                                                            \nonumber\\
& &\qquad + A_2 \langle D_{\mu}U^+U\chi^+D^{\mu}UU^+\chi\rangle \nonumber\\
& &\qquad + A_3 \langle D_{\mu}U^+U(\chi^+D^{\mu}\chi-D^{\mu}\chi^+\chi) +
          D_{\mu}UU^+(\chi D^{\mu}\chi^+ - D^{\mu}\chi\chi^+)\rangle
\nonumber\\
& &\qquad + A_4 \langle D_{\mu}U^+D^{\mu}U\rangle\langle\chi^+\chi\rangle
                                                                 \nonumber\\
& &\qquad + B_1 \langle D_{\mu}U^+D^{\mu}U (\chi^+U\chi^+U +
                                        U^+\chi U^+\chi)\rangle \nonumber\\
& &\qquad + B_2 \langle D_{\mu}U^+\chi D^{\mu}U^+\chi +
                        \chi^+D_{\mu}U\chi^+D^{\mu}U\rangle\nonumber\\
& &\qquad + B_4 \langle
D_{\mu}U^+D^{\mu}U\rangle\langle\chi^+U\chi^+U+U^+\chi
U^+\chi\rangle\nonumber\\
& &\qquad + C_1^S \langle D_{\mu}U\chi^+ + \chi D_{\mu}U^+\rangle
            \langle D^{\mu}U\chi^+ + \chi D^{\mu}U^+\rangle \lbl{L22}\\
& &\qquad + D^S \langle D_{\mu}U^+D^{\mu}U(\chi^+U+U^+\chi)\rangle
                               \langle\chi^+U+U^+\chi\rangle
+  \cdots \bigg\} \ .\nonumber
\ena
Finally, the tree-level contributions which behave as $O(m_{\rm quark}^4)$ in
the chiral limit are contained in ${\cal L}_{(0,4)}$, which reads
\bea
{\cal L}_{(0,4)} &=& {1\over{4}}F^2_0\bigg\{
E_1 \langle (\chi^+U)^4 + (U^+\chi)^4 \rangle \nonumber\\
& &\qquad + E_2 \langle \chi^+\chi(\chi^+U\chi^+U+U^+\chi U^+\chi) \rangle
                                                     \nonumber\\
& &\qquad + E_3 \langle \chi^+\chi U^+\chi\chi^+ U \rangle \nonumber\\
& &\qquad +
F_1^S \langle \chi^+U\chi^+U + U^+\chi U^+\chi \rangle^2 \nonumber\\
& &\qquad +
F_2^S \langle (\chi^+U)^3 + (U^+\chi)^3 \rangle\langle \chi^+U + U^+\chi\rangle
\nonumber\\
& &\qquad +
F_3^S \langle \chi^+\chi(\chi^+U + U^+\chi)\rangle
                           \langle\chi^+U + U^+\chi\rangle\nonumber\\
& &\qquad +
F_4^S \langle (\chi^+U)^2 + (U^+\chi)^2 \rangle\langle \chi^+\chi \rangle
\nonumber\\
& &\qquad +
F_5^{SS} \langle (\chi^+U)^2 + (U^+\chi)^2 \rangle
         \langle \chi^+U + U^+\chi \rangle ^2 \lbl{L04}\\
& &\qquad +
F_6^{SS} \langle \chi^+\chi \rangle \langle \chi^+U + U^+\chi \rangle ^2
+  \cdots \bigg\} \ . \nonumber
\ena
For ${\cal L}_{(2,2)}$ and ${\cal L}_{(0,4)}$, we have shown explicitly only
those terms which will actually contribute to $A(s\vert t,\, u)$. For the
complete list of counterterms which enter ${\cal L}_{(2,2)}$ and
${\cal L}_{(0,4)}$, we refer the reader to  \cite{K}. Notice also that in the
standard framework all these contributions would count as order $O({\rm
p}^6)$ and
order $O({\rm p}^8)$, respectively \footnote{The standard $O({\rm p}^6)$
effective lagrangian ${\cal
L}^{(6)}={\cal L}_{(6,0)}+{\cal L}_{(4,1)}+{\cal L}_{(2,2)}+{\cal L}_{(0,3)}$
has been worked out in Ref.  \cite{FeaSch}. }.

As far as the $\pi\pi$ amplitude is concerned, the $O({\rm p}^4)$ loop
corrections involve only graphs with a single or two vertices from
${\tilde{\cal L}}^{(2)}$:
\eq
{\tilde{\cal Z}}^{(4)}_{\rm 1\, loop} = {\tilde{\cal Z}}^{(4)}_{\rm tadpole}+
{\tilde{\cal Z}}^{(4)}_{\rm unitarity}+\cdots \lbl{Zloop}
\en
The divergent parts of these one loop graphs have been subtracted at a scale
$\mu$ in the same dimensional renormalisation scheme as described in
 \cite{GL1,GL2}. Accordingly, the low energy constants of ${\cal L}_{(4,0)}$,
${\cal L}_{(2,2)}$, and ${\cal L}_{(0,4)}$ stand for the renormalised
quantities, with an explicit logarithmic scale dependence ($X(\mu )$ denotes
generically any of these renormalised low-energy constants)
\eq
X(\mu )= X(\mu ') + {{\Gamma_X}\over{(4\pi )^2}}\,\cdot\ln({\mu '}/\mu )\ .
\lbl{scale}
\en
At order $O({\rm p}^4)$, the low-energy constants of
${\tilde{\cal L}}^{(2)}$ and
${\tilde{\cal L}}^{(3)}$ also need to be renormalised. The corresponding
counterterms, however, are of order $O(B_0^2)$ and $O(B_0)$,
respectively, and they are gathered in the  three last terms of Eq.  \rf{L4}:
in G$\chi$PT, renormalisation proceeds order by order in the
expansion in powers of $B_0$. Alternatively, one may think of Eqs.  \rf{L2}
and
 \rf{L3} as standing for the combinations
${\tilde{\cal L}}^{(2)} + B_0^2{\cal
L}'_{(0,2)}$ and ${\tilde{\cal L}}^{(3)} + B_0{\cal L}'_{(2,1)} + B_0{\cal
L}'_{(0,3)}$, respectively, with the corresponding low-energy constants
representing the renormalised quantities.
The full list of $\beta$-function coefficients $\Gamma_X$ can be found in
Ref.  \cite{K}, where the explicit expressions of ${\tilde{\cal
Z}}^{(4)}_{\rm tadpole}$ and of ${\tilde{\cal Z}}^{(4)}_{\rm unitarity}$ are
also displayed.

Computing the $O({\rm p}^4)$ expression of $A(s\vert t,\, u)$ is then a
straightforward exercise in field theory, and we shall merely quote the result,
expressed in terms of the loop functions $J^{r}_{PQ}$
and $M^{r}_{PQ}$ defined as in  \cite{GL1,GL2}:
\bea
&& A(s\vert t,\, u)\, =\, {{{\beta^r}}\over{F_{\pi}^2}}\left(
                                s-{4\over 3}M_{\pi}^2 \right) +
{\alpha^r}{{M_{\pi}^2}\over{3F_{\pi}^2}}\nonumber\\
&&\qquad\qquad \, +\, {4\over{F_{\pi}^4}}(2L_1^r + L_3 )(s-2M_{\pi}^2)^2
\, +\, {4\over{F_{\pi}^4}}L_2^r\left[ (t-2M_{\pi}^2)^2 + (u-2M_{\pi}^2)^2
\right] \nonumber\\
&&\ \, +\, {1\over{6F_{\pi}^4}}\, J^r_{\pi\pi}(s)\bigg\{
4\big[ s-{{M_{\pi}^2}\over 2} +10 {\widehat m}^2
(A_0 + 2Z_0^S)\big]^2
\, -\, \big[ s-2M_{\pi}^2 -8{\widehat m}^2 (A_0 +2Z_0^S)\big]^2
\bigg\}\nonumber\\
&&\ \, +\, {1\over{4F_{\pi}^4}}\, J^r_{\pi\pi}(t)\bigg[
t-2M_{\pi}^2 -8{\widehat m}^2 (A_0 +2Z_0^S)
\bigg]^2
\, +\, {1\over{4F_{\pi}^4}}\, J^r_{\pi\pi}(u)\bigg[
u-2M_{\pi}^2 -8{\widehat m}^2 (A_0 +2Z_0^S)
\bigg]^2\nonumber\\
&&\ \, +\, {1\over{8F_{\pi}^4}}\, J^r_{KK}(s)\bigg[
s+8{\widehat m}^2(1+r)(A_0 + 2Z_0^S)\bigg] ^2\nonumber\\
&&\ \, +\, {1\over{18F_{\pi}^4}}\, J^r_{\eta\eta}(s)\bigg[
M_{\pi}^2 +4{\widehat m}^2(1+2r)(A_0 + 2Z_0^S)
+8{\widehat m}^2 (1-r)(A_0 +2Z_0^P)\bigg] ^2\lbl{gAstu}\\
&&\ \, +\, {1\over{2\fpi^4}}\,
\bigg\{ (s-u)t\big[ 2M^r_{\pi\pi}(t) + M^r_{KK}(t)\big]
                  +(s-t)u\big[ 2M^r_{\pi\pi}(u) + M^r_{KK}(u)\big]\bigg\}
\ .\nonumber
\ena
The coefficients ${\alpha^r}$ and ${\beta^r}$
denote two combinations of the (renormalised, scale dependent) low-energy
constants of ${\tilde{\cal L}}^{(2)}+{\tilde{\cal L}}^{(3)}+
{\tilde{\cal L}}^{(4)}$:
\bea\lbl{beta^r}
{\beta^r} &=& 1 + 2{\widehat m}\xi + 4{\widehat m}{\tilde\xi}
\nonumber\\
&&\ + 2{\widehat m}^2\left[ 3A_2 + 2A_3 +4B_1 + 2B_2 + 8B_4 + 4C_1^S
+2(2+r)D^S\right. \lbl{bt}\\
&&\quad\qquad \left.
-2{\xi}^2 -4(2+r){\tilde\xi}^2 -2(2+r)(4+r)\xi{\tilde\xi}\right]
\ ,
\nonumber
\ena
and
\bea\lbl{alpha^r}
{{F_{\pi}^2}\over{F_0^2}}M_{\pi}^2{\alpha^r} &=&
\ 2{\widehat m}B_0 + 16{\widehat m}^2 A_0 +4(8+r){\widehat m}^2 Z_0^S
 -4{\widehat m}M_{\pi}^2 (\xi + 2{\tilde\xi})\nonumber\\
&+& {\widehat m}^3 \left[ 81{\rho}_1 + {\rho}_2 + 2(82+16r+r^2){\rho}_4
+(2+r^2){\rho}_5 + 12(2+r)(14+r){\rho}_7\right]\nonumber\\
&-& 8{\widehat m}^2 M_{\pi}^2\left[ 2B_1 -2B_2 +4B_4 -A_3 -4C_1^S +(4+r)D^S
\right]\nonumber\\
&+& 2{\widehat m}^4 \left[ 48 A_3(A_0 + 2Z_0^S) +128E_1 +8E_2
\right.\nonumber\\
&&\qquad +32(8+r^2)F_1^S +(272 +81r +r^3)F_2^S +(16+r+r^3)F_3^S +8(2+r^2)F_4^S
\nonumber\\
&&\qquad \left. +4(144 +82r +16r^2+r^3)F_5^{SS}
+2(16+2r+8r^2+r^3)F_6^{SS}\right]\nonumber\\
&~&\nonumber\\
&-& {\mu}_{\pi}\big[ 3M_{\pi}^2 + 116{\widehat m}^2 (A_0 +2Z_0^S) \big]
- 2{\mu}_K \big[ M_{\pi}^2 +4(7+r){\widehat m}^2 (A_0 +2Z_0^S) \big]
\lbl{at}\\
&-& {1\over 3}{\mu}_{\eta}\big[ M_{\pi}^2 +4(7+2r){\widehat m}^2 (A_0 +2Z_0^S)
+8(4-r){\widehat m}^2 (A_0 +2Z_0^P)\big]\ ,\nonumber
\ena
with ${\mu}_P = M_P^2/(32{\pi}^2 F_{\pi}^2)\ln (M_P^2/{\mu}^2)$, $P=\pi ,K,
\eta$. The deviation of the pion decay constant $F_{\pi}$ from its counterpart
in the chiral limit $F_0$ is given, at order $O({\rm p}^4)$, in Appendix A,
together with the corresponding expansion of the pion mass.

As mentioned above, both ${\alpha^r}$ and ${\beta^r}$ depend of the same
subtraction scale ${\mu}$ in terms of which the renormalised loop
functions $J^r_{PQ}$ and $M^r_{PQ}$ were defined. Using the results of
\hfill\break Ref.  \cite{K}, we find \footnote{Up to order $O({\rm
p}^3)$,  $\alpha^r$ and $\beta^r$ are scale independent and coincide
with the parameters $\alpha_{\pi\pi}$ and $\beta_{\pi\pi}$ which enter
the $O({\rm p}^5)$ expression of the $\gamma\gamma\to\pi^0\pi^0$
amplitude in generalised $\chi$PT
 \cite{ggpp}.}
\bea
{\alpha^r}(\mu ) &=& {\alpha^r}({\mu '}) \ +
\ {{M_{\pi}^2}\over{(4\pi F_{\pi})^2}} \ln ({\mu '}/\mu )\cdot
\bigg\{ \big[ 1 + (1+r){\widehat \epsilon}\big]\big[ 1 +3(1+r){\widehat
\epsilon}\big]\nonumber\\
&&\quad + {1\over 3}\big[ 1 + (1+2r){\widehat \epsilon} + 2 {{{\widehat
\Delta}_{GMO}}\over{1-r}}\big] ^2
+ \big[ 1 + 22{\widehat \epsilon} +33{\widehat \epsilon}^2 \big]
\bigg\}\lbl{atmu}
\ ,
\ena
and
\eq
{\beta^r}(\mu )\, =\, {\beta^r}(\mu ')\ +
\ {{M_{\pi}^2}\over{(4\pi F_{\pi})^2}} \ln ({\mu '}/\mu ) \,
\big[ 5+(11+r){\widehat \epsilon}\big]\ ,\lbl{btmu}
\en
where we have introduced
\eq
{\widehat\epsilon}\, =\, {{4{\widehat m}^2}\over{M_{\pi}^2}}\,(A_0 +2Z_0^S)\ ,
\lbl{epsi}
\en
and
\eq
{\widehat\Delta}_{GMO}\, =\, {{4({\widehat m} -m_s)^2}\over{M_{\pi}^2}}\,
(A_0 +2Z_0^P)\ . \lbl{GMO}
\en
{}From these formulae, one verifies that $A(s\vert t,\, u)$ is indeed
independent
of the subtraction scale $\mu$, thus providing us at the same time with a
nontrivial check of our calculation.

Using the $O({\rm p}^2)$ expressions for the pseudoscalar masses derived from
${\tilde{\cal L}}^{(2)}$  \cite{FSS1,FSS2,SSF,K}, one obtains
\bea
{\widehat\epsilon} &=& 2\,{{r_2 -r}\over{r^2 -1}}\,(1+2\zeta )\,
\left[ 1\, +\, O(m_{\rm quark})\,\right]\ ,\nonumber\\
\quad\\
{\widehat\Delta}_{GMO} &=& (3M_{\eta}^2 -4M_K^2 +M_{\pi}^2)/M_{\pi}^2 \,
\left[ 1\, + \, O(m_{\rm quark})\,\right]\ ,\nonumber
\ena
with $r_2 = 2M_K^2/M_{\pi}^2 - 1 \sim 26$, $\zeta = Z_0^S/A_0$, which is
expected to be small due to the Zweig rule, and $r = m_s/{\widehat m}$. In
G$\chi$PT, the latter quark mass ratio remains a free parameter.

On the other hand, in the standard case, $r$ is determined order by order in
the chiral expansion. For instance, at leading order, $r=r_2$ (and
${\widehat\Delta}_{GMO}=0$), while at order $O({\rm p}^4)$, one obtains
 \cite{GL2}
\bea
&&\qquad\qquad r_{\rm st} \, =\, r_2 - 2 {{M_K^2}\over{\mpi^2}}{\Delta_M}\ ,
\nonumber\\
&&\lbl{rst}\\
&& \Delta_M \, =\, -\mu_{\pi} + \mu_{\eta} + {8\over\fpi^2}(M_K^2 - \mpi^2)
(2L_8^r(\mu ) - L_5^r(\mu ))\ .\nonumber
\ena
Furthermore, the
contributions from ${\cal L}_{(0,3)}$, ${\cal L}_{(2,2)}$ and ${\cal
L}_{(0,4)}$ are relegated beyond order $O({\rm p}^4)$.
The remaining constants $A_0$,
$Z_0^S$, $Z_0^P$, $\xi$ and ${\tilde\xi}$ are related to the
low-energy constants $L_i$ of ${\cal L}^{(4)}$, introduced by Gasser and
Leutwyler  \cite{GL2}, by (both sides of these equations refer to
the {\it renormalised} quantities, defined at the same scale $\mu$):
\bea
&&\qquad L_4^r = {F_0^2\over 8B_0}
\,{\tilde\xi}\ ,\ \,
L_5^r = {F_0^2\over 8B_0}
\, \xi\ ,\nonumber\\
&&\lbl{Li}\\
&&L_6^r = {F_0^2\over 16B_0^2}
\, Z_0^S\ ,\ \,
L_7^r = {F_0^2\over 16B_0^2}
\, Z_0^P\ ,\ \,
L_8^r = {F_0^2\over 16B_0^2}
\, A_0\ .\nonumber
\ena
Thus, upon replacing $r$ by its standard expression $r_{\rm st}$ above, and
upon keeping only those contributions that are of order $O({\rm p}^4)$ in the
standard
counting, one recovers from Eqs.  \rf{gAstu},  \rf{Fpi},  \rf{Mpi} and
 \rf{Li}, {\it as a special case}, the
expression of the one-loop $\pi\pi$ amplitude in the $SU(3)_{\rm L}\times
SU(3)_{\rm R}$
framework of the standard chiral expansion  \cite{Mei}:
\bea
\left. A(s\vert t,\, u)\right\vert_{\rm st}
&=& {{\beta^r_{\rm st}}\over{F_{\pi}^2}}\left(
                                s-{4\over 3}M_{\pi}^2 \right) +
\alpha^r_{\rm st}{{M_{\pi}^2}\over{3F_{\pi}^2}}\nonumber\\
&+& {4\over{F_{\pi}^4}}(2L_1^r + L_3 )(s-2M_{\pi}^2)^2\, + \,
 {4\over{F_{\pi}^4}}L_2^r\left[ (t-2M_{\pi}^2)^2 + (u-2M_{\pi}^2)^2
\right]\nonumber\\
&+& {1\over{2F_{\pi}^4}} \left[ s^2 - \mpi^4\right]\,
J^r_{\pi\pi}(s)\nonumber\\
&+& {1\over{4F_{\pi}^4}} \left[ (t-2M_{\pi}^2)^2\, J^r_{\pi\pi}(t)
\ +\ (u-2M_{\pi}^2)^2\, J^r_{\pi\pi}(u)\right]\nonumber\\
&+& {1\over{8F_{\pi}^4}}\, s^2\, J^r_{KK}(s)
\ +\ {1\over{18F_{\pi}^4}}\, \mpi^4\, J^r_{\eta\eta}(s)\lbl{sAstu}\\
&+& {1\over{2\fpi^4}}\,
\left\{ (s-u)t\left[ 2M^r_{\pi\pi}(t) + M^r_{KK}(t)\right]
                  +(s-t)u\left[ 2M^r_{\pi\pi}(u) + M^r_{KK}(u)\right]\right\}
\ ,\nonumber
\ena
with
\bea
\beta^r_{\rm st} &=& 1 + {{8\mpi^2}\over{\fpi^2}}
\left( 2L_4^r + L_5^r \right)\ ,\lbl{btst}\\
&&\nonumber\\
&&\nonumber\\
\alpha^r_{\rm st} &=& 1 - {{16\mpi^2}\over{\fpi^2}}\left( 2L_4^r + L_5^r
\right) + {{48\mpi^2}\over{\fpi^2}}\left( 2L_6^r + L_8^r \right)\ .\lbl{atst}
\ena
The scale dependences of $\alpha^r_{\rm st}$ and $\beta^r_{\rm st}$ are readily
obtained, either from Eqs.  \rf{atmu} and  \rf{btmu}, upon taking
${\widehat\epsilon}$ and
${\widehat\Delta}_{GMO}$ equal to zero, or directly from the scale dependences
of the $L_i$'s as given in Ref.  \cite{GL2}. Both ways lead to the same result:
\bea
\alpha^r_{\rm st}(\mu ) &=& \alpha^r_{\rm st}(\mu ')\ +
\ {{7M_{\pi}^2}\over{3(4\pi F_{\pi})^2}} \ln ({\mu '}/\mu ) \ ,
\nonumber\\
&&\\
\beta^r_{\rm st}(\mu ) &=& \beta^r_{\rm st}(\mu ')\ +
\ {{5M_{\pi}^2}\over{(4\pi F_{\pi})^2}} \ln ({\mu '}/\mu ) \ .\nonumber
\ena
Starting from Eq.  \rf{sAstu} above, it is then straightforward to further
reduce this
expression to the $SU(2)_{\rm L}\times SU(2)_{\rm R}$ case, and thus
to recover the result, now expressed in terms of the constants ${\bar l}_i$,
$i=1,2,3,4$,
first obtained by Gasser and Leutwyler  \cite{GLlett,GL1}. We shall not pursue
this matter for the moment.

\section{The Use of Analyticity, Unitarity and \hfill\break
Crossing Symmetry in $\chi$PT}

\indent
\setcounter{equation}{0}

In the previous section, we have obtained the $\pi\pi$ amplitude to order
$O({\rm p}^4)$ in the generalised chiral expansion, and have compared it to its
standard counterpart. At this stage, two features are worth being kept in mind:
First, in both cases, the amplitude depends on only four independent
combinations of low-energy constants of the respective $O({\rm p}^4)$
truncations of the (same) effective lagrangian. Secondly, the essential
difference between the two cases arises from the fact that the quark mass ratio
$r=m_s/{\widehat m}$ remains a free parameter at each order of the generalised
expansion. Upon fixing this ratio to its value predicted by the standard
expansion, one recovers the $\pi\pi$ amplitude in the latter framework
as a special case. These observations lead to the expectation that the form of
the $\pi\pi$ amplitude might also be obtained beyond the $O({\rm p}^4)$
approximation independently of any prejudice concerning the size of the
quark-antiquark condensate. In this section we shall establish the expression
of
the $\pi\pi$ amplitude up to corrections of order $O({\rm p}^8)$. More
phenomenological aspects, such as how the actual value of $m_s/{\widehat m}$
might actually be inferred from low-energy $\pi\pi$ data will be the subject of
subsequent sections.

It has indeed been shown  \cite{SSF} that the Goldstone nature of the pion,
combined with analyticity, crossing symmetry and unitarity of the S-matrix,
determines the low-energy $\pi\pi$ scattering amplitude up to six arbitrary
constants (not fixed by these general requirements)
and up to corrections of the order $O\left( [{\rm p}/{\Lambda}_H]^8\right)$.
(As usual, $\Lambda_H$
denotes the mass scale at which particles other than Goldstone bosons are
created, $\Lambda_H\sim 1$GeV .) The G$\chi$PT one-loop $O({\rm p}^4)$
amplitude obtained
in the preceding section has to be a particular case of this general
six-parametrical
formula. Moreover, since the general low energy representation of the amplitude
extends up to and including order $O({\rm p}^6)$ (actually $O({\rm p}^7)$
in the generalised
counting), it must in fact contain the whole two-loop $\chi$PT amplitude both
in
the standard and in the generalised settings. In this section, we
shall first rederive
the $O({\rm p}^4)$ amplitude using dispersion relations, and subsequently
these same dispersive
methods will be used to work out the two-loop amplitude explicitly.

The starting point is the following Theorem, proved in Ref.  \cite{SSF}:
For pion momenta ${\rm p}\ll\Lambda_H$, the $\pi\pi$ scattering amplitude can
be expressed as
\bea
{1\over{32\pi}}\, A(s\vert t,\, u)
&=& {1\over 3}\left[ W_0(s) - W_2(s)\right] \nonumber\\
 &+& {1\over 2}\left[ 3(s-u)W_1(t) + W_2(t)\right] \lbl{Astu}\\
 &+& {1\over 2}\left[ 3(s-t)W_1(u) + W_2(u)\right]
+ O\left([{\rm p}/{\Lambda}_H ]^8\right) \ .\nonumber
\ena
The three functions $W_a(s)$, $a=0, 1, 2$, are analytic except for a cut
singularity at
$s > 4\mpi^2$.
Their discontinuities across this cut are given by
\eq
{\rm Im}\, W_a(s) = (s-4\mpi^2)^{-\varepsilon_a} {\rm Im}\, f_a(s)\,
\theta (s-4\mpi^2)\ ,\lbl{ImW}
\en
where
\eq
\varepsilon_0 = \varepsilon_2 = 0\ ,\,\ \varepsilon_1 = 1\ ,
\en
and with $f_a(s)$, $a=0, 1, 2$, denoting the three lowest partial-wave
amplitudes $f_l^{\rm I} (s)$ (the  I = 0 and I = 2 S-waves, and the P-wave),
\eq
f_0(s)\equiv f_0^0(s)\ ,\,\ f_1(s)\equiv f_1^1(s)\ ,\,\ f_2(s)\equiv f_0^2(s)\
{}.
\en
(Notation, normalisation and other conventions are summarised in Appendix A
of Ref. \cite{SSF}.)

The proof of this Theorem has been given in Ref.  \cite{SSF} with
all details and it will not be reproduced here. Let us just recall that
the main ingredient of this proof is the suppression of inelasticities and of
higher partial waves in the chiral limit ${\rm p}\to 0$ , $\mpi\to 0$, ${\rm p}
/\mpi$ fixed:
Below the threshold of production of non-Goldstone particles, one can write
\eq
{\rm Im}\,f_a(s) = {\sqrt{{s-4\mpi^2}\over s}} \,
\vert f_a(s) \vert^2 + O\left( [{\rm p}/\Lambda_H ]^8 \right)\ .\lbl{Imf}
\en
Similarly, the absorptive parts of higher partial waves are suppressed as
\eq
{\rm Im}\,f_l^{\rm I}(s) = O\left( [{\rm p}/\Lambda_H ]^8 \right)\ ,
\ \ l\ge 2\ .\lbl{ge2}
\en
As discussed in Ref.  \cite{SSF}, the properties  \rf{Imf} and  \rf{ge2} are
rigourous consequences of the fact that the whole amplitude $A(s\vert t,\, u)$
behaves dominantly as $O({\rm p}^2)$ in the chiral limit, combined with
analyticity, unitarity, crossing symmetry and chiral counting. At the same
time, these two equations explain why it is difficult to extend the
representation  \rf{Astu}
beyond two loops: At order $O({\rm p}^8)$, inelasticities and higher partial
waves set in.

Given ${\rm Im}\, f_a  (s)$, then the four functions  $W_a(s)$, and
consequently the
whole amplitude $A(s\vert t,\,u)$, are determined up to a polynomial. The
maximum degree
of this polynomial may be fixed according to the chiral order $O({\rm p}^8)$
of the neglected contributions: Without loss of generality, the functions
$W_a(s)$ can thus be further specified by the asymptotic conditions
\eq
\lim_{\vert s\vert\to\infty}\ s^{\varepsilon_a -4}\,W_a(s) = 0\ .\lbl{bd}
\en
It then follows that for a given set of imaginary parts ${\rm Im}\, f_a(s)$
consistent with the bounds  \rf{bd},
the amplitude $A(s\vert t,\, u)$ is determined up to six arbitrary constants.

It is worth stressing that in spite of Eq.  \rf{ImW}, the functions $W_a (s)$
do not coincide with the corresponding partial wave amplitudes $f_a(s)$. The
latter are obtained by the standard partial wave projections of  Eq.
 \rf{Astu},
\bea
f_l^0(s) &=& {1\over{32\pi}}\, \int_{4\mpi^2 -s}^0\,{{dt}\over{s-4\mpi^2}}\,
P_l\left( 1+{{2t}\over{s-4\mpi^2}}\right)\,
\left\{ 3A(s\vert t,\, u) + A(t\vert s,\,u) + A(u\vert s,\,t) \right\}
\nonumber\\
& &\nonumber\\
f_l^1(s) &=& {1\over{32\pi}}\, \int_{4\mpi^2 -s}^0\,{{dt}\over{s-4\mpi^2}}\,
P_l\left( 1+{{2t}\over{s-4\mpi^2}}\right)\,
\left\{ A(t\vert s,\,u) - A(u\vert s,\,t) \right\}\lbl{pwi}\\
& &\nonumber\\
f_l^2(s) &=& {1\over{32\pi}}\, \int_{4\mpi^2 -s}^0\,{{dt}\over{s-4\mpi^2}}\,
P_l\left( 1+{{2t}\over{s-4\mpi^2}}\right)\,
\left\{ A(t\vert s,\,u) + A(u\vert s,\,t) \right\}\ ,\nonumber
\ena
where the integrals are taken with $s$ kept fixed and with $u$ reexpressed as
$u = 4\mpi^2-s-t$.
The functions $W_a(s)$ have only a right-hand cut for $s>4\mpi^2$, whereas the
partial wave amplitudes $f_a(s)$  \rf{pwi} do have both a right-hand cut
($s>4\mpi^2$)
and a left-hand cut  ($s<0$), with the correct discontinuities, as dictated by
unitarity and crossing symmetry. Actually, the partial waves with $l\ge 2$
obtained from  \rf{Astu} are
real, in agreement with Eq.  \rf{ge2}.  The formula  \rf{Astu} may
be viewed as the most
general solution of the constraints imposed by analyticity, unitarity and
crossing
symmetry up to and including the chiral order $O({\rm p}^6)$ ($O({\rm p}^7)$ in
the generalised case). In practice, the representation  \rf{Astu} should be a
good approximation within a range of energies where the inelasticities in
 \rf{Imf} and the contributions of higher partial waves in  \rf{ge2} remain
small. Let us recall that the inelasticities, though in principle present
above 540 MeV, are actually observed to be very small even up to 1 GeV.
The same remark applies to the higher partial waves, $\ell \ge 2$.

\subsection{ First Iteration of the Unitarity Condition}

\indent

              The dispersion relation approach to $\chi$PT is based on the
iteration of the unitarity condition. Here, the latter will be considered  in
the elastic form  \rf{Imf}, i.e. below the ${\bar K}K$ and $\eta \eta$
thresholds. This does
not mean that we have to renounce  the advantages of an $SU(3)_{\rm L}\times
SU(3)_{\rm R}$ analysis of $\pi\pi$
scattering, but rather that the $K$ and $\eta$ loops will be expanded
in powers of $s/(2M_K)^2$ and $s/(2M_{\eta})^2$, respectively.
Since in practice $2M_K$ and
$2M_{\eta}$ are of the order of $\Lambda_H$, this should not affect the
accuracy of our results at low energies  $E\ll \Lambda_H$.

Chiral symmetry implies that the low-energy expansion of
the $\pi\pi$ amplitude starts at order $O({\rm p}^2)$.
Analyticity and crossing symmetry then restrict its form to
\eq
A(s\vert t,\,u) =  \alpha {{\mpi^2}\over{3\fpi^2}} +
{{\beta}\over{\fpi^2}}\left(s-{4\over{3}}\mpi^2\right) + O({\rm p}^4)\
,\lbl{Ap2}
\en
with $\alpha$ and $\beta$  two constants whose  dominant behaviour in the
chiral limit is of order $O(1)$.
In G$\chi$PT, where
odd chiral orders also occur, the constants $\alpha$ and $\beta$ in Eq.
 \rf{Ap2} consist
of both $O(1)$ and $O({\rm p})$ contributions, the latter being
proportional to the
first power of the quark masses. The Goldstone nature of the pions
together with the
general properties of their scattering amplitude, by
themselves, do not fix the
constants $\alpha$ and $\beta$. On the other hand, the values of
these constants are
actually at the center of our interest. They reflect the mechanism of
spontaneous chiral symmetry breakdown in
QCD and the corresponding chiral structure of the ground state: For instance,
in the limit of a large quark condensate ($B_0\sim\Lambda_H$,
the special case of standard
$\chi$PT), one has  $\alpha=1$, $\beta=1$, whereas in the opposite limit of a
vanishing condensate, $\alpha=4$ and $\beta =1$ at order $O({\rm p}^2)$.
It is important that the dispersive approach to
$\chi$PT and, in particular, the iteration of the unitarity condition, only
assumes
that $\alpha$ and $\beta$ are of the order $O(1)$ in the chiral counting,
but otherwise proceeds independently
of any particular values of these constants.

Equation  \rf{Ap2} implies for the three lowest partial wave $f_a(s)$
, $a=0,1,2$,
\eq
{\rm Re}\, f_a(s) = \varphi_a(s) + O({\rm p}^4)\ ,\,
\ {\rm Im}\, f_a(s) = O({\rm p}^4)\ ,\lbl{faLO}
\en
where
\bea
\varphi_0(s) &=& {1\over{96\pi\fpi^2}}\,
\left\{ 6\beta\, (s-{4\over 3}\mpi^2) + 5\alpha\mpi^2\right\}\nonumber\\
\varphi_1(s) &=& {1\over{96\pi\fpi^2}}\,\beta\, (s-4\mpi^2)
\lbl{phia}\\
\varphi_2(s) &=& {1\over{96\pi\fpi^2}}\,
\left\{ -3\beta\, (s-{4\over 3}\mpi^2) + 2\alpha\mpi^2\right\}\ .\nonumber
\ena
Inserting this information into the unitarity condition  \rf{Imf}, one obtains
\eq
{\rm Im}\,f_a(s) = {\sqrt{{s-4\mpi^2}\over s}}
\vert \varphi_a(s) \vert^2 + O({\rm p}^6)\ ,\lbl{Imf2}
\en
which are the discontinuities of the functions $W_a(s)$ up to
and including the order $O({\rm p}^{4-2\varepsilon_a})$\hfill\break
($O({\rm p}^{5-2\varepsilon_a})$ in G$\chi$PT).
Due to the polynomial
character of $\varphi_a(s)$, Eqs.  \rf{ImW} and  \rf{Imf2} allow one
to express the functions $W_a(s)$ as
\eq
W_a(s) = 16\pi\, (s-4\mpi^2)^{-\varepsilon_a}\,
\left[\varphi_a(s)\right]^2\,{\bar J}(s) + {\rm polynomial} + O({\rm
p}^{6-2\varepsilon_a})\ ,\lbl{W1}
\en
where ${\bar J}(s)$ denotes the dispersion integral
\eq
{\bar J}(s) = {s\over{16\pi^2}}\, \int_{4\mpi^2}^{\infty}\,
{{dx}\over x}\, {1\over{x-s}}\, \sqrt{{{x-4\mpi^2}\over x}}\ .\lbl{J}
\en
The reader will easily recognise in the analytic function ${\bar J}(s)$
the standard
scalar-loop integral $J_{\pi\pi}^r(s)$ of the previous Section,
subtracted at zero momentum
transfer, ${\bar J}_{\pi\pi}(s) = J^r_{\pi\pi}(s)-J^r_{\pi\pi}(0)$,
(see Appendix C).
The polynomial in Eq.  \rf{W1} is at most of degree $2-\varepsilon_a$ in $s$.
The coefficients
of higher powers of $s$ should be well behaved in the limit $\mpi\to 0$ ,
for $s$ fixed,
and consequently, such higher powers can be absorbed into the neglected
$O({\rm p}^{6-2\varepsilon_a})$ remainders.
Notice that in the case of Eq.  \rf{W1}, the asymptotic bounds  \rf{bd} are
not only satisfied, but they are
not even saturated (${\bar J}(s)$ grows as $\ln s$ ).
Returning to the formula  \rf{Astu},
the polynomial contributions in $W_a(s)$ beyond the leading contributions
 \rf{Ap2} can be collected into
\bea
\delta\, A(s\vert t,\, u) &=& \delta\alpha {{\mpi^2}\over{3\fpi^2}}
+ {{\delta\beta}\over{\fpi^2}}\left(s-{4\over{3}}\mpi^2\right)\nonumber\\
&+& {{\lambda_1}\over{\fpi^4}}(s-2\mpi^2)^2 + {{\lambda_2}\over{\fpi^4}}
\left[ (t-2\mpi^2)^2 + (u-2\mpi^2)^2 \right]
\ .\lbl{delA}
\ena
Here, in the chiral limit, the dominant behaviour of $\delta\alpha$,
$\delta\beta$, $\lambda_1$ and $\lambda_2$ is
\eq
\delta\alpha \ =\  O({\rm p}^2)\ ,\ \ \delta\beta\  =\ O({\rm p}^2)\ ,
\ \ \lambda_{1,2}\ =\ O(1)\ .
\en
It is convenient to absorb $\delta\alpha$ and $\delta\beta$ into the constants
$\alpha$ and $\beta$ which
characterise the leading order contribution  \rf{Ap2}. The constants $\alpha$
and $\beta$
so redefined by including higher-order contributions will keep their original
names. Notice that in the expression  \rf{W1} for $W_a(s)$, the redefinitions
$\alpha\to \alpha -\delta\alpha$, $\beta\to \beta -\delta\beta$ induce terms
of higher chiral
orders, which can be reabsorbed into the neglected $O({\rm
p}^{6-2\varepsilon_a})$ remainders.
Hence, the whole result of the first iteration of the unitarity condition can
be expressed in terms of four parameters, $\alpha$, $\beta$, $\lambda_1$ and
$\lambda_2$:
\bea
A(s\vert t,\,u)&=& {{\beta}\over{\fpi^2}}\left(s-{4\over{3}}\mpi^2\right) +
\alpha {{\mpi^2}\over{3\fpi^2}}\nonumber\\
&+&{{\lambda_1}\over{\fpi^4}}( s-2\mpi^2 )^2
+{{\lambda_2}\over{\fpi^4}}\left[ ( t-2\mpi^2 )^2 + (u-2\mpi^2 )^2 \right]
\lbl{amp}\\
&+& {\bar J}_{(\alpha, \beta )}(s\vert t,\,u) + O({\rm p}^6/\Lambda_H^6 )
\ ,\nonumber
\ena
where
\bea
{\bar J}_{(\alpha, \beta )}(s\vert t,\,u) &=& {1\over{6\fpi^4}}\left\{
4\left[ \beta (s-{4\over{3}}\mpi^2 ) + {5\over{6}}\alpha\mpi^2 \right]^2 -
\left[  \beta (s-{4\over{3}}\mpi^2 ) - {2\over{3}}\alpha\mpi^2 \right]^2
\right\} {\bar J}(s)\nonumber\\
&+& {1\over{12\fpi^4}}\left\{ 3\left[ \beta (t-{4\over{3}}\mpi^2 )
- {2\over{3}}\alpha\mpi^2 \right]^2 +
{\beta}^2 (s-u)(t-4\mpi^2 )\right\} {\bar J}(t)\lbl{Jab}\\
&+& {1\over{12\fpi^4}}\left\{ 3\left[ \beta (u-{4\over{3}}\mpi^2 )
- {2\over{3}}\alpha\mpi^2 \right]^2 +
{\beta}^2 (s-t)(u-4\mpi^2 )\right\} {\bar J}(u)\ .\nonumber
\ena

\subsection{ Comparing Perturbative $O({\rm p}^4)$
and Dispersive $\chi$PT Formulae}

\indent

The parameters $\alpha$, $\beta$, $\lambda_1$ and $\lambda_2$ in Eqs.  \rf{amp}
 and  \rf{Jab} have their own expansions in powers of quark masses and
they involve chiral logarithms. One has
\eq
\alpha = \sum_{n=0}^3 \alpha^{(n)},~~~ \
\beta = \sum_{n=0}^3 \beta^{(n)},~~~ \
\lambda_{1,2} = \sum_{n=0}^1 \lambda_{1,2}^{(n)}\ ,\lbl{mexp}
\en
where $\alpha^{(n)}$ denotes the $O({\rm p}^n)$ component of $\alpha$ and
likewise for the other constants. In the {\it standard} $\,\chi$PT, the odd
orders are absent, $\alpha^{(0)}=\beta^{(0)} =1$,
and the formula  \rf{amp} contains $O({\rm p}^2)$ and $O({\rm p}^4)$
parts only. In {\it generalised} $\chi$PT, the odd orders show up,
$\beta^{(0)}=1$, and $1\le \alpha^{(0)} \le 4$,
whereas the amplitude  \rf{amp} involves $O({\rm p}^2)$, $O({\rm p}^3)$,
$O({\rm p}^4)$ and $O({\rm p}^5)$
contributions before it reaches the (neglected) order $O({\rm p}^6)$. Detailed
discussion of the values of the constants $\alpha$, $\beta$, $\lambda_{1,2}$
will be given in Sections 4 and 5 .

It is now easy to show that the $O({\rm p}^4)$ result  \rf{gAstu} obtained
by a
direct G$\chi$PT evaluation of one-loop graphs, starting from the Lagrangian
${\tilde{\cal L}}^{(2)} + {\tilde{\cal L}}^{(3)} + {\tilde{\cal L}}^{(4)}$,
is indeed of the form of Eqs.  \rf{amp} and  \rf{Jab}.
To see this, one should first reexpress the scale dependent
renormalised pion loop integrals $J^r_{\pi\pi}(s)$ and $M^r_{\pi\pi}(s)$
in terms of the scale independent function ${\bar J}(s)$:
\bea
J^r_{\pi\pi}(s) &=& {\bar J}(s)\, -\, {1\over{16\pi^2}}\left( \ln\,
{{\mpi^2}\over{\mu^2}} + 1 \right)\ ,\nonumber\\
& &\\
M^r_{\pi\pi}(s) &=& {1\over{12s}}\, (s-4\mpi^2)\, {\bar J}(s)\, -\,
{1\over{192\pi^2}}\left( \ln\,{{\mpi^2}\over{\mu^2}} + {1\over 3}\right)\ .
\nonumber
\ena
Next, the $K$ and  $\eta$ loop contributions $J^r_{KK}(s)$, $M^r_{KK}(s)$ and
$J^r_{\eta\eta}(s)$ have to
be expanded in powers of  $s/(2M_K)^2$ and $s/(2M_{\eta})^2$, respectively.
Keeping only terms that contribute up to and including the order $O({\rm
p}^4)$, one has
\bea
J^r_{PP}(s) &=& - {1\over{16\pi^2}}\left( \ln\,
{{M_P^2}\over{\mu^2}} + 1 \right)\, +\, O(s/4M_P^2)\ ,\ \,
P=K,\, \eta\ ,\nonumber\\
& &\lbl{exp}\\
M^r_{KK}(s) &=& - {1\over{192\pi^2}}\left( \ln\,
{{M_K^2}\over{\mu^2}} + 1 \right)\, +\, O(s/4M_K^2)\ .\nonumber
\ena
Let us stress again at this point that it is perfectly possible to
recover the full
$K$ and
$\eta$ loop contributions within the dispersive approach: for this, it is
sufficient to include the ${\bar K}K$ and $\eta \eta$ intermediate states
into the unitarity
condition. Here we do not present this extended analysis for reasons of
simplicity,
and because performing the expansions  \rf{exp} in Eq.  \rf{gAstu}
should be a fairly good approximation at low energies.

Finally, in order to compare the two expressions, all
contributions beyond the order $O({\rm p}^4)$ in Eqs.  \rf{amp} and
 \rf{Jab}, {\it i.e.} all
$O({\rm p}^5)$ contributions, should be dropped.
In particular, only the $O({\rm p}^4)$ part
of the loop function ${\bar J}_{(\alpha,\beta)}(s\vert t,\, u)$ should be
maintained. This amounts to replacing ${\bar J}_{(\alpha,\beta)}(s\vert t,\,
u)$ by ${\bar J}_{(\alpha^{(0)},\beta^{(0)})}(s\vert t,\, u)$, where
\eq
\alpha^{(0)} = 1 + 3{\widehat \epsilon}\ ,\ \, \beta^{(0)} = 1\ ,\lbl{lead}
\en
are the leading parts of the constants $\alpha$ and $\beta$,
as can be inferred from Eqs.  \rf{bt},  \rf{at}.
Then one finds that Eqs.  \rf{gAstu} and  \rf{amp} are indeed the same,
upon making the following identifications
(${\widehat\epsilon}$ and ${\widehat\Delta}_{GMO}$
were defined in Eqs.  \rf{epsi} and  \rf{GMO})
\bea
\alpha &=& {\alpha^r}(\mu )
\ -\ {{M_{\pi}^2}\over{32\pi^2\fpi^2}}
\left( \ln\, {{M_K^2}\over{\mu^2}}\, + 1\right)
\,  \left[ 1 + (1+r){\widehat \epsilon}\right]\left[ 1 +3(1+r){\widehat
\epsilon}\right]\nonumber\\
&&\quad\quad  -\ {{M_{\pi}^2}\over{96\pi^2 \fpi^2}}
\left( \ln\, {{M_{\eta}^2}\over{\mu^2}}\, + 1\right)
\, \left[ 1 + (1+2r){\widehat \epsilon} + 2 {{{\widehat
\Delta}_{GMO}}\over{1-r}}\right] ^2 \lbl{a}\\
&&\quad\quad -\ {{M_{\pi}^2}\over{32\pi^2 \fpi^2}}
\left( \ln\, {{M_{\pi}^2}\over{\mu^2}}\, + 1\right)
\, \left[ 1 + 22{\widehat \epsilon} +33{\widehat \epsilon}^2 \right]\ ,
\nonumber \\
\nonumber\\
\beta &=& {\beta^r}(\mu )
\ -\ {{M_{\pi}^2}\over{16\pi^2\fpi^2}}
\left( \ln\, {{M_{\pi}^2}\over{\mu^2}}\, + 1\right)
\left[ 2+ 5{\widehat \epsilon}\right]\nonumber\\
&&\quad\quad -\ {{M_{\pi}^2}\over{32\pi^2 \fpi^2}}
\left( \ln\, {{M_{K}^2}\over{\mu^2}}\, + 1\right)
\left[ 1+ (1+r){\widehat \epsilon}\right]\ , \lbl{b}
\ena
and
\bea
\lambda_1 \ =\ \lambda_1^{(0)} &=& 4\,(2L_1^r(\mu )+L_3 )
-{1\over{48\pi^2}}\,\left\{\ln{{M_{\pi}^2}\over{\mu^2}} +
{1\over{8}}\ln{{M_K^2}\over{\mu^2}} + {35\over{24}}\right\}\ ,\nonumber\\
& &\lbl{lambda}\\
\lambda_2 \ =\ \lambda_2^{(0)} &=& 4\,L_2^r(\mu )
-{1\over{48\pi^2}}\,\left\{\ln{{M_{\pi}^2}\over{\mu^2}} +
{1\over{8}}\ln{{M_K^2}\over{\mu^2}} + {23\over{24}}\right\}\ .\nonumber
\ena
Upon using Eqs.  \rf{atmu} and  \rf{btmu} and the scale dependences of the
renormalised constants $L_1^r$ and $L_2^r$, it is straightforward to
ascertain the
scale independence of $\alpha$, $\beta$, $\lambda_1$ and $\lambda_2$.

At the end of Section 2 it was shown how the standard
$SU(3)_{\rm L}\times SU(3)_{\rm R}$
$O({\rm p}^4)$ amplitude appears as a particular case
of Eq.  \rf{gAstu}: It
may be expressed in terms of the standard $O({\rm p}^4)$ constants $L_1$,...
$L_8$.
The expressions  \rf{lambda} for
$\lambda_1$ and for $\lambda_2$ remain unchanged, while one finds that $\alpha$
and $\beta$ become ($\alpha_{\rm st}^{(0)}=\beta_{\rm st}^{(0)}=1$)
\bea
\alpha_{\rm st}\, =\, 1 &-& 16\,{{M_\pi^2}\over{F_\pi^2}}(2L_4^r + L_5^r ) +
                48\,{{M_\pi^2}\over{F_\pi^2}}(2L_6^r + L_8^r )\nonumber\\
 &-&{1\over{32\pi^2}}\,{{M_\pi^2}\over{F_\pi^2}}
\left\{ \ln {{M_\pi^2}\over{\mu^2}} + \ln {{M_K^2}\over{\mu^2}} +
{1\over 3}\,\ln {{M_\eta^2}\over{\mu^2}} + {7\over 3}\right\}\lbl{ast}\\
\beta_{\rm st}\,=\, 1 &+& 8\, {{M_\pi^2}\over{F_\pi^2}}(2L_4^r + L_5^r )
\nonumber\\
&-&{1\over{32\pi^2}}\,{{M_\pi^2}\over{F_\pi^2}}
\left\{ 4\,\ln {{M_\pi^2}\over{\mu^2}} + \ln {{M_K^2}\over{\mu^2}} + 5 \right\}
\ .\lbl{bst}
\ena

If we use for the constants $L_i$ the values currently given in the literature
 \cite{BCG} and take $\fpi = 92.4$ MeV, $\mpi = 139.57$ MeV, we obtain,
for the corresponding scale independent constants $\alpha$, $\beta$,
$\lambda_1$ and $\lambda_2$
the following values at order $O({\rm p}^4)$ in the standard
case:
\bea
\alpha_{\rm st} &=& 1.04\pm 0.15\nonumber\\
\beta_{\rm st} &=& 1.08\pm 0.03\nonumber\\
&&\lbl{stnum}\\
\lambda_{1,{\rm st}} &=& (-6.4\pm6.8)\times 10^{-3}\nonumber\\
\lambda_{2,{\rm st}} &=& (10.8\pm 1.2)\times 10^{-3}\ .\nonumber
\ena
Similarly, the standard $SU(2)_{\rm L}\times SU(2)_{\rm R}$ $O({\rm p}^4)$
amplitude originally given by Gasser and Leutwyler  \cite{GLlett,GL1}
in terms of the four scale independent parameters ${\bar l}_1$, ${\bar l}_2$,
${\bar l}_3$ and ${\bar l}_4$ can be put into the
form  \rf{amp} and  \rf{Jab}, with the identifications
\bea
\alpha_{{\rm GL}} &=& 1 + {1\over{32\pi^2}}\, {{\mpi^2}\over{\fpi^2}}\,
\left( 4{\bar l}_4 - 3{\bar l}_3 -1 \right)\ , \lbl{aGL} \\
& &\nonumber\\
& &\nonumber\\
\beta_{{\rm GL}} &=& 1 + {1\over{8\pi^2}}\, {{\mpi^2}\over{\fpi^2}}\,
\left( {\bar l}_4 - 1 \right)\ , \lbl{bGL}
\ena
and
\eq
\lambda_{1,{\rm GL}} \, =\, {1\over{48\pi^2}}\,
\left( {\bar l}_1 -{4\over 3} \right)
\ ,\ \,
\lambda_{2,{\rm GL}} \, =\, {1\over{48\pi^2}}\,\left(
{\bar l}_2 -{5\over 6} \right)\lbl{lGL}
\ .
\en
Using the numerical values recently updated by Gasser in  \cite{Ghbk}
\eq
{\bar l}_1 = -2.15 \pm 4.29\ ,\ \,
{\bar l}_2 = 5.84 \pm 1.72\ ,\ \,
{\bar l}_3 = 2.9 \pm 2.4\ ,\ \,
{\bar l}_4 = 4.55 \pm .29\ ,
\en
one obtains
\bea
\alpha_{{\rm GL}} &=& 1.06 \pm 0.06 \nonumber\\
\beta_{{\rm GL}} &=& 1.103 \pm 0.008 \nonumber\\
& &\lbl{GLnum}\\
\lambda_{1,{\rm GL}} &=& (-7.35 \pm 9.06)\times 10^{-3}\nonumber\\
\lambda_{2,{\rm GL}} &=& (10.57 \pm 3.63)\times 10^{-3}\ .\nonumber
\ena
It thus appears clearly that the situation of a large quark
condensate, {\it i.e.} the standard version of $\chi$PT, is characterised by
the values of the parameters $\alpha$ and $\beta$ remaining close to unity.

\subsection{ The Two-Loop $\pi\,\pi$ Amplitude}

\indent

We now proceed to the second iteration of the unitarity
condition  \rf{Imf}.
The starting point is the result of the previous step, {\it i.e.}
the one-loop amplitude  \rf{amp}.
First, we use Eqs.  \rf{pwi} to
calculate the S- and P-wave projections of the amplitude  \rf{amp}.
All the integrals are elementary and the result can be represented as
\eq
{\rm Re}\, f_a(s) = \varphi_a(s) + \psi_a(s) + O({\rm p}^6)\ ,\lbl{faNLO}
\en
where $\varphi_a(s)$ are the polynomials  \rf{phia} and the new contributions
$\psi_a(s)$, which
will be given shortly, are dominantly of order $O({\rm p}^4)$.
The unitarity condition  \rf{Imf} now
allows us to push the knowledge of ${\rm Im}\, f_a(s)$ one step further:
\eq
{\rm Im}\,f_a(s) = {\sqrt{{s-4\mpi^2}\over s}}\left\{
\vert \varphi_a(s) \vert^2 + 2 \varphi_a(s)\psi_a(s) \right\} + O({\rm p}^8)
\ .\lbl{Imf4}
\en
It is convenient to express the functions $\psi_a(s)$ arising from the partial
wave projections in the following way   ($s> 4\mpi^2$):
\eq
\psi_a(s) = {{\mpi^4}\over{\fpi^4}}\,
\sqrt{{s\over{s-4\mpi^2}}}\, \sum_{n=0}^4\,
\xi_a^{(n)}(s)\,  k_n(s)\ .\lbl{psia}
\en
Here, $\xi_a^{(n)}(s)$ are {\it polynomials} in $s$
of at most second degree, with coefficients given in terms of
$\alpha$, $\beta$, $\lambda_1$ and $\lambda_2$. They are listed in Appendix B.
$k_n(s)$ represent the following set of elementary functions
\bea
&&\quad k_0(s) = {1\over{16\pi}}\, \sqrt{{{s-4\mpi^2}\over s}}\ ,\ \,
        k_1(s) = {1\over{8\pi}}\, L(s)\ ,\nonumber\\
&&\nonumber\\
&&\quad k_2(s) = {1\over{8\pi}}\,\left( 1-{{4\mpi^2}\over s}\right)\,L(s)\ ,\
\, k_3(s) = {3\over{16\pi}}\,{{\mpi^2}\over{\sqrt{s(s-4\mpi^2)}}}\,L^2(s)\ ,
\lbl{kn}\\
&&\nonumber\\
&& k_4(s) = {1\over{16\pi}}\,{{\mpi^2}\over{\sqrt{s(s-4\mpi^2)}}}\,
\left\{ 1 + \sqrt{{s\over{s-4\mpi^2}}}\, L(s) + {{\mpi^2}\over{s-4\mpi^2}}\,
L^2(s)\right\}\ ,\nonumber
\ena
where
\eq
L(s) = \ln {{1- \sqrt{1- {{4\mpi^2}\over s}}}\over
                     {1+ \sqrt{1- {{4\mpi^2}\over s}}}}\ \, ,\ \ s>4\mpi^2\ .
\lbl{L}
\en
Notice that in the chiral limit ($s\to 0$, $\mpi^2\to 0$, $s/\mpi^2$ fixed)
all functions $k_n(s)$ are of order $O(1)$.
Furthermore,
they are all asymptotically bounded by $\ln s$.
Eqs.  \rf{faNLO} and  \rf{Imf4} now lead
to the  following expressions for ${\rm Im}\, W_a(s)$
\eq
{\rm Im}\, W_a(s) = \sum_{n=0}^4\, w_a^{(n)}(s)\, k_n(s) + O({\rm
p}^{8-2\varepsilon_a})\ ,
\en
where
\eq
w_a^{(n)}(s) = (s - 4\mpi^2)^{-\varepsilon_a}\,
\left\{\, 16\pi\, \left[\varphi_a(s)\right] ^2\,\delta_{n0}\, +
\, {{2\mpi^4}\over{\fpi^4}}\varphi_a(s)
\xi_a^{(n)}(s)\right\}\lbl{wa}
\en
are polynomials of degree $3 - \varepsilon_a$ in $s$.
(Remember that $\varphi_1(s)$ is proportional to $s-4\mpi^2$.)
Consequently,
\eq
W_a(s) = \sum_{n=0}^4 w_a^{(n)}(s) {\bar K}_n(s) +
{\rm polynomial} + O({\rm p}^{8-2\varepsilon_a})\ ,\lbl{W2}
\en
where ${\bar K}_n(s)$ denote the dispersion integrals
\eq
{\bar K}_n(s) = {s\over{\pi}}\, \int_{4\mpi^2}^{\infty}\,
{{dx}\over x}\, {{\ \ k_n(x)\ \ }\over{x-s}}\ .\lbl{Kn}
\en
The formula  \rf{W2} contains and extends the one-loop result  \rf{W1}: By
definition,
\eq
{\bar K}_0(s) = {\bar J}(s) \ ,
\en
and the leading $O({\rm p}^4)$ part in $w_a^{(0)}(s)$ which contributes to Eq.
 \rf{W1} is
manifest in the expression  \rf{wa}.

As for the polynomial part of Eq.  \rf{W2}, its degree in $s$
can  be at most $3-\varepsilon_a$: The coefficient of any power of $s$
higher than $3-\varepsilon_a$
should not blow up in the limit $\mpi \to 0$, $s$ fixed,
and consequently, any such
power can be absorbed into the neglected contribution $O({\rm
p}^{8-2\varepsilon_a})$.

The functions $W_a(s)$
then satisfy the bounds  \rf{bd}, since the functions ${\bar K}_n(s)$  grow
at most as
$(\ln s)^2$. This means that Eq.  \rf{W2} determines the amplitude
$A(s\vert t,\, u)$ up
to a general crossing symmetric polynomial $\delta\, A(s\vert t,\, u)$
of at most cubic order.
The most general polynomial of this type contains six arbitrary parameters
and it can be written as
\bea
\delta\, A(s\vert t,\, u) &=& \delta\alpha {{\mpi^2}\over{3\fpi^2}}
+ {{\delta\beta}\over{\fpi^2}}\left(s-{4\over{3}}\mpi^2\right)\nonumber\\
&+& {{\delta\lambda_1}\over{\fpi^4}}(s-2\mpi^2)^2
+ {{\delta\lambda_2}\over{\fpi^4}}
\left[ (t-2\mpi^2)^2 + (u-2\mpi^2)^2 \right]\lbl{delA2}\\
&+& {{\lambda_3}\over{\fpi^6}}(s-2\mpi^2)^3
+ {{\lambda_4}\over{\fpi^6}}
\left[ (t-2\mpi^2)^3 + (u-2\mpi^2)^3 \right]\ . \nonumber
\ena
This formula represents the polynomial contribution to $A(s\vert t,\, u)$
beyond the
one-loop order  \rf{amp}. In the chiral limit, the six constants involved in
Eq.  \rf{delA2} {\it dominantly} behave as
\eq
\delta\alpha ,\ \delta\beta\ = \ O({\rm p}^4)\ ,\ \,
\delta\lambda_{1,2}\ =\ O({\rm p}^2)\ ,\ \,
\lambda_{3,4}\ =\ O(1)\ .
\en
Furthermore $\delta\alpha$, $\delta\beta$, $\delta\lambda_1$
and $\delta\lambda_2$ can
be absorbed into the
parameters $\alpha$, $\beta$, $\lambda_1$ and $\lambda_2$, respectively, which
characterise the one loop result  \rf{amp}.
This redefinition simply extends the
expansion  \rf{mexp} in powers of quark masses and
consequently the parameters $\alpha$, $\beta$ and $\lambda_{1,2}$ so
redefined will be referred to by their
original names. Notice that in the expressions  \rf{W2} for the functions
$W_a(s)$, the
redefinitions $\alpha \to \alpha - \delta\alpha$, $\beta \to \beta
-\delta\beta$ and $\lambda_{1,2} \to \lambda_{1,2} -\delta\lambda_{1,2}$
induce terms of higher chiral order which can be reabsorbed into the neglected
$O({\rm p}^{8-2\varepsilon_a})$ remainders.
This implies that the whole $\pi\pi$ scattering amplitude up
to and including two loops can be expressed in terms of only six parameters,
and the result reads:
\bea
A(s\vert t,\,u)&=& {{\beta}\over{\fpi^2}}\left(s-{4\over{3}}\mpi^2\right) +
\alpha {{\mpi^2}\over{3\fpi^2}}\nonumber\\
&+&{{\lambda_1}\over{\fpi^4}}( s-2\mpi^2 )^2
+{{\lambda_2}\over{\fpi^4}}\left[ ( t-2\mpi^2 )^2 + (u-2\mpi^2 )^2 \right]
\lbl{amp2}\\
&+&{{\lambda_3}\over{\fpi^6}}( s-2\mpi^2 )^2
+{{\lambda_4}\over{\fpi^6}}\left[ ( t-2\mpi^2 )^3 + (u-2\mpi^2 )^3 \right]
\nonumber\\
&+& {\bar K}(s\vert t,\,u) + O({\rm p}^8/\Lambda_H^8 )\ , \nonumber
\ena
where
\bea
{\bar K}(s\vert t,\, u) &=& 32\pi\, \sum_{n=0}^4 \, \left\{
{1\over 3}\left[ w_0^{(n)}(s)-w_2^{(n)}(s)\right]\,
{\bar K}_n(s)\right. \nonumber\\
&+& {1\over 2}\left[ w_2^{(n)}(t) + 3(s-u)w_1^{(n)}(t) \right]
\,{\bar K}_n(t)\lbl{Kstu}\\
&+& \left. {1\over 2}\left[ w_2^{(n)}(u) + 3(s-t)w_1^{(n)}(u) \right]
\,{\bar K}_n(u)
\right\}\ .\nonumber
\ena

The one-loop contribution ${\bar J}_{(\alpha ,\beta )}(s\vert t,\, u)$,
{\it cf.} Eq.  \rf{Jab}, is hidden in the $n=0$
part of the sum  \rf{Kstu} : It arises from the first terms, proportional to
$\fpi^{-4}$, in  \rf{wa}. The terms proportional to
$\fpi^{-6}$ in the polynomials $w_a^{(n)}(s)$ represent the $O({\rm p}^6)$
(and $O({\rm p}^7)$) contributions:
The parts linear in $\lambda_1$ and $\lambda_2$ correspond to
single loop graphs with
one $O({\rm p}^4)$ vertex, while the remaining parts, which are cubic
in $\alpha$ and $\beta$, give rise to the genuine two-loop contributions.

             If one wishes to identify a given chiral order in the general
formula  \rf{amp2}, the parameters $\alpha$, $\beta$, $\lambda_i$,
$i=1,2,3,4$, have to be expanded in
powers of quark masses. The only difference with the previous expansion
 \rf{mexp},
is that now more terms have to be included, until one reaches the first
neglected order $O({\rm p}^8)$. Hence
\eq
\alpha = \sum_{n=0}^5 \alpha^{(n)}\, ,\
\beta = \sum_{n=0}^5 \beta^{(n)}\, ,\
\lambda_{1,2} = \sum_{n=0}^3 \lambda_{1,2}^{(n)}\, ,\
\lambda_{3,4} = \sum_{n=0}^1 \lambda_{3,4}^{(n)}\ ,\lbl{mexp2}
\en
where $\alpha^{(n)} = O({\rm p}^n)$ and likewise for the other constants.
The particular case of
the standard $\chi$PT is again characterised by the absence of odd chiral
orders.

              Finally, we turn to the dispersion integrals ${\bar
K}_n(s)$  \rf{Kn}
which occur in the two loop formula  \rf{amp2},  \rf{Kstu}.
It is remarkable that all these
integrals are elementary and that they can be expressed in terms of powers of
the standard one-loop function ${\bar J}(s)$  \rf{J}:
\pagebreak
\bea
{\bar K}_0(s) &=& {\bar J}(s)\nonumber\\
{\bar K}_1(s) &=& {1\over{16\pi^2}}\,{s\over{s-4\mpi^2}}\,
\left[ 16\pi^2{\bar J}(s) - 2\right]^2\nonumber\\
{\bar K}_2(s) &=& {{s-4\mpi^2}\over{s}}\,{\bar K}_1(s) - {1\over{4\pi^2}}
\lbl{Kbar}\\
{\bar K}_3(s) &=& {1\over{16\pi^2}}\,{{\mpi^2}\over{s-4\mpi^2}}\,
\left\{ {s\over{s-4\mpi^2}}\,\left[ 16\pi^2{\bar J}(s) - 2\right]^3 +
\pi^2\left[ 16\pi^2{\bar J}(s) - 2\right]\right\} - {1\over{32}}\nonumber\\
{\bar K}_4(s) &=& {1\over{16\pi^2}}\,{{\mpi^2}\over{s-4\mpi^2}}\left\{
16\pi^2{\bar J}(s) - 2 + 8\pi^2{\bar K}_1(s) + {{16\pi^2}\over 3}\,{\bar
K}_3(s) + {{\pi^2}\over 3}\right\} - {1\over{32\pi^2}} + {1\over {192}}
\ .\nonumber
\ena
The proof of these formulae and a discussion of some properties of the
functions ${\bar K}(s)$ can be found in Appendix C.

\section{Phenomenological Aspects}

\indent
\setcounter{equation}{0}

The preceding discussion has led us to the expression  \rf{amp2} for the
$\pi\pi$ amplitude at two-loop order, which depends on the six parameters
$\alpha$, $\beta$ and $\lambda_i$, $i=1,2,3,4$. {\it A priori}, neither
the one-loop formula
 \rf{amp} nor the two-loop expression  \rf{amp2}, by itself, contains
information about these parameters. On the other hand, Eqs.  \rf{a},  \rf{b} or
 \rf{lambda} relate them to the low-energy constants, such as ${\widehat
m}B_0$,
${\widehat m}^2A_0$, ${\widehat m}\xi$, $L_1$, $L_2$,...,
which appear in the chiral
expansion of the effective action ${\cal Z}^{\rm eff}$. To lowest chiral
orders, these same low-energy constants are also present in the expansion of
other observables, such as the masses of the pseudoscalars, $\fpi/F_K$, the
scalar form-factor of the pion, the $K_{l4}$ form-factors, etc... This kind of
information, which reflects the Ward identities obeyed by the QCD correlation
functions, is rather crucial and has always played a central role in
applications of chiral perturbation theory. In particular, it allows one to pin
down approximate values of the parameters $\beta$, $\lambda_1$ and $\lambda_2$
from independent experiments, and it provides a clear interpretation of the
{\it leading order value} of $\alpha$ in terms of the mechanism of spontaneous
breakdown of chiral symmetry in QCD, by relating this value to the quark
mass ratio $r=m_s/{\widehat m}$, {\it cf.} Eq.  \rf{lead}. It is also clear,
however, that the {\it higher orders} in the expansions of $\alpha$, $\beta$
and the $\lambda_i$'s in powers of quark masses involve a rapidly increasing
number of new low-energy constants from ${\cal Z}^{\rm eff}$. The task of
estimating them from independent measurements of different observables, and of
re-evaluating the values of the previous low-energy constants taking into
account the higher-order corrections, seems rather out of reach at the moment.

Fortunately, it turns out to be possible to pin down four of
the parameters ($\lambda_1$, $\lambda_2$, $\lambda_3$ and $\lambda_4$),
which enter the two-loop expression  \rf{amp2} of the $\pi\pi$
scattering amplitude,
using sum rules and available $\pi\pi$ data at medium energies. The
derivation of these sum rules and their evaluations are sketched in the
following subsection (a more detailed account will be postponed to a
forthcoming publication). Next, we consider a few applications.
In particular, we
discuss how the low-energy data constrain the remaining two parameters
$\alpha$
and $\beta$.  We remind the reader that $\alpha$ is the key parameter
for deciding whether the chiral symmetry is broken according to
the standard scenario, with a large $\langle {\bar q}q \rangle$ condensate,
or not.

\subsection{Sum Rule Evaluation of the $\lambda_i$'s}

\indent

Consider the s-channel isospin I=0,1,2 amplitudes $F^{\rm I}(s,t)$ defined as
($u=4\mpi^2-t-s$)
\bea
F^0(s,t) &=& {1\over{32\pi}}\, \left\{ 3A(s\vert t,\, u) + A(t\vert s,\, u) +
A(u\vert s,\, t)\right\}\nonumber\\
F^1(s,t) &=& {1\over{32\pi}}\, \left\{ A(t\vert s,\, u) -
A(u\vert s,\, t)\right\}\lbl{F^I}\\
F^2(s,t) &=& {1\over{32\pi}}\, \left\{ A(t\vert s,\, u) +
A(u\vert s,\, t)\right\}\ .\nonumber
\ena
Because of the Froissart bound, they
satisfy twice-subtracted, fixed-$t$,  dispersion relations.
These form the starting point for the derivation of the sum rules for the
constants $\lambda_i$.
The next, and
somewhat tedious, step consists in transforming
this set of fixed-$t$ dispersion relations
into a form which allows direct comparison with the two-loop chiral
expression. This is performed by first identifying and rejecting the
contributions which are of chiral order $O({\rm p}^8)$ or more.
In order to do so, one has to split the integration ranges of the dispersive
integrals into
two regions, a low-energy region $[4\mpi^2, E^2]$ and a high energy
region $[E^2, \infty]$.
In the first region, one may drop the contributions from the partial
waves with $l\ge2$, while in the latter region one may expand in powers of
$s,t,u$ divided by $E^2$ and stop at the third power.
The borderline energy $E$ corresponds,
roughly speaking, to the point where the chiral expansion
starts to break down, {\it i.e.} $E$ is in the range 500-600 MeV.
Secondly, one has to impose
crossing symmetry. In practice, one proceeds in close analogy
with subsection III-C of Ref.  \cite{SSF}. There, starting from
thrice-subtracted dispersion relations, it was shown how, up to corrections
of chiral order
$O({\rm p}^8)$, crossing symmetry entirely determines the form of the
amplitude except for six arbitrary parameters.  If, instead, one
uses twice-subtracted dispersion relations one finds that two
parameters remain undetermined ($\alpha$ and $\beta$),
while the four others ($\lambda_1$,...$\lambda_4$)
obey well-convergent sum rules \footnote{As a matter of fact, if one
assumes conventional Regge
asymptotics for the t-channel I=1 amplitude, one can derive
a slowly convergent sum rule for $\beta$ as well, which is equivalent
to the one derived long ago by Olsson  \cite{olsson}. }.
Details will be provided elsewhere. Here,
we shall only give the final result:
\bea
&&{\lambda_1\over\fpi^4}={1\over 3}\left( {\cal I}_0^{(2)} -{\cal I}_2^{(2)}
\right) + 2\mpi^2\left( {\cal I}_0^{(3)}
-{\cal I}_2^{(3)}\right) -3{\cal I}_1^{(1)} -6\mpi^2{\cal I}_1^{(2)}+
{{16\pi}\over 3}\left( h_0 - 4h_2\right)
\nonumber\\
&&{\lambda_2\over\fpi^4}={1\over 2}{\cal I}_2^{(2)}
+3\mpi^2 {\cal I}_2^{(3)} + {3\over 2}{\cal I}_1^{(1)}
+3\mpi^2{\cal I}_1^{(2)} -{16\pi\over3}\left( h_0-h_2\right)
\nonumber\\
&&{\lambda_3\over\fpi^6}={1\over 3}\left( {\cal I}_0^{(3)} -{\cal I}_2^{(3)}
\right)+{\cal I}_1^{(2)}+{32\pi\over9} \left( -9h_1+h'_0-h'_2\right)
\nonumber\\
&&{\lambda_4\over\fpi^6}={1\over 2}{\cal I}_2^{(3)}
-{1\over 2}{\cal I}_1^{(2)}+{32\pi\over9} \left( h'_0-h'_2\right)
\lbl{lambdas} \ .
\ena
In these formulae, the quantities ${\cal I}_a^{(k)}$ are given
in terms of
integrals over the low-energy range $[4\mpi^2,E^2]$,
while $h_n$ and $h'_n$ involve integrals
from $E^2$ to infinity (detailed formulae will be given below).
Since experimental data on $\pi\pi$ phase shifts exist down to roughly
500 MeV, we can calculate the latter integrals essentially using
experimental input. In the low-energy integrals we shall use the two-loop
chiral expression for the phase shifts. Stability of the
result with respect to variations of the energy $E$ within the range indicated
above measures the extent
to which the $\chi$PT phases do match to the experimental ones
at this energy. Note that since the $\chi$PT phases themselves depend
on the parameters $\lambda_i$, the set of relations  \rf{lambdas} must
actually be solved in a self-consistent way.

Let us now give the explicit expressions for the various entries
in Eqs.  \rf{lambdas}. First, we define the ${\cal I}_a^{(k)}$'s as
\eq
{\cal I}_a^{(k)} = {{32\pi}\over k!}{d\over ds^k}
\left\{ {s^{2-\varepsilon_a}\over{\pi}}\int_{4M^2}^{E^2}
{dx\over x^{2-\varepsilon_a}}\,
{{{\rm Im} W_a}\over{x-s}}\ - \ \sum_{n=0}^4 w_a^{(n)}(s){\bar K_n}(s)
\right\} _{s=0}\ ,\lbl{Iak}
\en
where the polynomials $w_a^{(n)}(s)$ and the functions
${\bar K}_n(s)$ are those which occur in the expression of the two-loop
amplitude of the preceding section.
The quantities $h_a$ and $h'_a$ are defined in terms of the following
high-energy integrals
\eq
h_0(t)={-1\over\pi}\int_{E^2}^\infty\,dx{2x+t-4M^2\over x^2(x+t-4M^2)^2}
{\rm Im} \left[ {1\over3}F^0(x,t)+ F^1(x,t)+{5\over3} F^2(x,t)\right] \ ,
\lbl{h0t}
\en
and
\eq
h_2(t)={-1\over\pi}\int_{E^2}^\infty\,dx{2x+t-4M^2\over x^2(x+t-4M^2)^2}
{\rm Im} \left[ {1\over3}F^0(x,t)-{1\over2} F^1(x,t)+
{1\over6} F^2(x,t)\right] \ .
\lbl{h2t}
\en
Then, $h_a$ and $h'_a$ are given simply by
\eq
h_a\equiv h_a(0)\ ,\qquad h'_a\equiv h'_a(0)\ , \qquad a=0,2\ .
\en
Finally, $h_1$ is given by
\eq
h_1={-1\over\pi}\int_{E^2}^\infty{dx\over x(x-4M^2)^2}
{\rm Im} \left[ {1\over3}F^0(x,0)+{1\over2} F^1(x,0)-
{5\over6} F^2(x,0)\right] \ .
\lbl{h1t}
\en
Notice that the three integrals  \rf{h0t},  \rf{h2t} and  \rf{h1t}
involve the {\it t-channel}
isospin ${\rm I}_t$=0, 2 and 1 combinations, respectively.
In addition to the constraints it imposes on the low-energy parts, crossing
symmetry also implies the following relation between these high-energy
integrals:
\eq
9h_1=2h'_0-5h'_2 \ .
\lbl{sternrel}
\en

In practice, the most reliable source of information on the
$\pi\pi$
phase shifts remains the old high-statistics CERN-Munich
experiment \cite{hyams} (see {\it e.g.} the reviews by Ochs
 \cite{ochsrev} and by Morgan and Pennington  \cite{mprev} for  critical
discussions of more recent experiments). In order to estimate the
errors in the evaluation of the high-energy integrals, we have varied
the experimental data within their error bars and we have also compared
the results obtained using  different analyses of the production
data of Ref.  \cite{hyams}. The errors in the low-energy integrals, which
reflect the presence of $O({\rm p}^8)$ contributions, were estimated
from the sensitivity to variations in the value of the matching
energy $E$ and by
trying different ans\"atze for ${\rm Im} f_a$ (which all differ only by
$O({\rm p}^8)$ contributions)
in equations  \rf{Iak}. Furthermore, a small dependence of the
result upon the remaining two parameters $\alpha$ and $\beta$ was
absorbed in the error. More details of this analysis will be given
elsewhere. We end up with the following estimates for the values
of $\lambda_1$ and $\lambda_2$:
\eq
\lambda_1= (-5.3\pm 2.5)\times 10^{-3}\qquad\lambda_2= (9.7\pm1.0)\times
10^{-3}\lbl{lam12} \ .
\en
These numbers are in agreement with the result of the analysis of
Bijnens {\it et al.}  \cite{BCG}, which incorporates constraints from
sum rules for D-wave threshold parameters as well as from $K_{l4}$
decays. An independent estimate was recently made
by Pennington and Portol\`es  \cite{pp},
who derived simple, but approximate, formulae
by setting the isospin I=2 contributions in both direct
and crossed channels exactly equal to zero. From their analysis it follows that
\hfill\break $\lambda_1=
(-4.3\pm 2.3)\times 10^{-3}$ and $\lambda_2=(8.8\pm1.1)\times 10^{-3}$
for $\fpi$=92.4 MeV. Our result
for $\lambda_2$ lies  between those of these two references. Finally,
for the remaining two parameters, we obtain
\eq
\lambda_3=(2.9\pm 0.9)\times 10^{-4}\qquad\lambda_4= (-1.4\pm0.2)
\times 10^{-4}\lbl{lam34} \ .
\en

\subsection{ Low-energy Phase Shifts}

\indent

It is well known that the chiral expansions of the $\pi\pi$
partial-wave amplitudes
do not satisfy unitarity exactly, but rather in a perturbative sense.
The deviations from unitarity provide an estimate for the energy range
where the chiral expansion, at a given order, can be trusted. We show
in Fig. 1 the Argand diagram for the amplitude $f^0_0$ computed at
one-loop and at two-loop order. The figure shows that there is a
considerable improvement with respect to unitarity at low energies in going
from
the one-loop result to the two-loop result.
On the other hand, it also suggests that one should
expect sizeable $O({\rm p}^8)$ corrections starting at 450 MeV.

In spite of the fact that the unitarity relation is not exactly
satisfied, it is possible to define a chiral expansion for the
phase shifts (see, for instance, the discussion in  \cite{gm}).
At lowest order, and corresponding to the amplitude  \rf{Ap2},
the partial waves for $l<2$ are defined as
($a=0,1,2$)
\eq
\delta_a = \sqrt{s-4M^2\over s}\,\varphi_a \ +\ O({\rm p}^4)\ .
\lbl{phase2}
\en
The one-loop form  \rf{amp} of the amplitude extends
the above definition of the phases to
\eq
\delta_a = \sqrt{s-4M^2\over s}\,[\varphi_a +\psi_a ]\ +
\ O({\rm p}^6)\ ,\lbl{phase4}
\en
whereas the two-loop expression  \rf{amp2} leads to
\eq
\delta_a = \sqrt{s-4M^2\over s}\,\left[{\rm Re }f_a +
{2\over3}(\varphi_a)^3\right]\ +\ O({\rm p}^8)\ ,\lbl{phase6}
\en
where in this last formula Re$f_a$ must be evaluated from the full
two-loop amplitude  \rf{amp2}.
The higher partial waves receive contributions starting only at the one-loop
approximation, otherwise the formulae are exactly the same.
The result for the phases as computed from the successive approximations
 \rf{phase2},  \rf{phase4} and  \rf{phase6} are shown in Figs. 2a and 2b.
For simplicity, we have used
the same values for $\alpha$ and $\beta$ ($\alpha=2$, $\beta=1.08$ )
in all three cases (strictly speaking, of course, one
should use the expansions  \rf{mexp2}
of $\alpha$ and $\beta$ at the given order).
Surprisingly, for I $=l=0$ (Fig. 2a) the one-loop and two-loop phases  remain
fairly close at energies where one would expect (say from the Argand
diagram) that $O({\rm p}^8)$ terms should become important
(approximately above 500 MeV). This is
somewhat fortuitous, and does not happen for the isospin I=2 wave
or the P-wave. In the latter case, it is
interesting to observe how the $\rho$ resonance is gradually built
up as the order of the perturbation increases. Finally, Fig. 4a shows the
behaviour, at increasing orders of the chiral expansion, of the difference
$\delta_0^0 - \delta_1^1$.

We investigate next the sensitivity of the phase shifts on the
values of the parameters $\alpha$ and $\beta$. Results are shown in
Figure 3, representing $\delta_0^0$, $\delta_0^2$ and $\delta_1^1$
respectively, and Fig. 4b which shows the difference $\delta_0^0-
\delta_1^1$ in the energy range accessible in
$K_{l4}$ decays. We have taken values of $\alpha$ ranging from 1 to 3.
Figure 4b suggests that the data of Rosselet {\it et al.} \cite{rosselet} are
approximately compatible with such a range. The figures show that
the sensitivity upon $\alpha$ is most significant at very low energies.
An important issue for us is to assess the extent to which the
availability of a new generation of data on $K_{l4}$ decays
from {\it e.g.} the DA$\Phi$NE facility will allow
a really precise determination of these parameters $\alpha$ and
$\beta$. For this purpose we show in Fig. 5 how the errors on
$\lambda_1$,...,$\lambda_4$ affect the prediction for
$\delta_0^0-\delta_1^1$. The figure shows that the
situation where $\alpha=1$ and the one where $\alpha=2$ can be
distinguished, provided sufficiently precise data
are available in the energy range below
$E=340$ MeV.
Let us now examine in a more quantitative way how the presently
available data constrain the values of $\alpha$ and $\beta$, using the
two-loop form of the amplitude and our estimates of the values of
the parameters $\lambda_i$. Since there are only five data points
provided in Ref.  \cite{rosselet}, it is useful to look for an additional
constraint between $\alpha$ and $\beta$. Such a relation can indeed be obtained
from the ``Morgan-Shaw Universal Curve"  \cite{universal}. The latter
represents
a correlation between the I = 0 and I = 2 S-wave scattering lengths,
obtained by fitting the data in the region between 600 MeV
and 1 GeV (i.e. outside
of the range of applicability of the two-loop chiral expansion)
using the Roy equations (see  \cite{bfp}, \cite{petersen} ).
We reproduce here the form
quoted by Petersen in Ref.  \cite{nagels}:
\eq
2a_0^0-5a_0^2=0.692\pm0.027 +0.9\,(a_0^0-0.3)+1.2\,(a_0^0-0.30)^2\ .
\lbl{univc}
\en
Upon using the explicit expressions for the scattering lengths
(see Appendix D), this equation transcribes into the desired relation
between $\alpha$ and $\beta$, which is plotted in Fig. 6.
We have performed a chi-square minimisation using this
relation in addition to the data of Rosselet {\it et al.}  \cite{rosselet}
and to the
results  \rf{lam12},  \rf{lam34}. We find
\eq
\alpha=2.16\pm0.86\ ,\qquad \beta=1.074\pm0.053\ ,\lbl{fit}
\en
with a $\chi^2$ equal to 2.01 for four degrees of freedom.
The threshold parameters will be discussed in detail
below. Let us give  here the corresponding value for $a_0^0$,
\eq
a_0^0=0.263\pm0.052\ .
\en
Interestingly enough, this result is in agreement with the one quoted
in Ref.  \cite{nagels}, which is based on a fit  using the Roy equations.

A physically interesting quantity which is related to the $\pi\pi$
phase shifts is the phase of the CP violation parameter $\epsilon'$
\eq
\phi(\epsilon')=90 +\delta_0^2-\delta_0^0 \quad ({\rm degrees})\ ,
\en
where the phases are to be evaluated at the energy $E=M_{K^0}=
497.7 $ MeV. The preceding discussion shows that it makes sense
to use the two-loop approximation at this energy and, indeed,
Gasser and Mei{\ss}ner have discussed $\phi(\epsilon')$ at $O({\rm p}^4)$
in the standard chiral expansion and find $\phi=(45\pm6)^\circ$.
Using the values of $\alpha$ and $\beta$  \rf{fit} derived from the fit
discussed above, we obtain
\eq
\phi=(43.5\pm2\pm6)^\circ\ ,
\en
where we have shown separately the error coming from the uncertainties in the
fit values  \rf{fit} of $\alpha$ and $\beta$ ($\pm 2^\circ$), and the error due
to the uncertainties in our determinations  \rf{lam12} and  \rf{lam34} of the
$\lambda_i$'s ($\pm 6^\circ$).  At this energy, a significant gain in
precision is achieved by going through a numerical solution of the Roy
equations, rather than by using the chiral expansion.  This has been done by
Ochs
\cite{ochsrev}, who found a result very similar to ours but with a much
smaller error: $\phi=(46\pm3)^\circ$ .

\subsection{Threshold Parameters}

\indent

{}From the two-loop amplitude  \rf{amp2}, it is straightforward to obtain
explicit expressions for the scattering lengths $a_l^{\rm I}$ and the slope
parameters $b_l^{\rm I}$. For the reader's convenience, we have gathered, in
Appendix D, a selection of formulae relevant for the discussion that follows.

\begin{table}
\begin{center}
\begin{tabular}{|c|c||c|c|c|c|} \hline
$\alpha$&$\beta$& $a_0^0$ & $b_0^0$ &--10$a_0^2$ &--10$b_0^2$ \\
\hline
  1.04  & 1.08  &   0.210 &   0.26  &       0.43 &      0.78
\\ \hline
        & 1.127 &   0.264 &   0.27  &       0.32 &      0.83
\\ \cline{2-6}
   2    & 1.071 &   0.255 &   0.25  &       0.29 &      0.78
\\ \cline{2-6}
        & 1.016 &   0.246 &   0.24  &       0.26 &      0.74
\\ \hline
 2.16  & 1.074 &   0.263 &   0.25  &       0.27 &      0.79
\\ \hline
 2.5    & 1.085 &   0.283 &   0.25  &       0.22 &      0.80
\\ \hline
 3.0    & 1.102 &   0.311 &   0.25  &       0.16 &      0.82
\\ \hline
 3.5    & 1.122 &   0.342 &   0.25  &       0.10 &      0.85
\\ \hline
\multicolumn{2}{|c||}{Error bars}
      & $\pm 0.006\ \ $ & $\pm 0.02\ \ $ & $\pm 0.02\ \ $ & $\pm 0.06\ \ $
\\ \hline\hline
\multicolumn{2}{|c||}{Experiment  \cite{nagels}}
      & 0.26$\pm$0.05 & 0.25$\pm$0.03 & 0.28$\pm$0.12 & 0.82$\pm$0.08
\\ \hline
\end{tabular}
\vskip 1.5 true cm
\begin{tabular}{|c|c||c|c|c|c|c|} \hline
$\alpha$&$\beta$& $10a_1^1$ & $10^2b_1^1$ & $10^2a_2^0$ &
                                                 $10^3a_2^2$ & $10^4a_3^1$
\\ \hline
  1.04  & 1.08  &    0.37   &     0.60    &     0.17    &
                                                    0.08     &   0.56
\\ \hline
        & 1.127 &    0.39   &     0.55    &     0.17    &
                                                    0.13     &   0.65
\\ \cline{2-7}
   2    & 1.071 &    0.37   &     0.55    &     0.17    &
                                                    0.14     &   0.63
\\ \cline{2-7}
        & 1.016 &    0.35   &     0.55    &     0.18    &
                                                    0.14     &   0.62
\\ \hline
 2.16  & 1.074 &    0.37   &     0.54    &     0.17    &
                                                    0.15     &   0.65
\\ \hline
 2.5    & 1.085 &    0.37   &     0.53    &     0.17    &
                                                    0.17     &   0.68
\\ \hline
  3     & 1.102 &    0.38   &     0.50    &     0.17    &
                                                    0.20     &   0.74
\\ \hline
 3.5    & 1.122 &    0.38   &     0.48    &     0.17    &
                                                    0.23     &   0.80
\\ \hline
\multicolumn{2}{|c||}{Error bars}
     & $\pm 0.01\ \ $  & $\pm 0.15\ \ $  & $\pm 0.05\ \ $  & $\pm 0.28\ \ $ &
                                                        $\pm 0.18\ \ $
\\ \hline\hline
\multicolumn{2}{|c||}{Experiment  \cite{nagels}}
      & 0.38$\pm$0.02 & $\qquad\qquad$  &
               0.17$\pm$0.03 &$0.13\pm 0.30$ &$\ 0.6\pm 0.2\ $
\\ \hline
\end{tabular}
\vskip 0.5 true cm
{\bf Table 1:}{{\it Threshold parameters for} $\fpi = 92.4$ MeV, {\it in units
of} $M_{\pi^+}=139.57$ MeV for different values of $\alpha$ and of $\beta$.}
\end{center}
\end{table}

Table 1 shows the numerical values of the threshold parameters at two-loop
precision for different
values of $\alpha$ and $\beta$, with the $\lambda_i$'s fixed at their central
values as given by Eqs.  \rf{lam12} and  \rf{lam34} above. The second line of
the
table gives the two-loop threshold parameters for the values of
$\alpha$ and $\beta$
corresponding to the $O({\rm p}^4)$ predictions  \rf{stnum} of the standard
case \footnote{Of course, the two-loop values of $\alpha_{\rm st}$ and
$\beta_{\rm st}$ need not coincide with their one-loop evaluations  \rf{stnum}.
However, the changes are expected to be small, and we feel that the second line
of Table 1 gives a realistic description of the standard two-loop predictions
within the error bars quoted in the penultimate line.}
within the $SU(3)_{\rm L}\times SU(3)_{\rm R}$ analysis,
while the fourth line corresponds
to the central values  \rf{fit}, obtained from the fit of the two-loop phases
to
the Rosselet data {\it cum} Universal Curve constraint. For the remaining
entries, we have, for each value of $\alpha$, chosen for $\beta$ the central
value given by the Universal Curve  \rf{univc},
except for $\alpha = 2$, where we have also
displayed the results corresponding to the maximal and minimal values of
$\beta$ as allowed by  \rf{univc}. This variation of $\beta$ at fixed $\alpha$
only affects the values of the S- and P-wave scattering
lengths and the S-wave slopes, which are the only threshold parameters to
depend on $\beta$ already at leading order. The effect is strongest in $a_0^2$
and is of the order of about $\pm10\%$ in average.
The S-wave scattering lengths $a_0^0$ and $a_0^2$ are clearly the most
sensitive to variations in the value of $\alpha$. The penultimate line of Table
1 gives the error bars (which are independent on both $\alpha$ and $\beta$ at
the level of numerical accuracy we consider here) on the threshold parameters.
They only include the effects of varying the values of the $\lambda_i$'s within
their ranges as given by Eqs.  \rf{lam12} and  \rf{lam34}. The
last line gives the values, as quoted in  \cite{nagels}, obtained from
Roy-equation  analyses of the $K_{l4}$ and production data.

Figure 7 shows the S-wave scattering lengths as functions of $\alpha$. The
shaded area gives the uncertainties coming from the errors on the
$\lambda_i$'s, as given by Eqs.  \rf{lam12} and  \rf{lam34}, whereas the band
delimited by the solid lines includes the error on the $\alpha$-dependence of
$\beta$ inferred from the Universal Curve  \rf{univc}. The present experimental
value $a_0^0 = 0.26\pm 0.05$ allows $\alpha$ to lie between 1 and 3.4.
In this whole range of $\alpha$, corrections to $a_0^0$ represent a
25\% increase with
respect to tree level, and the two-loop corrections are about 5\% of the
one-loop result.
(see also the discussion at the end of
Appendix D). The observed
convergence rate leads one to expect that higher order effects should be within
the error bars quoted in Table 1.
In addition, we show
(Fig. 7) the combination $\vert a_0^0 - a_0^2\vert$ which can be extracted
from the measurement of the $\pi^+\pi^-$-atom lifetime planned at CERN
 \cite{dirac}.

\section{Testing the Strength of Quark Condensation}

\indent
\setcounter{equation}{0}

In Section 4.2 it has been shown how the two-loop expression for the
scattering amplitude can be exploited to extract values of the
parameters $\alpha$ and $\beta$ from available low-energy data.  The
lack of precision in the result, {\it viz.} $\alpha = 2.16 \pm 0.86$,
merely reflects the large error bars in currently existing data, rather
than the uncertainty attached to the parameters $\lambda_1, \ldots,
\lambda_4$.  More accurate data are awaited, and with them a rather
precise measurement of $\alpha$ and $\beta$ should become possible.  We have
already stressed the key role played by the parameter $\alpha$ in the
measurement of the amount of quark condensation in the QCD vacuum:  In
the standard, large condensate alternative of symmetry breaking,
$\alpha$ should remain close to 1 (cf. Eqs.  \rf{stnum} and  \rf{GLnum}),
whereas the case of a marginal quark condensation is characterised by a
considerably higher value, $\alpha \sim 3 - 4$.  In both cases, $\beta$
should stay close to 1.

In this section, we attempt a more quantitative interpretation of the
low-energy constants $\alpha$ and $\beta$, relating them to fundamental
QCD parameters such as the quark mass ratio $r = m_s/\widehat{m}$ and
the condensate parameter $2\widehat{m}B_0$.  To leading order, this
relation has been given before  \cite{FSS2,SSF}.  The complexity of this
relationship rapidly increases with increasing chiral order to which
$\alpha$ and $\beta$ are expanded.  Here, we restrict our discussion to
one-loop order $O({\rm p}^4)$, using the corresponding perturbative
expressions  \rf{bt} and  \rf{at} for $\beta$ and $\alpha$.  Let us
stress that this is a perfectly consistent procedure:  It is conceivable
that two-loop accuracy in the expansion of the amplitude is important
for a precise {\em measurement} of $\alpha$ and $\beta$ using, in
addition to other input, energy-dependent phase shift data above
threshold.  On the other hand, to make use of the results of
such a measurement, one-loop accuracy in the expansion of
$\alpha$ and $\beta$ in powers of quark masses could be sufficient to
distinguish a value of $r \sim 25$ from,
say, $r \sim 10$, or the case $2\widehat{m}B_0/\mpi^2 \sim 1$ from
$2\widehat{m}B_0/\mpi^2 \sim 0$.

\subsection{Dimensional Analysis and Order of Magnitude Estimates}

\indent

Before proceeding to a more refined analysis, it might be useful to
recall some crude and simple order of magnitude estimates of various
contributions to $\alpha$ and $\beta$ that are exhibited in Eqs.
 \rf{bt} and   \rf{at}.  A constant of ${\cal L}^{\rm eff}$ multiplying an
invariant that is made up from $n$ covariant derivatives and $m$ powers
of the quark mass is, for $n + m \ge 2$, expected to be of order $F_0^2
\Lambda_H^{2-n-m}$, unless it is suppressed by the Zweig rule.  Indeed,
this is the expected order of magnitude of a (massless) QCD correlation
function that consists of $n$ vector and/or axial-vector and of  $m$
scalar and/or pseudoscalar current densities, at low external momenta,
with the singularities arising from exchanges of Goldstone bosons
subtracted.  This estimate is obtained by saturating correlation
functions that are smooth enough at short distances by the lowest {\em
massive} bound states ($\rho, a_1, f_0, \ldots$), which have masses of
the order
of $\Lambda_H \sim $ 1 GeV.  ($F_0$ is a scale characterizing the coupling
of mesons $M$ to currents $J$: $\langle 0 | J | M\rangle \sim F_0
\Lambda_H$.)
This argument is a special case of the ``naive dimensional analysis" of
Ref.  \cite{Georgi}, except that we leave open the question as to whether
it may be extended to the ``one-point function" ($n = 0, m=1$)
representing the quark condensate.  The reason for this restriction is
quark confinement, which prevents us from analysing the quark condensate
by the same dispersion techniques as in the case of $(n+m)$-point
functions for $n+m \ge 2$.

Factoring out $F_0^2$ from all terms of ${\cal L}^{\rm eff}$, {\it cf.} Eqs.
 \rf{L2},  \rf{L3},  \rf{L22} and  \rf{L04}, the
$\tilde{\cal{L}}^{(2)}$ constants $A_0$ and $Z_0^P$ are two-point
functions divided by $F_0^2$, and consequently they are expected to be
of the order of 1.  (A more detailed estimate, using QCD sum rules,
gives $A_0 \sim 3 \pm 2$.)  Similarly, the constants $\xi$, $\rho_1$,
$\rho_2$ are expected to be of the order of $1/\Lambda_H$, whereas the
$\tilde{{\cal L}}^{(4)}$ constants $A_1$, $A_2$, $A_3$, $B_1$, $B_2$, $E_1$,
$E_2$, $E_3$ should be of the order $1/\Lambda_H^{2}$.  The
constants $Z_0^S$, $\tilde{\xi}$, $\rho_4$,
$\rho_5$ and $\rho_7$ violate the Zweig rule, and they are therefore
expected to be suppressed compared to the corresponding dimensional estimates.

It should be stressed that for $n+m \ge 2$ these estimates are
independent of any chiral counting or of the importance of the quark
condensate.  On the other hand, attempts to estimate the value of the
$u$ and $d$ running quark masses \footnote{All QCD running quantities
will be understood to be normalised at the scale 1 Gev $\sim \Lambda_H$.
The combinations of constants times powers of quark masses appearing in
the expansions of $\alpha$ and $\beta$ are QCD
renormalisation group invariant.}, especially $\widehat{m} = (m_u + m_d)/2$,
depend
in one way or another on the presumed mechanisms of chiral symmetry breakdown
in QCD.  A model-independent evaluation of corresponding QCD sum rules is
at present problematic, due to the complete absence of experimental
information on the size of the spectral function of the divergence of
the axial current beyond the one-pion contribution.  The existing
evaluations of $\widehat{m}$ usually normalise  \cite{DdR,BPdR}
the unknown spectral function by
its low-energy behaviour as predicted by the standard $\chi$PT.  However,
in generalised $\chi$PT the latter prediction is modified by an
enhancement factor
which can be as large as 13.5, depending on the actual importance of the
quark condensate  \cite{Mar,HBK1}. The issue can and should be settled
experimentally, by measuring the magnitude of the divergence of the
axial current in hadronic $\tau$ decays  \cite{Mar}. Fortunately, the
existing determinations of the quark {\em mass difference} $m_s - m_u$
(see {\it e.g.} the recent analyses in  \cite{JaMu,Dom,Nar}, from
which earlier references can
be traced back) are on less speculative grounds.  They are based on sum rules
which involve the two-point function of the divergence of the vector
current $\bar{s}\gamma_\mu u$; the corresponding spectral function
can be normalised using experimental data in a rather model-independent
way.  For definiteness, we shall adopt the corresponding result of Ref.
 \cite{JaMu}, {\it viz.} $m_s - m_u$ =
(184$\pm$32) MeV, which is compatible, within the error bars, with the values
obtained by the authors of Refs.  \cite{Dom} and  \cite{Nar}. For the
sake of our order of magnitude estimates
we shall thus use for ${\widehat m}$ a value which depends on the
ratio $r = m_s/\widehat{m}$ in the following way

\eq\lbl{m(r)}
\widehat{m} = \frac{(184 \pm 32)}{r-1}\, {\rm MeV}\ ,
\en
keeping in mind that in generalised $\chi$PT, $r$ is not fixed from the outset.
Even for $r$ as small as $r\sim 8$, the corresponding value of
$\widehat{m}$ would be
\eq\lbl{m8}
\widehat{m} \sim{\rm (26 \pm 4.6)~MeV}\ ,
\en
which would keep the expansion parameter $\widehat{m}/\Lambda_H$
reasonably small,
although somewhat larger than in the standard $\chi$PT.  Given the above
value for $m_s - m_d$, Eq. \rf{m8} represents the upper bound for
$\widehat{m}$ in G$\chi$PT.

We are now in a position to estimate orders of magnitude of various
terms contributing to $\alpha$, dividing Eq.  \rf{at} by $\mpi^2$.  For
definiteness, we set $\Lambda_H \simeq M_\rho$ and we first consider the
low $B_0$ alternative, taking, {\it e.g.}, $r \simeq$ 10.  In this case, both
terms $2\widehat{m}B_0/\mpi^2$ and $16\widehat{m}^2A_0/\mpi^2$ should be of
order unity. The $\tilde{\cal L}^{(3)}$ contributions $4\widehat{m}\xi$ and $81
\widehat{m}^3 \rho_1/\mpi^2$ should both reach the ten-percent level,
whereas the ${\cal L}_{(2,2)}$ terms $\widehat{m}^2B_1$, $\widehat{m}^2 A_3$
and
the ${\cal L}_{(0,4)}$ terms, including their large coefficients --
{\it viz.}, $256 \widehat{m}^4 E_1/\mpi^2$ -- should hardly reach
the one-percent level.
Chiral logarithms and the Zweig rule violating, (explicitly)
$r$-dependent, terms will be discussed shortly.

In the standard large $B_0$ alternative ($r \sim 26$), the whole
expression  \rf{at} is dominated by the Gell-Mann--Oakes--Renner (GOR)
ratio $2\widehat{m}B_0/\mpi^2$, which remains close to 1.  All remaining
terms should contribute at most at the one-percent level, reflecting the
smaller value of $\widehat{m} \sim $ 6 MeV.

\subsection{Constraints from Pseudoscalar Meson Masses \hfill\break
and Decay Constants}

\indent

More information on the low-energy constants contained in $\tilde{\cal
L}^{(2)}$ and $\tilde{\cal L}^{(3)}$ can be obtained from the expansion
of the Goldstone boson masses and decay constants $\fpi$ and $F_K$.
These expansions, up to and including order $O({\rm p}^4)$ contributions in
generalised $\chi$PT, are
collected in Appendix A.  To leading, $O({\rm p}^2)$
order, $\widehat{m}^2A_0$ and
$\widehat{m}^2Z_0^P$ can be expressed as functions of $r = m_s/\widehat{m}$
and of pseudoscalar meson masses:

\bea\lbl{A0Z0P}
\frac{4 \widehat{m}^2 A_0}{\mpi^2} &=& 2\, \frac{r_2 - r}{r^2 -1} \equiv
\epsilon(r)\nonumber\\
&&\\
\frac{8 \widehat{m}^2 Z_0^P}{\mpi^2} &=& - \epsilon(r) +
\frac{{\widehat \Delta}_{GMO}}{(r-1)^2}\ ,\nonumber
\ena
where (using $\mpi = 135$ MeV, $M_K = 495.7$ MeV, $M_{\eta} = 547.5$ MeV )

\bea\lbl{r2Delta}
 r_2 &=& 2\frac{M_K^2}{\mpi^2} - 1 = 26 \nonumber \\
&&\\
{\widehat \Delta}_{GMO}&=&  (3 M_\eta^2 + \mpi^2 - 4M_K^2)/\mpi^2
= -3.6\ .\nonumber
\ena
The ratio $2\widehat{m}B_0/\mpi^2$ depends, in addition, on the Zweig rule
violating constant $Z_0^S$:

\eq\lbl{GOR-ratio}
\frac{2\widehat{m}B_0}{\mpi^2} = 1 - \epsilon(r) [ 1 + (r+2) \zeta]\ ,
\en
where
\eq\lbl{zeta}
\zeta \equiv \frac{Z_0^S}{A_0} = \frac{L_6}{L_8}
\en
is expected to be small.  For $r = r_2 + O(m_{\rm quark}/\Lambda_H)$, the
ratio
 \rf{GOR-ratio} is close to 1, whereas the constants  \rf{A0Z0P} are
relegated to higher orders, provided, of course, that ${\widehat \Delta}_{GMO}
=
O(m_{\rm quark}/\Lambda_H)$.  This is the standard scenario, in which the quark
condensate is large enough to dominate the expansion of Goldstone boson
masses as well as the expansion  \rf{at} of the parameter $\alpha$.  As
$r$ decreases, the  ratio  \rf{GOR-ratio}
becomes smaller than 1 and (neglecting $\zeta$)
it vanishes for
\eq\lbl{r1}
r = r_1 = 2 \frac{M_K}{\mpi} -1 = 6.3\ .
\en
Notice that $r$ cannot be smaller than this critical value, since vacuum
stability does not allow the ratio  \rf{GOR-ratio}
to be negative.  For $r < r_2$,
the ratio $\epsilon(r)$  \rf{A0Z0P} gradually rises towards 1, in
agreement with the dimensional analysis mentioned above:  Assuming the quark
mass ${\widehat m}$ to be given by Eq.  \rf{m(r)}, and taking, for instance,
$r = 10$,
Eq.  \rf{A0Z0P} implies $\epsilon \simeq 0.3$ and $A_0 \simeq 3.4$.  In
generalised $\chi$PT, the leading order contribution to $\alpha$ is obtained by
summing up the first three terms in the expression  \rf{at}:
\eq
\alpha^{(0)} = 1 + 6\, \frac{r_2 - r}{r^2 - 1}\, (1 + 2 \zeta) \ ,
\lbl{alpha0}
\en
whereas
\eq
\beta^{(0)}  = 1\ .\lbl{beta0}
\en
Hence, the quark mass ratio $r$, instead of being determined by
pseudoscalar meson masses, may be used as a measure of the amount of
quark condensation.  It determines the leading order value of the
parameter $\alpha$,
which lies in the range $1 \lapprox \alpha^{(0)} \lapprox 4$.

Let us now consider the constants of $\tilde{\cal L}^{(3)}$ characterizing the
next to the leading order $O({\rm p}^3)$.  To that order, the
constant $\widehat{m}\xi$ can be inferred from the splitting of the decay
constants $F_K$ and $\fpi$, {\it cf.} Appendix A (we take the determination
$F_K/\fpi = 1.22\pm 0.01$ from Ref.  \cite{LRoos}),
\eq\lbl{mxi}
\widehat{m}\xi = \frac{1}{r-1} \left(\frac{F_K^2}{\fpi^2} -1 \right) =
\frac{ 0.488 \pm 0.024}{r-1}\ .
\en
This value agrees with the dimensional analysis estimate within a factor
of 2.  The next to the leading order contribution to $\beta$ is then
\eq
\beta = 1 + \frac{2}{r-1} \left(\frac{F_K^2}{\fpi^2} -1 \right)\left(1 +
2\frac{\tilde{\xi}}{\xi}\right) + \ldots
\en
The $O({\rm p}^3)$ corrections affect the leading order formulae  \rf{A0Z0P},
 \rf{GOR-ratio} and  \rf{alpha0}.  The most important
effect arises precisely from the splitting $F_K \neq \fpi$.  It amounts
to replacing $\epsilon(r)$ everywhere by $\epsilon^*(r)$, where
\eq\lbl{Op3corr}
\epsilon^*(r) = 2\, \frac{r_2^* -r}{r^2 -1}\ ,\ \
r_2^* = 2\left(\frac{F_KM_K}{\fpi\mpi}\right)^2 -1\ .
\en
The relative importance of this correction is greatest for $r \sim r_2$
($r_2$ = 26, whereas $r_2^*$ = 39).  In the low $B_0$ alternative,
this ${\cal L}_{(2,1)}$ correction shows up mainly as a slight increase
in the critical value of $r$ for which the quark condensate vanishes: This
critical values moves from $r=r_1\simeq 6.3$ towards $r=r_1^*$, with
\eq
r_1^* = 2\,\frac{F_KM_K}{\fpi\mpi} -1
\simeq 8\ .
\en

Finally, we discuss the ${\cal L}_{(0,3)}$ parameters $\rho_i$,
especially $\rho_1$ and $\rho_2$ which are not suppressed by the Zweig
rule.  The magnitude of one of them -- the parameter $\rho_2$ -- can be
inferred from isospin breaking effects induced by $m_d \neq m_u$.  It
has been pointed out in  \cite{FSS2} that (in the absence
of electromagnetism) there exists a particular
{\em linear combination} of the $K^+, K^0$ and $\pi^+$ masses, namely
\eq\lbl{Dashen}
\Delta(r,R) = M_K^2 - \mpi^2 - R \Delta M_K^2 + (M_K^2 -
\frac{r+1}{2} \mpi^2)\, \frac{r-1}{r+1}\ ,
\en
where
\eq
R = \frac{m_s - m}{m_d - m_u}\ ,
\ \ \Delta M_K^2 = (M_{K^0}^2 - M_{K^+}^2)_{QCD}\ ,
\en
which has remarkable properties:  Neglecting terms quadratic in $m_d-m_u$,
all contributions to $\Delta(r,R)$ of the type $m_{\rm quark}
B_0,~ m_{\rm quark}^2$ and even $m_{\rm quark}^2 \ln m_{\rm quark}$ cancel
out.  Consequently, in the standard $\chi$PT,~ $\Delta(r,R)|_{m_d=m_u}
 = O(m_{\rm quark}^3)$.  In generalised $\chi$PT,
$\Delta(r,R)$ also vanishes at leading order; however, it receives
$O({\rm p}^3)$ contributions from $\tilde{\cal L}^{(3)}$.  In both cases,
\eq\lbl{DeltaOp3}
\Delta(r,R)|_{m_d=m_u} = \left( \frac{F_K^2}{\fpi^2} -1\right)
\left(M_K^2 - \frac{r+1}{2}\mpi^2 \right) + (r-1)^2(r+1) \widehat{m}^3 \rho_2
+ O(m_{\rm quark}^3 B_0, m_{\rm quark}^4)\ .
\en

Let us write, as usual,
\eq
\Delta M_K^2 = (M_{K^0}^2 - M_{K^+}^2)_{\rm exp} +
\gamma (M_{\pi^+}^2 - M_{\pi^0}^2),
\en
where the coefficient $\gamma-1$ measures the violation of Dashen's
theorem. We shall take
\eq
\gamma = 1.8 \pm 0.4\  ,
\en
based on the recent estimates  \cite{MaKo}.
Taking for the ratio $R$ the value  \cite{GL3}
\eq
R = 43.7 \pm 2.7
\en
inferred from baryon mass splittings and from $\omega - \rho$ interference,
one obtains
\eq
\frac{R\Delta M_K^2}{\mpi^2} = 14.4 \pm 2.1\ .
\en
Upon inserting this value into Eq.  \rf{DeltaOp3}, one finds that the central
value of $\rho_2$ is largely dominated by the uncertainty:  For $r < 15$,
$\rho_2$ is compatible with zero; more precisely,
\eq
\frac{\widehat{m}^3\rho_2}{\mpi^2} \simeq \frac{-0.6 \pm 2.1}{(r-1)^2 (r+1)}\ .
\en
In comparison, for $r=25$, the above numerator would be
$\rm -2.9 \pm 2.1$.  It is interesting to notice
that the dimensional analysis of Section 5.1 gives (the error bars come from
the uncertainty on $m_s$)
\eq\lbl{DArho}
\left\vert\frac{\widehat{m}^3 \rho_i}{\mpi^2}\right\vert_{DA} =
\frac{0.4 \pm 0.2}{(r-1)^3}\ ,\ \ i=1,2\ .
\en
The constants $\rho_1$ and $\rho_2$ also describe the leading
$SU(3)_{\rm V}$-breaking effects in two-point functions
of the scalar and pseudoscalar densities, which might lead to additional
information. We have however not performed a complete analysis along these
directions so far. In the
sequel, we shall use the estimate  \rf{DArho}, since it is compatible
with the more direct determination from $\Delta(r,R)$ in the case of $\rho_2$.

\subsection{$\alpha$ and $\beta$ as Functions of the Quark Mass Ratio
$r=m_s/{\widehat m}$}

\indent

Up to now, nothing has been said about chiral logarithms. In generalised
$\chi$PT, they start to contribute at the order $O({\rm p}^4)$, {\it i.e.} in
the expansions of $\alpha$ and of $\beta$, they first appear in the
next-to-next-to leading order corrections $\alpha^{(2)}$ and $\beta^{(2)}$. As
already pointed out, one particular feature of generalised $\chi$PT is that
$O({\rm p}^4)$ loops renormalise the $O({\rm p}^2)$ constants $A_0$ and $Z_0^S$
by higher order terms proportional to $B_0^2$ (remember that $B_0$ counts as
$O({\rm p})$). Similarly, the constants $\xi$, $\tilde\xi$ and $\rho_i$ of
$\tilde{\cal L}^{(3)}$ are also subject to $O({\rm p}^4)$ renormalisations,
proportional to $B_0$. The constants of $\tilde{\cal L}^{(2)}$ and of
$\tilde{\cal L}^{(3)}$, which appear in the expressions  \rf{at} and  \rf{bt}
for
$\alpha^r$ and for $\beta^r$ (or in the formulae of Appendix A), are the
renormalised constants at scale $\mu$, with the $O({\rm p}^4)$ $B_0$-dependent
counterterms exhibited in Eq.  \rf{L4} included. For a quantitative discussion
of $O({\rm p}^4)$ effects, it is useful to explicitly split off these
$\mu$-dependent parts from the constants that are dominantly of order $O({\rm
p}^2)$ or $O({\rm p}^3)$. This separation is needed in particular when dealing
with the Zweig rule violating constants $Z_0^S$, $\tilde\xi$, $\rho_4$,
$\rho_5$ and $\rho_7$: While it makes perfect sense to
declare these constants to be small at
their respective $O({\rm p}^2)$ and $O({\rm p}^3)$ tree levels, they will,
however, be generated at the $O({\rm p}^4)$ accuracy by Zweig rule
violating chiral loops. In what follows, we shall {\it assume} that there
exists
a scale $\Lambda$ (not related to the renormalisation scale $\mu$) at which
these constants vanish. Upon using the coefficients $\Gamma_X$,
Eq.  \rf{scale}, of the corresponding beta functions \footnote{ For $Z_0^S$ and
for $\tilde\xi$, they can be inferred from the renormalisations of the
constants $L_6$ and $L_4$, respectively, as given in Ref.  \cite{GL2}, whereas
for the $\rho_i$'s, see Ref.  \cite{K}.}, this assumption amounts to writing
\bea\lbl{zweig}
{{4{\widehat m}^2Z_0^S}\over{\mpi^2}} &=& - {{11\mpi^2}\over{288\pi^2\fpi^2}}\,
x^2\,\ln {{\mu^2}\over{\Lambda^2}}\nonumber\\
2{\widehat m}{\tilde\xi} &=& - {\mpi^2\over{32\pi^2\fpi^2}}\,x\,
\ln {{\mu^2}\over{\Lambda^2}}\nonumber\\
{{{\widehat m}^3\rho_4}\over{\mpi^2}} &=& -{{\mpi^2}\over{192\pi^2\fpi^2}}\,
x\left( {11y\over 3} - 2z\right)\,\ln {{\mu^2}\over{\Lambda^2}}\nonumber\\
{{{\widehat m}^3\rho_5}\over{\mpi^2}} &=& -{{\mpi^2}\over{64\pi^2\fpi^2}}\,
x\left( y - {4z\over 3}\right)\,\ln {{\mu^2}\over{\Lambda^2}}\\
{{{\widehat m}^3\rho_7}\over{\mpi^2}} &=& -{{\mpi^2}\over{576\pi^2\fpi^2}}\,
xz\,\ln {{\mu^2}\over{\Lambda^2}}\nonumber\ ,
\ena
where we have introduced the notation
\eq\lbl{xyz}
x = {{2{\widehat m}F_0^2B_0}\over{\fpi^2\mpi^2}}\ ,
\ y = {{4{\widehat m}^2F_0^2A_0(\Lambda )}\over{\fpi^2\mpi^2}}\ ,
\ z = {{4{\widehat m}^2F_0^2Z_0^P}\over{\fpi^2\mpi^2}}\ .
\en
The vanishing at a common scale $\Lambda$ of all the $\tilde{\cal L}^{(2)}$
and $\tilde{\cal L}^{(3)}$
Zweig-rule violating constants is by
no means obvious. Here it is assumed mainly for reasons of simplicity.
An experimental determination
of the sizes of $Z_0^S(\Lambda )$ and of ${\tilde\xi}(\Lambda )$ is in
principle possible from a simultaneous analysis of $K_{l4}$ form factors and of
the low-energy $\gamma\gamma\to\pi^0\pi^0$ cross-section, when more precise
data on these quantities become available \footnote{An attempt to estimate the
possible effects of non-vanishing $Z_0^S$ and $\tilde\xi$ has been made in a
preliminary version of this work  \cite{HBK2}. See in particular Fig. 1 in this
reference.}.

It is now rather natural to use the scale $\Lambda$ in order to separate the
induced $O({\rm p}^4)$ components of the constants $A_0(\mu )$ and $\xi(\mu )$.
In terms of the notation introduced in Eq.  \rf{xyz}, one has
\bea\lbl{A0xi}
{{4{\widehat m}^2F_0^2A_0(\mu )}\over{\fpi^2\mpi^2}} &=&
y - {{5\mpi^2}\over{96\pi62\fpi^2}}\, x^2\, \ln{{\mu^2}\over{\Lambda^2}}
\nonumber\\
&&\\
2{\widehat m}\xi &=& 2{\widehat m}\xi(\Lambda )
- {{3\mpi^2}\over{32\pi^2\fpi^2}}\, x\, \ln{{\mu^2}\over{\Lambda^2}}\ .
\nonumber
\ena
Similarly, the constants $\rho_1(\mu )$ and $\rho_2(\mu )$ of $\tilde{\cal
L}^{(3)}$ which are not suppressed by the Zweig rule may be decomposed as
\bea\lbl{rho12}
{{{\widehat m}^3F_0^2\rho_1}\over{\fpi^2\mpi^2}} &=&
{\widehat\rho}_1 - {{\mpi^2}\over{64\pi^2\fpi^2}}\left( {xy\over 6} +xz\right)
\ln {{\mu^2}\over{\Lambda^2}}\nonumber\\
&&\\
{{{\widehat m}^3F_0^2\rho_2}\over{\fpi^2\mpi^2}} &=&
{\widehat\rho}_2 - {{\mpi^2}\over{64\pi^2\fpi^2}}\left( {xy\over 6} -3
xz\right)
\ln {{\mu^2}\over{\Lambda^2}}\nonumber\ ,
\ena
where
\eq
{\widehat \rho}_i = {{{\widehat m}^3 F_0^2 \rho_i (\Lambda
)}\over{\fpi^2\mpi^2}}\ \ \ ,\ i=1,\,2\ .
\en

We are now in a position to reexpress various contributions to $\alpha$
and $\beta$ in terms of observables, including $O(p^4)$ accuracy.  From
the expressions of $\fpi^2/F_0^2$ and $F_K^2/F_0^2$ given in Appendix A,
one obtains
\bea\lbl{2mxi}
2 \widehat{m} \xi(\Lambda) &=& \frac{2}{r-1} \left(F_K^2/\fpi^2 -1 \right)
      +  \frac{4}{(r-1)^2} \left(F_K^2/\fpi^2 -1 \right)^2
      - {4\widehat{m}^2 \over{r-1}}\,\left[ \delta_{F,K}^{(2,2)} -
                              \delta_{F,\pi}^{(2,2)}\right]\nonumber\\
   &+&  \frac{3\mpi^2}{32 \pi^2\fpi^2}\, x \, \ln (\mu^2/\Lambda^2)
    + \frac{1}{r-1} \, (-5\mu_\pi + 2\mu_K + 3\mu_\eta) + O({\rm p}^3),
\ena
where  $\delta_{F,K}^{(2,2)} - \delta_{F,\pi}^{(2,2)}$
collects all $O({\rm p}^4)$ tree contributions arising from ${\cal L}_{(2,2)}$.
Of course, the entire expression  \rf{2mxi} is $\mu$-independent.  All
results collected so far in this Section can now be used to reexpress
the {\em full} parameter $\alpha$, {\it i.e.} $\alpha^r(\mu)$ as given by Eq.
 \rf{at}, plus the logarithmic unitarity corrections displayed in Eq.
 \rf{a}, as follows:
\eq
\alpha = x + 4 y - \frac{4}{r-1} \left(\frac{F_K^2}{\fpi^2} -1 \right)
      + 81 \widehat{\rho}_1 + \widehat{\rho}_2 + \Delta_\alpha(\mu)
      + \omega_\alpha(x,y,r) + O({\rm p}^3).
\en
Here, $\Delta_\alpha(\mu)$, which is of order $O({\rm p}^2)$, collects
all the tree
contributions arising from ${\cal L}_{(2,2)}$ and ${\cal L}_{(0,4)}$. It
can be read off from Eqs.  \rf{at},  \rf{2mxi} and the formulae collected in
Appendix A; however, its
explicit expression would be of little use here.  $\omega_\alpha(x,y,r)$
involves all the logarithms arising both from tadpoles and from the
unitarity corrections:
\bea\lbl{omegaalpha}
\omega_\alpha(x,y,r) &=& - \frac{\mpi^2}{192 \pi^2\fpi^2}
\left[\frac{1}{3}(1945 + 352r + 31r^2)xy \right.\nonumber\\
&&\quad + (10-4r^2)xz + \left. \frac{2}{3}(148 + 11r)x^2 - 24x \right]
\ln (\mu^2/\Lambda^2)\nonumber\\
&+& \mu_\pi\left[\frac{10}{r-1} - 3 - 29y \right]
- 2\mu_K \left[\frac{2}{r-1} +1 +(r+7)y\right]\nonumber\\
&-& {1\over 3}\mu_\eta\left[\frac{18}{r-1} +1 + 15y -4(r-4)z\right]\\
&-& \frac{\mpi^2}{32\pi^2\fpi^2}(1 + 22y + 33y^2)
[\ln (\mpi^2/\mu^2) +1]\nonumber\\
&-& \frac{\mpi^2}{32\pi^2\fpi^2}[1 + (r+1)y][1+3(r+1)y]
[\ln (M_K^2/\mu^2) +1]\nonumber\\
&-& \frac{\mpi^2}{32\pi^2\fpi^2}\left[1 + (2r+1)y -
\frac{2}{r-1}{\widehat \Delta}_{GMO}\right]^2
[\ln (M_{\eta}^2/\mu^2) +1]\ .\nonumber
\ena
Similarly, the {\em full} parameter $\beta$, {\it i.e.} $\beta^r$ of Eq.
 \rf{bt}, plus unitarity corrections given in Eq.  \rf{b}, reads
\eq\lbl{beta(r)}
\beta = 1 + \frac{2}{r-1}\left(\frac{F_K^2}{\fpi^2} -1\right) +
     \beta_{\rm tree}^{(2)}(\mu) +
\beta_{\rm loop}^{(2)}(\mu) + O({\rm p}^3) \ .
\en
Here again, $\beta^{(2)}_{\rm tree}(\mu)$, which is of order $O({\rm p}^2)$,
collects all the tree contributions from ${\cal L}_{(2,2)}$, whereas
$\beta^{(2)}_{\rm loop}(\mu)$ contains the logarithms:
\bea
\beta^{(2)}_{\rm loop}(\mu)
&=& - \frac{\mpi^2}{16\pi^2\fpi^2}\,x\,\ln(\mu^2/\Lambda^2)\nonumber\\
&+& \frac{1}{r-1} (-5\mu_\pi + 2\mu_K + 3\mu_\eta)\nonumber\\
&-& \frac{\mpi^2}{16\pi^2\fpi^2}(2+5y)[\ln (\mpi^2/\mu^2) +1]\nonumber\\
&-& \frac{\mpi^2}{32\pi^2\fpi^2}[1+(1+r)y][\ln (M_K^2/\mu^2) +1]\ .
\ena
The parameters $x$ and $y$  \rf{xyz} satisfy the following two equations,
which may be obtained by multiplying the expansion formulae for $\fpi^2\mpi^2$
and $F_K^2M_K^2$ (see App. A) by the factor $F_0^2/\fpi^2\mpi^2$:
\bea\lbl{xy}
x+y &=& 1 - 9\widehat{\rho}_1 - \widehat{\rho}_2 + \Delta_\pi(\mu)
     + \omega_\pi(x,y,mr)\ , \nonumber\\
&&\\
x + \frac{r+1}{2}y &=& \frac{2}{r+1} \left(\frac{F_K M_K}{\fpi\mpi}\right)^2
     -3(r^2 +r +1)\widehat{\rho}_1 - (r^2 -r +1)\widehat{\rho}_2\nonumber\\
&+&\Delta_K(\mu) + \omega_K(x,y,r) \ .\nonumber
\ena
As before, the $\mu$-dependence of the tree contributions
$\Delta_{\pi,K}(\mu)$, which are $O({\rm p}^2)$, is
compensated by the collections
$\omega_{\pi,K}$ of all chiral logarithms, which read
\bea
\omega_\pi(x,y,r) &=& \frac{\mpi^2}{576\pi^2\fpi^2}\nonumber\\
&\times& \left[ (74+22r)x^2 +(253+88r+31r^2)xy -(42+12r^2)xz\right]
     \ln (\mu^2/\Lambda^2) \, ,\nonumber\\
&+& \mu_\pi (3x+8y) + 2\mu_K [x + (r+2)y] +
{1\over 3}\mu_\eta[x+4y+4(r-1)z]\nonumber\\
\omega_K(x,y,r) &=& \frac{\mpi^2}{192\pi^2\fpi^2}\nonumber\\
&\times& \left[\frac{1}{3}(59+37r)x^2 + (52+45r+27r^2)xy
-(16-10r+12r^2)xz \right]      \ln (\mu^2/\Lambda^2)\nonumber\\
&+& \frac{3}{4}\mu_\pi[2x+(5+r)y] + \frac{3}{2}\mu_K [2x +3(r+1)y]\nonumber\\
&+& \frac{1}{12} \mu_\eta[10x + (5+17r)y +8(r-1)z]\ .
\ena

In order to get the full expressions of the parameters $\alpha$ and $\beta$ as
functions of $r$ up to and including the $O(\widehat{m}^2)$ terms, it is
sufficient to solve the system of equations  \rf{xy} for $x$ and $y$.  This
can be done perturbatively:  $x,y$ and $z$ in the functions $\omega_\pi,
\omega_K$ can be replaced by their leading order values $x_0, y_0, z_0$,
\eq
x_0 = 1-y_0\ ,~~~~y_0 = \epsilon(r)\ ,~~~~z_0 = -\frac{\epsilon(r)}{2}
    + \frac{{\widehat\Delta}_{GMO}}{2(r-1)^2}\ ,
\en
since they multiply terms of order $O({\rm p}^2)$.  The final expression for
$\alpha$ as a function of $r$ then reads
\eq\lbl{alpha(r)}
\alpha(r) = 1 + 3 \epsilon^*(r) - \frac{4}{r-1}
\left( \frac{F_K^2}{\fpi^2} -1 \right) + 18(2-r)\widehat{\rho}_1
- 6r\widehat{\rho}_2 + \alpha^{(2)}(r) + O({\rm p}^3) \ ,
\en
where $\epsilon^*(r)$ is defined in Eq.  \rf{Op3corr} and the $O(p^2)$ part
$\alpha^{(2)}$ is given by $\alpha^{(2)}=\alpha^{(2)}_{\rm tree} +
\alpha^{(2)}_{\rm loop}$, with
\bea
\alpha^{(2)}_{\rm tree} &=& \frac{6}{r-1} (\Delta_K - \Delta_\pi) +
\Delta_\alpha
+ \Delta_\pi \ , \\
\alpha^{(2)}_{\rm loop} &=& \frac{6}{r-1} [\omega_K(x_0,y_0,r) -
\omega_\pi(x_0,y_0,r)]
+ \omega_\alpha(x_0,y_0,r) + \omega_\pi(x_0,y_0,r)\ .
\ena

The various contributions to $\alpha$ and $\beta$, displayed in  Eqs.
\rf{alpha(r)} and \rf{beta(r)}, are shown in Figures 8 and 9,
respectively.  The leading contribution $\alpha^{(0)}$ is given by Eq.
\rf{alpha0}, ($\beta^{(0)}$ = 1).  The $O({\rm p})$ correction
$\alpha^{(1)}$ consists of terms proportional to $F_K/\fpi - 1$ and of
the $\widehat{\rho}_1$ and $\widehat{\rho}_2$ terms, which are enhanced by
factors of $r$.  Using the  estimate \rf{DArho}, the uncertainty in
$\alpha^{(0)} + \alpha^{(1)}$  induced by the latter terms is shown as
the doubly-hatched area in Fig.  8a.  Notice that $\beta^{(0)} +
\beta^{(1)}$ (Fig. 9a) is free from such an uncertainty -- {\it cf.} the
first two terms in Eq. \rf{beta(r)}.   Considering the $O({\rm p^2})$
components $\alpha^{(2)}$ and  $\beta^{(2)}$, two facts should be kept
in mind.  First, there is at  present no direct way to determine the
constants of ${\cal L}_{(2,2)}$ and ${\cal L}_{(0,4)}$ entering the
corresponding tree contributions  $\alpha^{(2)}_{\rm tree}(\mu)$ and
$\beta^{(2)}_{\rm tree}(\mu)$.   Next, the  splitting of $\alpha^{(2)}$
and $\beta^{(2)}$ into the ($\mu$-dependent) tree and loop parts is a
matter of convention.  At present, the only  available information on
$\alpha^{(2)}$ and $\beta^{(2)}$ that is  unambiguous concerns the {\it
variation} of the loop contributions  $\alpha^{(2)}_{\rm loop}(\mu)$ and
$\beta^{(2)}_{\rm loop}(\mu)$ with the renormalisation scale $\mu$ (see
Figs. 8b and 9b).  We use this variation between $\mu = M_\eta$ and
$\mu$ = 1 GeV as an estimate of the size of the contributions of
$\alpha^{(2)}$ and $\beta^{(2)}$.  The hatched region of Figs 8a and 9a
then shows the total $\alpha$ and $\beta$ respectively, with
uncertainties arising from the $\rho$'s and from the $O({\rm p^2})$
contributions added.  (It should be noted that the final $\alpha(r)$ and
$\beta(r)$ displayed in Figs. 1a and 1b of Ref. \cite{HBK2} have been
obtained in a slightly different manner:  the contributions
$\alpha^{(2)}_{\rm loop}(\mu)$ and $\beta^{(2)}_{\rm loop}(\mu)$ have
been simply added to $\alpha^{(0)} + \alpha^{(1)}$ and $\beta^{(0)} +
\beta^{(1)}$ respectively.  While in the case of $\alpha$ the two
procedures practically coincide, the curve $\beta(r)$ shown in Fig. 1
of Ref. \cite{HBK2} should be considered as slightly overestimated.)
The final uncertainties in $\alpha(r)$ and $\beta(r)$ could be reduced
when an independent estimate of the constants from  ${\cal L}_{(2,2)}$
and ${\cal L}_{(0,4)}$ becomes available.

\subsection{The $SU(2)_{\rm L}\times SU(2)_{\rm R}$ GOR Ratio}

The above expression for the parameters $\alpha$ and $\beta$ as functions of
the quark mass ratio $r = m_s/\widehat{m}$ is characteristic of the
$SU(3)_{\rm L}\times SU(3)_{\rm R}$ $\chi$PT and
it is subject to the corresponding
uncertainty.  One may ask whether there is a direct relation between
$\alpha$ and $\beta$ and the condensate defined in the
$SU(2)_{\rm L}\times SU(2)_{\rm R}$ symmetry limit, which
would not be based on an
expansion in $m_s$ of a quantity such as $M_K^2$.  The existence of such
a relationship can be indeed be inferred already from the standard
$SU(2)_{\rm L}\times SU(2)_{\rm R}$ one-loop analysis
of Gasser and Leutwyler  \cite{GLlett,GL1}. In
this case, the parameters $\alpha$ and $\beta$ are given by Eqs.
 \rf{aGL} and  \rf{bGL} in terms of the constants $\bar{\ell}_3$ and
$\bar{\ell}_4$.  It turns out that $\bar{\ell}_3$ also measures the
deviation from the $SU(2)_{\rm L}\times SU(2)_{\rm R}$ GOR relation
\cite{GL1}:
\eq
\frac{2\widehat{m}\bar{B}}{\mpi^2} = 1 + \frac{\mpi^2}{32\pi^2\fpi^2}
\bar{\ell}_3 + O(\mpi^4) \ ,
\en
where
\eq\lbl{Bbarlim}
\bar{B} = - \lim_{\stackrel{m_u,m_d \rightarrow 0}{m_s~fixed}}
\langle\Omega|\bar{u}u|\Omega\rangle/\fpi^2\ .
\en
Eliminating the constants $\bar{\ell}_3$ and $\bar{\ell}_4$, one obtains
\eq\lbl{2mBbar}
\frac{2\widehat{m}\bar{B}}{\mpi^2} = 1 - \frac{\alpha - \beta}{3}
 + \frac{\mpi^2}{32\pi^2\fpi^2} + O(\mpi^4)\ .
\en
This formula clearly
shows that the  ratio on the left hand side should be indeed
considerably less than 1, provided $\alpha$ is as large as indicated by the
central value of existing data.  However, Eq.  \rf{2mBbar} is based on the
standard $\chi$PT and there is no need for it to hold if either $\alpha$
or $\beta$ differ from 1 by more than an amount of order
$O(\mpi^2/\Lambda_H^2)$.
It is convenient to rewrite Eq.  \rf{2mBbar} in terms of the GOR ratio
\eq\lbl{xGORdef}
x_{GOR} = -\frac{2\widehat{m}}{\fpi^2\mpi^2}
\lim_{\stackrel{m_u,m_d \rightarrow 0}{m_s~fixed}}
\langle \Omega | \bar{u}u | \Omega \rangle \nonumber\\
= \frac{2\widehat{m}\bar{B}\bar{F}^2}{\fpi^2\mpi^2}\ ,
\en
where $\bar{F} = \lim_{m_u,m_d \rightarrow 0}\fpi$.  Using the standard
$\chi$PT relation
\eq
\frac{\fpi^2}{\bar{F}^2} = \beta + \frac{\mpi^2}{8\pi^2\fpi^2}\ ,
\en
Eq.  \rf{2mBbar} can be reexpressed as
\eq\lbl{xab}
x_{GOR} = 2 - \frac{\alpha + 2\beta}{3} - \frac{3\mpi^2}{32\pi^2\fpi^2}\ .
\en
Within the standard expansion, Eqs.  \rf{2mBbar} and  \rf{xab} are strictly
equivalent. It
turns out, however, that Eq.  \rf{xab} is of a more general validity: It
remains
true at the leading order of G$\chi$PT, {\it i.e.} irrespective of how much
$\alpha$ does deviate from 1. Here, we are going to discuss the $O({\rm p})$
and $O({\rm p}^2)$ corrections to Eq.  \rf{xab}.

The $SU(2)_{\rm L}\times SU(2)_{\rm R}$
condensate  \rf{Bbarlim} can be obtained from Appendix A as follows
\eq
{{{\bar F}^2}\over{F_0^2}}\,{\bar B} =
\lim_{\stackrel{{\widehat m}\rightarrow
0}{m_s~fixed}}{{\fpi^2\mpi^2}\over{2{\widehat m}F_0^2}}\ .
\en
The result is given in the form of an expansion in powers of $m_s$,
\bea\lbl{Bbar}
{{{\bar F}^2}\over{F_0^2}}\,{\bar B} &=&
B_0 + 2m_s Z_0^S + m_s^2\left( \rho_4 +{1\over 2}\rho_5 + 6\rho_7\right)
\nonumber\\
&+& m_s^3\left( F_2^S + F_3^S + 4F_5^{SS} + 2F_6^{SS}\right)
\nonumber\\
&-& 2{\bar \mu}_K\left( B_0 + 2m_sA_0 + 6m_sZ_0^S\right)\nonumber\\
&-& {1\over 3}{\bar \mu}_{\eta}\left( B_0 + 10m_sZ_0^S - 8m_sZ_0^P\right)\ ,
\ena
where ${\bar M}_K$ and ${\bar M}_{\eta}$ denote the kaon and eta masses in the
$SU(2)_{\rm L}\times SU(2)_{\rm R}$ chiral limit, and
\eq
{\bar \mu}_{K, \eta} = \lim_{\stackrel{{\widehat m}\rightarrow 0}{m_s~fixed}}
\mu_{K, \eta} = {{{\bar M}_{K, \eta}}\over{32\pi^2{\bar F}^2}}
\ln \left({{{\bar M}^2_{K,\eta}}\over{\mu^2}}\right)\ .
\en
The formula  \rf{Bbar} collects Zweig-rule violating contributions to the
$SU(2)_{\rm L}\times SU(2)_{\rm R}$ condensate ${\bar B}$
which are induced by the
non-vanishing strange quark mass. It includes, in particular, $K$ and $\eta$
loops, which, being proportional to a power of $m_s$ (or of $r$), may be rather
large and sensitive to the subtraction scale $\mu$. This does not mean,
however, that the difference ${\bar F}^2{\bar B}-F_0^2B_0$ of the two-massless
flavour and three-massless flavour condensates should be expected to be large,
but rather that the chiral logarithms alone can hardly be used here in order to
estimate the size of this difference.

One may now proceed as follows. First, one subtracts from
$\fpi^2\mpi^2\alpha_r$ given by Eq.  \rf{alpha^r} the expansion of
$4\fpi^2\mpi^2$ displayed in Appendix A. This eliminates ${\widehat m}^2A_0$.
Then, one replaces
$F_0^2B_0$ by ${\bar F}^2{\bar B}$, using Eq.  \rf{2mBbar}. One finds that the
dangerous Zweig-rule violating terms enhanced by higher powers of $r$ cancel.
Next, one eliminates ${\widehat m}(\xi + 2{\tilde\xi})$ in favour of $\beta^r$,
using Eq.  \rf{beta^r}. Finally, one replaces $\alpha^r$ and $\beta^r$ by
$\alpha$ and
$\beta$, introducing the unitarity logarithms displayed in Eqs.  \rf{a}
and
 \rf{b}. As usual ({\it cf.} the discussion in Section 11 of Ref. \cite{GL2}),
the remainders of $K$ and $\eta$ loops are
absorbed into the $O({\rm p}^4)$ tree-level constants. In this way, one arrives
at the desired relation
\eq\lbl{xGORSU2}
x_{GOR} = 2 - {{\alpha + 2\beta}\over{3}} + 15{\widehat \rho}_1 - {\widehat
\rho}_2 +\Delta x (\mu ) + x_{GOR}^{\rm loop}(\mu )\ .
\en
Here, $\Delta x (\mu ) = O({\widehat m}^2)$ collects all (redefined) tree
contributions arising from ${\cal L}_{(2,2)}$ and ${\cal L}_{(0,4)}$. The
$O({\rm p}^2)$ pion-loop contribution reads
\bea
x_{GOR}^{\rm loop} &=& -{{\mpi^2}\over{192\pi^2\fpi^2}}\left({311y\over 3} +
22z\right)\,\ln{{\mu^2}\over{\Lambda^2}}\nonumber\\
&& - \mu_{\pi}\left(17y + 11y^2\right) -{{\mpi^2}\over{32\pi^2\fpi^2}}
\left( 3 + 14y + 11y^2\right)\ ,
\ena
where the parameter $y$ may be replaced by its leading order expression in the
$SU(2)_{\rm L}\times SU(2)_{\rm R}$ expansion:
\eq
y = -2z = 1 - x_{GOR}\vert_{\rm lead} = {{\alpha + 2\beta}\over 3} - 1\ .
\en
Eq.  \rf{xGORSU2} gives the $SU(2)_{\rm L}\times SU(2)_{\rm R}$
GOR ratio  \rf{xGORdef} as a function of
$(\alpha + \beta )/3$. It is seen that for $\alpha$ and $\beta$ close to 1, one
recovers the standard $\chi$PT formula  \rf{xab} up to and including
$O({\rm p}^2)$ accuracy.  Notice that, using the estimate \rf{DArho}, the
$O({\rm p})$ contribution  $15 \widehat{\rho}_1$ - $\widehat{\rho}_2$ to
$x_{GOR}$ becomes negligible in the whole range of $r$.  Fig. 10a shows the
GOR ratio \rf{xGORSU2} with the $O({\rm p^2})$ uncertainty obtained from
$x_{GOR}^{\rm loop}(\mu)$ as described at the end of Subsection 5.3.

\section{Summary and Conclusions}

\indent
\setcounter{equation}{0}

In the present article, we have obtained the explicit expression of the
$\pi\pi$ scattering amplitude $A(s\vert t,\,u)$ to two-loop accuracy
in the chiral expansion.
This constitutes our main result, and it is contained in
Eqs.  \rf{amp2},  \rf{Kstu} and
 \rf{Kbar}. At this order, the $\pi\pi$ amplitude is entirely
determined by the general
requirements of analyticity, crossing symmetry, unitarity and the
Goldstone-boson nature of the pion, up to the six independent parameters
$\alpha$, $\beta$, $\lambda_i$, $i=1,2,3,4$, which are
not fixed by chiral symmetry.

For the four constants $\lambda_i$, we have established a set of sum rules
which are evaluated using available information on $\pi\pi$ interaction at
medium and high energies ($E\gapprox 600$ MeV). Details about the derivation of
these sum rules and the corresponding numerical analysis will be published
separately. The resulting values of the $\lambda_i$'s
are given in Eqs.  \rf{lam12} and  \rf{lam34}. For $\lambda_1$ and
$\lambda_2$, they are compatible with previous determinations, but with a much
smaller error bar in the case of $\lambda_1$.
The constants $\lambda_3$ and $\lambda_4$ are genuine
$O({\rm p}^6)$ contributions. To the best of our knowledge, they
have not been discussed before.

The most important improvement when going from one loop to two loops concerns
unitarity: The latter is only satisfied in the perturbative sense by the
amplitude $A(s\vert t,\,u)$. A study of the Argand plots for the lowest partial
waves ({\it cf.} the
partial wave $f_0^0$ in Figure 1)
shows that violations of the unitarity bounds become
important only at energies greater than 450 MeV. From the size of the two-loop
corrections, we also expect that higher orders are under control. Typically,
for
{\it e.g.} the S-wave scattering lengths, the one-loop corrections
represent a 25\% effect. The two-loop contributions, however, do not modify the
one-loop result by more than 5\%, independently of the value of $\alpha$.
(At this level of accuracy, isospin-breaking effects due to $m_d \neq m_u$
and to electromagnetism could start to play a role.  We intend to discuss
these effects elsewhere.)

As far as the two remaining parameters, $\alpha$ and $\beta$, are concerned,
let us stress once more the particular importance attached to the first one:
$\alpha$ measures the amount of explicit chiral symmetry breaking in the
$\pi\pi$ amplitude due to the non-vanishing of the light quark masses.
As such,
it is intimately correlated to the ratio $x=-2{\widehat m}\langle 0\vert
{\bar q}q\vert 0\rangle/\fpi^2\mpi^2$
or equivalently, to the quark mass ratio $r=m_s/{\widehat m}$.
The commonly
accepted picture that spontaneous breakdown of chiral symmetry results from
a strong condensation of quark-antiquark pairs in the QCD vacuum requires that
$x\sim 1$  (or $r\sim$25), and it is only compatible with values of
$\alpha$ and $\beta$ close to unity, as predicted by standard $\chi$PT.
A sizeable deviation of $\alpha$ from 1, say {\it e.g.} $\alpha\gapprox 2$,
even if not ``a major earthquake"  \cite{Leut2}, would nevertheless shake our
confidence in the validity of the standard lore.
These connections between $\alpha$ and the ratios $x$ or $r$
are well established at leading, $O({\rm p}^2)$,
and at next to leading, $O({\rm p}^3)$, orders, {\it cf.} Eqs.
 \rf{xGORSU2}, \rf{alpha(r)} and  \rf{beta(r)} and Figs. 8, 9, and 10.
Our estimates of $O({\rm p}^4)$ corrections
suggest that these correlations do not suffer major changes
even at this order.  In the particular case of the standard $\chi$PT
($r \sim 25$), higher order corrections to $\alpha$ are under even tighter
control:  The large-condensate hypothesis can hardly be compatible with
$\alpha > 1.2$.
Thus, an accurate determination of
$\alpha$ and $\beta$ would confirm or contradict the standard
picture of the chiral structure of the QCD vacuum.
Unfortunately, the result of a fit to the presently
available $K_{l4}$ data  \cite{rosselet} remains inconclusive in this respect.
We obtain
$$
\alpha = 2.16\pm 0.86\ ,\ \ \beta = 1.074\pm 0.053\ .
$$
Clearly, additional information, which would make such fits more accurate
is needed. This may be provided by new high statistics $K_{l4}$ experiments,
that could be done {\it e.g.}, at DA$\Phi$NE, or by a precise measurement of
the lifetime of $\pi^+\pi^-$-atoms at CERN.

It is, however, interesting to stress that the above fit result
leads to values for the threshold parameters which are in perfect agreement
with
the results obtained some time ago from analyses of $K_{l4}$ data
based on the Roy equations.
For instance, the fitted value of the I=0 S-wave scattering length, $a_0^0 =
0.263\pm 0.052$, reproduces the result quoted in  \cite{nagels}. Similar
conclusions hold for the other threshold parameters, see Table 1. It seems thus
quite reasonable to conclude that in the energy range accessible in $K_{l4}$
decays, our two-loop chiral result contains all the
relevant information on the $\pi\pi$ interaction which is already
encoded in the Roy equation (for instance, the constraints
coming from the production data at higher energies are satisfactorily
reproduced by our determination of the $\lambda_i$'s).

The fact that the two-loop expression of the $\pi\pi$ scattering amplitude was
obtained, in Section 3, in a rather direct and easy way from general S-matrix
properties, and without any reference
to the effective lagrangian ${\cal L}^{\rm eff}$ or to Feynman diagrams, should
not be surprising. After all, the effective lagrangian is merely a technical
device which collects the general properties of transition amplitudes among
Goldstone bosons. All the results obtained with the help of the effective
lagrangian are, in principle, reproducible using directly Ward Identities,
analyticity, crossing symmetry and unitarity within a systematic low-energy
expansion of QCD correlation functions.
It just happens that in the case of the $\pi\pi$ amplitude, it is
technically simpler to use the direct S-matrix method than to evaluate all
two-loop graphs generated by the effective lagrangian. In any case, the results
have to be identical in both approaches. We have shown that this is indeed the
case at the $O({\rm p}^4)$ level (see Section 3.2), and since we understand
that work towards an effective lagrangian computation of $A(s\vert t,\, u)$ to
order $O({\rm p}^6)$ in
the standard framework is in progress  \cite{col,BCEG}, we hope that it will
soon become possible to check this statement to two loops
by a direct comparison at least in this particular case. Undertaking a similar
enterprise in the generalised case would be considerably more difficult,
due to the proliferation of low-energy constants in ${\cal L}^{\rm eff}$
to that order.  For the same reason, it is not clear that it would be
particularly useful.

\section*{Acknowledgments}

This work was supported in part by the EEC Human Capital and Mobility
Progam, EEC-Contract No. CHRX-CT920026 (EURODA$\Phi$NE).  Useful
discussions with H. Bijnens, G. Colangelo,  G. Ecker, J. Gasser, J.
Kambor, H. Leutwyler, L. Montanet, D. Morgan, L. Nemenov, M. R.
Pennington and J. Portol\`es are greatly acknowledged.

\newpage

\noindent
\appendix{\large\bf Appendix A $\quad$Expansions of Masses
and of Decay Constants \\
\mbox{\hspace{4.8cm}} to Order $O({\rm p}^4)$}

\indent
\renewcommand{\theequation}{A.\arabic{equation}}
\setcounter{equation}{0}

{}From the expression of the effective action at order $O({\rm p}^4)$ one
computes
\eq
{{F_{\pi}^2}\over{F_0^2}} = 1 + 2{\widehat m}[\xi +(2+r){\tilde\xi}]
+ 2{\widehat m}^2 {\delta}_{F,\pi}^{(2,2)} -4{\mu}_{\pi} -2{\mu}_K\ ,
\lbl{Fpi}
\en
and
\eq
{{F_K^2}\over{F_0^2}} = 1 + {\widehat m}[(r+1)\xi +2(2+r){\tilde\xi}]
+ 2{\widehat m}^2 {\delta}_{F,K}^{(2,2)} -{3\over 2}{\mu}_{\pi} -3{\mu}_K
-{3\over 2}\mu_{\eta}\ ,
\lbl{Fka}
\en
where $\delta_{F,\pi}^{(2,2)}$ and $\delta_{F,K}^{(2,2)}$
contain the contributions from ${\cal
L}_{(2,2)}$, {\it cf.} Eq.  \rf{L22},
\bea
{\delta}_{F,\pi}^{(2,2)} &=& A_1 +{1\over 2}A_2 +2A_3 +
{1\over 2}(2+r^2)(A_4 +2B_4)
+ B_1 - B_2 + 2(2+r)D^S\ ,\nonumber\\
&&\\
{\delta}_{F,K}^{(2,2)} &=& {1\over 2}(1+r^2)(A_2 + A_3 + B_1) +
{r\over 2}(A_2 + 2A_3 - 2B_2) \nonumber\\
&+& {1\over 2}(2+r^2)(A_4 +2B_4) + (r+1)(2+r)D^S\ .\nonumber
\ena
Similarly, the expansion of the pion and kaon masses read
\bea
{{\fpi^2}\over{F_0^2}}\mpi^2
&=&  2{\widehat m}B_0 + 4{\widehat m}^2 A_0
+ 4(2+r){\widehat m}^2 Z_0^S \nonumber\\
&+& 2{\widehat m}^3\delta_{M,\pi}^{(0,3)} + 2{\widehat
m}^4\delta_{M,\pi}^{(0,4)} + 4{\widehat m}^2\mpi^2A_3\nonumber\\
&-& \mu_{\pi}\left[ 3\mpi^2 +20{\widehat m}^2 (A_0 + 2Z_0^S)\right]
\lbl{Mpi}\\
&-& 2\mu_K\left[ \mpi^2 + 4(1+r){\widehat m}^2 (A_0 +2Z_0^S)\right]\nonumber\\
&-& {1\over 3}\mu_{\eta}\left[ \mpi^2 + 4(1+2r){\widehat m}^2
(A_0 + 2Z_0^S) + 8(1-r){\widehat m}^2 (A_0 + 2Z_0^P)\right]\ ,\nonumber
\ena

\bea
{{F_K^2}\over{F_0^2}}M_K^2
&=&  (r+1){\widehat m}B_0 + (r+1)^2{\widehat m}^2 A_0
+ 2(r+1)(2+r){\widehat m}^2 Z_0^S \nonumber\\
&+& (r+1){\widehat m}^3\delta_{M,K}^{(0,3)} + (r+1){\widehat
m}^4\delta_{M,K}^{(0,4)} + (r+1)^2{\widehat m}^2 M_K^2A_3\nonumber\\
&-&{3\over 2}\mu_{\pi}\left[ M_K^2 +4(r+1){\widehat m}^2
(A_0 + 2Z_0^S)\right]
\lbl{Mka}\\
&-& 3\mu_K\left[ M_K^2 + 2(1+r)^2{\widehat m}^2 (A_0 +2Z_0^S)\right]\nonumber\\
&-& {1\over 6}\mu_{\eta}\left[ 5M_K^2 + 4(r+1)(1+2r){\widehat m}^2
(A_0 + 2Z_0^S) - 4(1-r^2){\widehat m}^2 (A_0 + 2Z_0^P)\right]\ .\nonumber
\ena
\newpage
The $O({\rm p}^3)$ tree-level contributions $\delta_{M,P}^{(0,3)}$, $P=\pi , K$
are given as follows
\bea
\delta_{M,\pi}^{(0,3)}&=&{9\over 2}\rho_1 + {1\over 2}\rho_2 +
                         (10+4r+r^2)\rho_4 + {1\over 2}(2+r^2)\rho_5 +
                         6(2+r)^2\rho_7\ ,\nonumber\\
&&\\
\delta_{M,K}^{(0,3)}&=&{3\over 2}(1+r+r^2)\rho_1 + {1\over 2}(1-r+r^2)\rho_2 +
                         3(2+2r+r^2)\rho_4 + {1\over 2}(2+r^2)\rho_5 +
                         6(2+r)^2\rho_7\ .\nonumber
\ena
The $O({\rm p}^4)$ tree level contributions from ${\tilde{\cal L}}^{(0,4)}$
contained in $\delta_{M,\pi}^{(0,4)}$ and in $\delta_{M,K}^{(0,4)}$ are not
displayed explicitly.

\newpage

\noindent
\appendix{\large\bf Appendix B $\quad$The Polynomials $\xi_a^{(n)}(s)$}

\indent
\renewcommand{\theequation}{B.\arabic{equation}}
\setcounter{equation}{0}

We give here the list of polynomials $\xi_a^{(n)}(s)$ which define the one-loop
partial-wave projections in Eqs.  \rf{faNLO} and  \rf{psia}. The center-of-mass
momentum $q$ of the pions is given (in units of $\mpi$) by
\eq
q = \sqrt{{{s-4\mpi^2}\over{4\mpi^2}}}\ .
\en

\bea
\xi_0^{(0)}(s) &=& {1\over{144\pi^2}}\,\left[ 35\alpha^2 +
80\alpha\beta + 134\beta^2\right] + 10\,(\lambda_1 +
2\lambda_2)\nonumber\\
&+& \,
\left\{ {1\over{72\pi^2}}(60\alpha + 209\beta )\beta + 16(2\lambda_1 +
3\lambda_2)\right\}\, q^2\nonumber\\
&+& \,
\left\{ {311\over{108\pi^2}}\beta^2 + {8\over 3}(11\lambda_1 + 14\lambda_2)
\right\}\, q^4\nonumber\\
&&\nonumber\\
\xi_0^{(1)}(s) &=& {1\over{192\pi^2}}\,\left[ 5\alpha^2 + 4\beta^2
\right] +
{1\over{9\pi^2}}\,
{\beta^2}\, q^2 +
{7\over{36\pi^2}}\, \beta^2\, q^4\nonumber\\
&&\\
\xi_0^{(2)}(s) &=& {1\over{1152\pi^2}}\,\left[ 5\alpha + 16\beta
+ 24\,\beta\, q^2\right] ^2\nonumber\\
&&\nonumber\\
\xi_0^{(3)}(s) &=& {1\over{288\pi^2}}\, \left[ -5\alpha^2
+4\beta^2\right] + {1\over{12\pi^2}}\,
\beta^2\, q^2\nonumber\\
&&\nonumber\\
\xi_0^{(4)}(s) &=& 0\nonumber\\
&&\nonumber\\
&&\nonumber\\
\xi_2^{(0)}(s) &=& {1\over{288\pi^2}}\,\left[ 31\alpha^2 -
122\alpha\beta + 220\beta^2\right] + 4 \,(\lambda_1 + 2\lambda_2)\nonumber\\
&+& \,
\left\{ {1\over{144\pi^2}}(-69\alpha + 268\beta )\beta + 8(\lambda_1 +
3\lambda_2)\right\}\, q^2\nonumber\\
&+& \,
\left\{ {265\over{216\pi^2}}\beta^2 + {16\over 3}(\lambda_1 + 4\lambda_2)
\right\}\, q^4\nonumber\\
&&\nonumber\\
\xi_2^{(1)}(s) &=& {1\over{576\pi^2}}\,\left[ 9\alpha^2
-42\alpha\beta + 60\beta^2 \right] \nonumber \\
&+& {1\over{144\pi^2}}\,
(-9\alpha + 37\beta )\,\beta\, q^2 \\
&+& {{11}\over{72\pi^2}}\,\beta^2\, q^4\nonumber
\ena
\newpage
\bea
\xi_2^{(2)}(s) &=& {{1}\over{288\pi^2}}\,\left[ \alpha - 4\beta
- 6\beta\, q^2\right] ^2\nonumber\\
&&\nonumber\\
\xi_2^{(3)}(s) &=& {1\over{288\pi^2}}\, \left[ -3\alpha^2
+2\alpha\beta -12\beta^2\right]
- {1\over{24\pi^2}}\,
\beta^2\, q^2\nonumber\\
&&\nonumber\\
\xi_2^{(4)}(s) &=& 0\nonumber\\
&&\nonumber\\
&&\nonumber\\
\xi_1^{(0)}(s) &=& {1\over{576\pi^2}}\,\left\{ 5\alpha^2 -
80\alpha\beta + 10\beta^2\right\}\nonumber\\
&+&  \left\{ {1\over {432\pi^2}}\left[ 55\alpha
 - 68\beta\right]\,\beta  - {8\over 3}\left( \lambda_1 - \lambda_2\right)
\right\}\, q^2\nonumber\\
&-& \left\{ {{\beta^2}\over{108\pi^2}} + {8\over 3}\left( \lambda_1 -
\lambda_2\right)\right\}\, q^4\nonumber\\
&&\nonumber\\
\xi_1^{(1)}(s) &=& {1\over{288\pi^2}}\,\left( -5\alpha
+ 7\beta \right)\,\beta\,  + {1\over{144\pi^2}}
\left( 5\alpha - 3\beta\right)\,\beta\, q^2 -
{1\over{72\pi^2}}\,\beta^2\, q^4\nonumber\\
&&\nonumber\\
\xi_1^{(2)}(s) &=& {1\over{72\pi^2}}
\,\beta^2\, q^4\\
&&\nonumber\\
\xi_1^{(3)}(s) &=& {1\over{864\pi^2}}\, \left[ -5\alpha^2
+10\alpha\beta +28\beta^2\right] + {1\over{24\pi^2}}
\,\beta^2\, q^2\nonumber\\
&&\nonumber\\
\xi_1^{(4)}(s) &=& -{{5}\over{144\pi^2}}\left[ \alpha^2 +
4\alpha\beta - 2\beta^2\right]\nonumber
\ena

\newpage

\noindent
\appendix{\large\bf Appendix C $\quad$The Loop Integrals ${\bar K}_n(s)$}

\indent
\renewcommand{\theequation}{C.\arabic{equation}}
\setcounter{equation}{0}

In this Appendix, we discuss some properties of the functions ${\bar K}_n(s)$,
{\it cf.} Eqs.  \rf{Kn},  \rf{kn} and  \rf{Kbar}. Let us start with the
function
$\bar{J}(s)$ which denotes the usual $d$-dimensional one-loop integral
\eq
J(p^2)= -i \int \frac{d^dq}{(2\pi)^d} \ \frac{1}{(M_{\pi}^2 - q^2)
[M_{\pi}^2 - (q-p)^2]}
\en
subtracted at $p^2=0$.  $\bar{J}(s)$ is finite for $d=4$ and may be
expressed in the standard parametrical form  \cite{GL1,GL2}
\eq
\bar{J}(s) = -\frac{1}{16 \pi^2} \int_0^1 d\lambda \ln \left[1 -
\frac{s}{M_{\pi}^2}\lambda(1-\lambda)\right]\ .
\en
For $s < 4M_{\pi}^2$, the latter formula can be integrated by parts and,
after introducing a new integration variable
\eq
x = \frac{M_{\pi}^2}{\lambda(1-\lambda)}\ ,
\en
it can be transformed into the dispersive representation  \rf{J}.  For
real $s$, one has
\eq
16\pi^2 \bar{J}(s) = \left \{
\begin{array}{ll}
2 + \sigma(\ln\frac{1-\sigma}{1+\sigma} +
i\pi),& \mbox{if $s \ge 4M_{\pi}^2$}\\
2 - 2\left({{4\mpi^2-s}\over s}\right)^{1\over 2}
\, {\rm arctg}\left({s\over{4\mpi^2-s}}\right)^{1\over 2},
&\mbox{if $0 \le s \le 4M_{\pi}^2$}\\
2 + \sigma \ln\frac{\sigma -1}{\sigma + 1},
&\mbox{if $s \le 0$},
\end{array}
\right.
\en
where
\eq
\sigma = \left( 1 - \frac{4M_{\pi}^2}{s}\right)^{1/2}\ .
\en
The function $\bar{J}(s)$ was known at times as the Chew-Mandelstam
function  \cite{chewman}.

In order to simplify the discussion of the two-loop integrals, it is
convenient to introduce the notation
\eq
F(s) \equiv 16 \pi^2 \bar{J}(s) - 2\ .
\en

The function $F(s)$ vanishes at threshold; more precisely, for
$s-4M_{\pi}^2 \rightarrow 0^+$, one has the expansion
\eq
F(s) = i\pi\sigma - 2\sigma^2 - \frac{2}{3}\sigma^4 -
\frac{2}{5}\sigma^6 + \ldots ,
\en
whereas for $s \rightarrow 0$,
\eq
F(s) = -2 + \frac{1}{6}\frac{s}{M_{\pi}^2} + \ldots.
\en
The dispersion integrals  \rf{Kn} can now be directly
determined by algebraic manipulations with the functions $F(s)$, without
calculating a single integral.  It is sufficient to construct functions
that are analytic except for a branch-cut singularity on the positive real
half-axis starting at $s = 4 M_{\pi}^2$, with the
discontinuities given by the functions $k_n(s)$, Eqs.  \rf{kn}:
\eq
{\rm Im} \bar{K}_n(s) \equiv \frac{1}{2i}[\bar{K}_n(s+i\epsilon) -
\bar{K}_n(s-i\epsilon)] = k_n(s) \theta(s - 4M_{\pi}^2)\ .
\en
Such analytic functions are, of course, only defined up to a
polynomial.  The latter, however, can be unambiguously fixed, using the
boundary conditions
\eq
\lim_{|s| \rightarrow \infty} s^{-1} \bar{K}_n(s) = 0\ ,~~~~~ \bar{K}_n(0)=0\ ,
\en
which follow from Eqs.  \rf{kn} and  \rf{Kn}.  The main ingredients of this
construction are the analytic functions $F(s)$, $F^2(s)$, and $F^3(s)$,
with the discontinuities
\begin{eqnarray}
{\rm Im} F(s) &=& \pi \sqrt{\frac{s-4M_{\pi}^2}{s}}\, \theta (s-4M_{\pi}^2)\\
{\rm Im} F^2(s) &=& 2\pi \left(\frac{s-4M_{\pi}^2}{s}\right) L(s)
\,\theta (s-4M_{\pi}^2)\\
{\rm Im} F^3(s) &=& \left(\frac{s-4M_{\pi}^2}{s}\right)^{3/2}
( 3\pi L^2(s) - \pi^3 )\,\theta (s-4M_{\pi}^2)\ ,
\end{eqnarray}
where $L(s)$ is the logarithmic function  \rf{L}.  Since
$F(4M_{\pi}^2)=0$, new analytic functions with the desired
discontinuities can be obtained upon dividing by $(s-4M_{\pi}^2)$.  This
leads to
\eq
{\rm Im}\, \frac{F(s)}{s-4M_{\pi}^2} = \frac{\pi}{\sqrt{s(s-4M_{\pi}^2)}}
\,\theta(s-4M_{\pi}^2)\ ,
\en
\eq
{\rm Im}\, \frac{sF^2(s)}{s-4M_{\pi}^2} = 2\pi L(s)\,\theta(s-4M_{\pi}^2)\ .
\en
This procedure can be extended further:  Subtracting from the
function $sF^2(s)/(s-4M_{\pi}^2)$ its threshold value
\eq
\lim_{s \rightarrow 4M_{\pi}^2}\, \frac{sF^2(s)}{s-4M_{\pi}^2} = - \pi^2\ ,
\en
one obtains from Eq. (C.15)
\eq
{\rm Im}\, \frac{1}{s - M_{\pi}^2} \left[ \frac{sF^2(s)}{s - M_{\pi}^2}
   + \pi^2 \right] = \frac{2\pi L(s)}{s - M_{\pi}^2}\,\theta(s-4M_{\pi}^2)\ .
\en
The functions $k_0(s)$, $k_1(s)$ and $k_2(s)$ can be recognised on the
right-hand sides of Eqs. (C.11), (C.15) and (C.12) respectively.
Similarly, Eqs. (C.14) and (C.17) reproduce the first two terms in the
expression of
$k_4(s)$.  The remaining terms, quadratic in $L(s)$, can be inferred from
Eq. (C.13):  The function $F^3(s) + \pi^2 (s - 4M_{\pi}^2)F(s)/s$
exhibits a double zero at $s=4M_{\pi}^2$.  Consequently, one can write
\eq
{\rm Im}\, \frac{s}{(s - 4M_{\pi}^2)^2} \left [ F^3(s) + \frac{\pi^2}{s}
(s - 4M_{\pi}^2)F(s) \right] =
\frac{3\pi L^2(s)}{\sqrt{s(s - 4M_{\pi}^2})}\,\theta(s-4M_{\pi}^2)\ .
\en
Finally, subtracting from the analytic function on the left-hand side of
the last equation its threshold value, one obtains
\begin{eqnarray}
{\rm Im}\, \frac{1}{s - 4M_{\pi}^2} \left \{ \frac{M_{\pi}^2}{(s -
4M_{\pi}^2)^2}
\left[sF^3(s) + \pi^2 (s - 4M_{\pi}^2)F(s) \right]
- \pi^2 \right \} = \nonumber \\
\frac{3\pi M_{\pi}^2}{s - 4M_{\pi}^2}\cdot \frac{L^2(s)}{\sqrt{s(s -
4M_{\pi}^2)}}\,\theta(s-4M_{\pi}^2)\ .
\end{eqnarray}
All the elements needed to write down the functions ${\bar K}_n(s)$ are
now collected.  Eqs.(C.11), (C.15), (C.12), (C.18) imply, in succession,
\begin{eqnarray}
\bar{K}_0(s) &=& \frac{1}{16 \pi^2} F(s) + \kappa_0(s) \nonumber \\
\bar{K}_1(s) &=& \frac{1}{16 \pi^2} \frac{s}{s - 4M_{\pi}^2}F^2(s) +
\kappa_1(s) \nonumber \\
\bar{K}_2(s) &=& \frac{1}{16 \pi^2} F^2(s) + \kappa_2(s) \\
\bar{K}_3(s) &=& \frac{1}{16 \pi^2} \frac{sM_{\pi}^2}{(s - 4M_{\pi}^2)^2}
\left \{ F^3(s) + \pi^2 \frac{s - 4M_{\pi}^2}{s} F(s) \right \} +
\kappa_3(s)\ ,\nonumber
\end{eqnarray}
and from Eqs. (C.14), (C.17) and (C.19) one reconstructs $\bar{K}_4(s)$:
\begin{eqnarray}
\bar{K}_4(s) &=& \frac{M_{\pi}^2}{16 \pi^2} \frac{F(s)}{s - 4M_{\pi}^2}
\nonumber \\
&+&\frac{M_{\pi}^2}{32 \pi^2}\frac{1}{s-4M_{\pi}^2} \left[\frac{s}{s
- 4M_{\pi}^2}F^2(s) + \pi^2 \right]  \\
&+&\frac{M_{\pi}^2}{48 \pi^2}\frac{1}{s-4M_{\pi}^2}
\left\{\frac{M_{\pi}^2s}{(s - 4M_{\pi}^2)^2}\left[F^3(s) +
\pi^2 \frac{s-4M_{\pi}^2}{s} F(s)\right] - \pi^2 \right\} + \kappa_4(s)\ .
\nonumber
\end{eqnarray}
In Eqs. (C.20) and (C.21), $\kappa_n(s)$ denote sofar arbitrary polynomials.
They may be fixed using the boundary conditions (C.10).  Since
$F(s)$ is bounded by $\ln s$ for $|s|\rightarrow\infty$, the asymptotic
condition (C.10) implies that all $\kappa_n(s)$ must be constant.
Their values are then determined by the conditions $\bar{K}_n(0) = 0$:
\eq
\kappa_0 = \frac{1}{8\pi^2},~~\kappa_1 = 0,~~\kappa_2 = -\frac{1}{4\pi^2},
{}~~\kappa_3 = -\frac{1}{32},~~\kappa_4 = \frac{1}{132} - \frac{1}{32\pi^2}.
\en
Eqs. (C.20) and (C.21) now coincide with the formula  \rf{Kbar} given in
Section 3 without proof.  It is rather interesting that all two-loop
contributions to the $\pi\pi$ scattering amplitude can be obtained by
algebraic manipulations with the Chew-Mandelstam function, without having
to evaluate a single integral.

\newpage

\noindent
\appendix{\large\bf Appendix D $\quad$Threshold Parameters}

\indent
\renewcommand{\theequation}{D.\arabic{equation}}
\setcounter{equation}{0}

In this Appendix, we display the explicit two-loop expressions of the threshold
parameters $a_l^{\rm I}$ and $b_l^{\rm I}$. They are defined from the low-$q^2$
expansions of the real parts of the corresponding partial waves $f_l^{\rm
I}(s)$
\eq
{\rm Re}f_l^{\rm I}(s)\, =\, q^{2l}\,\left\{\, a_l^{\rm I}\, +\, b_l^{\rm
I}\,q^2\,+\,\cdots\right\}\ ,
\en
with $q^2=(s-4\mpi^2)/4\mpi^2$. The partial waves are defined by taking the
corresponding projections  \rf{pwi} of the two-loop amplitude  \rf{amp2},
 \rf{Kstu}.

\bea
a_0^0 &=& {1\over{96\pi}}{{\mpi^2}\over{\fpi^2}} (5\alpha + 16 \beta )
          + {5\over{8\pi}}{{\mpi^4}\over{\fpi^4}} (\lambda_1 + 2\lambda_2 )
          + {1\over{4608\pi^3}}{{\mpi^4}\over{\fpi^4}} (5\alpha + 16\beta )^2
\nonumber\\
&+& {1\over{4\pi}}{{\mpi^6}\over{\fpi^6}}(\lambda_3 -6\lambda_4 )
    + {5\over{192\pi^3}}{{\mpi^6}\over{\fpi^6}}(\lambda_1 + 2\lambda_2 )
                                               (5\alpha +16\beta )
\nonumber\\
&+& {25\over{221184\pi^5}} {{\mpi^6}\over{\fpi^6}}\, (23 - 2\pi^2 ){\alpha}^3 +
    {5\over{13824\pi^5}} {{\mpi^6}\over{\fpi^6}}\, (33 - 2\pi^2 )
{\alpha}^2\beta
\nonumber\\
&+& {5\over{55296\pi^5}} {{\mpi^6}\over{\fpi^6}}\, (198 - \pi^2 ) \alpha\beta^2
+ {1\over{3456\pi^5}} {{\mpi^6}\over{\fpi^6}}\, (70 - \pi^2) \beta^3
\\
&&\nonumber\\
&&\nonumber\\
%( --------------------------------------------------- )
b_0^0 &=& {1\over{4\pi}}{{\beta}\over{\fpi^2}}
         + {1\over\pi}{{\mpi^2}\over{\fpi^4}}( 2\lambda_1 + 3\lambda_2 )
         + {1\over{3\pi^3}}{{\mpi^2}\over{\fpi^4}}
 ({{\beta^2}\over{3}} + {5\over{96}}\alpha\beta - {5\over{256}}\alpha^2 )
\nonumber\\
&+& {3\over\pi} {{\mpi^2}\over{\fpi^4}}\, (\lambda_3 -\lambda_4 ) +
{5\over{216\pi^3}} {{\mpi^4}\over{\fpi^6}}\, \lambda_1
({47\over 4}\alpha + 67\beta ) +
{5\over{27\pi^3}}  {{\mpi^4}\over{\fpi^6}}\, \lambda_2
({29\over 16}\alpha + 13\beta )
\nonumber\\
&-& {5\over{331776\pi^5}} {{\mpi^4}\over{\fpi^6}}\,
(161 + {61\over 24}\pi^2 )\alpha^3 +
{5\over{9216\pi^5}} {{\mpi^4}\over{\fpi^6}}\,
({13\over 2} - {37\over 9}\pi^2 )\alpha^2\beta
\nonumber\\
&+& {5\over{20736\pi^5}} {{\mpi^4}\over{\fpi^6}}\,
(89 - {85\over 4}\pi^2 )\alpha\beta^2 +
{1\over {41472\pi^5}} {{\mpi^4}\over{\fpi^6}}\,
({4793\over 2} - {823\over 3}\pi^2 )\beta^3
\\
&&\nonumber\\
&&\nonumber\\
%( --------------------------------------------------- )
a_2^0 &=&  {1\over{30\pi}}{1\over{\fpi^4}} (\lambda_1 + 4\lambda_2 )
          + {1\over{34560\pi^3}}{1\over{\fpi^4}} (\alpha^2 - 48\beta^2 )
\nonumber\\
&-& {1\over{5\pi}} {{\mpi^2}\over{\fpi^6}}\, (\lambda_3 + 4\lambda_4 )
- {1\over {36\pi^3}} {{\mpi^2}\over{\fpi^6}}\, \lambda_1
( {7\over 72}\alpha - {53\over 225}\beta )
- {1\over {36\pi^3}} {{\mpi^2}\over{\fpi^6}}\, \lambda_2
( {7\over 36}\alpha + {53\over 225}\beta )
\nonumber\\
&+& {11\over{7464960\pi^5}} {{\mpi^2}\over{\fpi^6}}\,
({41\over 4} - \pi^2 )\alpha^3 -
{1\over{138240\pi^5}} {{\mpi^2}\over{\fpi^6}}\,
( {121\over 18} - \pi^2 )\alpha^2\beta
\nonumber\\
&-& {1\over{248832\pi^5}} {{\mpi^2}\over{\fpi^6}}\,
(89 - {46\over 5}\pi^2 )\alpha\beta^2 +
{1\over {9331200\pi^5}} {{\mpi^2}\over{\fpi^6}}\,
({1679\over 2} - 329\pi^2 )\beta^3
\ena
\newpage
\bea
%( --------------------------------------------------- )
a_1^1 &=& {1\over{24\pi}}{1\over{\fpi^2}} \beta
          - {1\over{6\pi}}{{\mpi^2}\over{\fpi^4}} (\lambda_1 - \lambda_2 )
          + {1\over{41472\pi^3}}{{\mpi^2}\over{\fpi^4}} (5\alpha^2
                           - 40\alpha\beta - 16\beta^2 )
\nonumber\\
&+& {1\over{2\pi}} {{\mpi^4}\over{\fpi^6}}\, (\lambda_3 - \lambda_4 ) +
{1\over{144\pi^3}} {{\mpi^4}\over{\fpi^6}}\, \lambda_1
({5\over 12}\alpha - \beta ) +
{1\over{81\pi^3}} {{\mpi^4}\over{\fpi^6}}\, \lambda_2
({5\over 32}\alpha - \beta )
\nonumber\\
&+& {5\over{497664\pi^5}} {{\mpi^4}\over{\fpi^6}}\,
(1- {{\pi^2}\over 6})\alpha^3 +
{5\over{331776\pi^5}} {{\mpi^4}\over{\fpi^6}}\,
(5 - {{\pi^2}\over 3})\alpha^2\beta
\nonumber\\
&-& {5\over{497664\pi^5}} {{\mpi^4}\over{\fpi^6}}\,
({374\over 3} - 13\pi^2 )\alpha\beta^2 -
{5\over {746496\pi^5}} {{\mpi^4}\over{\fpi^6}}\,
( {73\over 2} + 5\pi^2 )\beta^3
\\
&&\nonumber\\
&&\nonumber\\
%( --------------------------------------------------- )
b_1^1 &=& - {1\over{6\pi}}{1\over{\fpi^4}}( \lambda_1 - \lambda_2 )
         + {1\over{4320\pi^3}}{1\over{\fpi^4}}
 ({47\over{3}}\beta^2 - {65\over{6}}\alpha\beta - {5\over{24}}\alpha^2 )
\nonumber\\
&+& {1\over{\pi}} {{\mpi^2}\over{\fpi^6}}\, (\lambda_3 - \lambda_4 ) +
{1\over{48\pi^3}} {{\mpi^2}\over{\fpi^6}}\, \lambda_1
({{\alpha}\over 2} -{53\over 15}\beta ) +
{1\over{162\pi^3}} {{\mpi^2}\over{\fpi^6}}\, \lambda_2
({{\alpha}\over 2} - {31\over 5}\beta )
\nonumber\\
&-& {1\over{746496\pi^5}} {{\mpi^2}\over{\fpi^6}}\,
({107\over 8} - \pi^2 )\alpha^3 -
{1\over{27648\pi^5}} {{\mpi^2}\over{\fpi^6}}\,
({113\over 36} - \pi^2 )\alpha^2\beta
\nonumber\\
&+&{1\over{41472\pi^5}} {{\mpi^2}\over{\fpi^6}}\,
({851\over 9} - {47\over 4}\pi^2)\alpha\beta^2 -
{1\over{1244160\pi^5}} {{\mpi^2}\over{\fpi^6}}\,
({4601\over 2} - {589\over 3}\pi^2 )\beta^3
\\
&&\nonumber\\
&&\nonumber\\
%(----------------------------------------------------)
a_3^1 &=& {1\over{13230\pi^3\mpi^2\fpi^4}}\, ( {{\alpha^2}\over 64} +
{17\over 32}\alpha\beta + \beta^2 )
\nonumber\\
&+&{1\over{35\pi\fpi^6}}\, (\lambda_3 - \lambda_4 ) +
{1\over{294\pi^3\fpi^6}}\, \lambda_1 ({23\over 240}\alpha - \beta ) +
{1\over {3528\pi^3\fpi^6}}\, \lambda_2 (\alpha - {211\over 10}\beta )
\nonumber\\
&+& {1\over{20321280\pi^5\fpi^6}}\, ({163\over 12} - \pi^2 )\alpha^3 +
{1\over{8128512\pi^5\fpi^6}}\, ({311\over 10} - \pi^2 )\alpha^2\beta
\nonumber\\
&-& {1\over{2903040\pi^5\fpi^6}}\, ({2537\over 14} - 17\pi^2 )\alpha\beta^2 +
{1\over{3386880\pi^5\fpi^6}}\, ({8011\over 36} - {13\over 5}\pi^2 )\beta^3
\\
&&\nonumber\\
&&\nonumber\\
%( --------------------------------------------------- )
a_0^2 &=& {1\over{48\pi}}{{\mpi^2}\over{\fpi^2}} (\alpha - 4 \beta )
          + {1\over{4\pi}}{{\mpi^4}\over{\fpi^4}} (\lambda_1 + 2\lambda_2 )
          + {1\over{1152\pi^3}}{{\mpi^4}\over{\fpi^4}} (\alpha - 4\beta )^2
\nonumber\\
&-& {1\over{2\pi}}{{\mpi^6}\over{\fpi^6}}\, \lambda_3 +
{1\over{48\pi^3}} {{\mpi^6}\over{\fpi^6}} (\lambda_1 + 2\lambda_2 )
(\alpha -4\beta )
\nonumber\\
&+& {1\over{18432\pi^5}} {{\mpi^6}\over{\fpi^6}}\,
({29\over 3} - \pi^2 )\alpha^3 -
{1\over{9216\pi^5}} {{\mpi^6}\over{\fpi^6}}\,
(37 - {23\over 6}\pi^2 )\alpha^2\beta
\nonumber\\
&+& {1\over{1536\pi^5}} {{\mpi^6}\over{\fpi^6}}\,
(17 - {31\over 18}\pi^2 )\alpha\beta^2 -
{1\over{384\pi^5}} {{\mpi^6}\over{\fpi^6}}\,
({47\over 9} - {{\pi^2}\over 2})\beta^3
\ena
\newpage
\bea
%( --------------------------------------------------- )
b_0^2 &=& - {1\over{8\pi}} {{\beta}\over{\fpi^2}}
         + {1\over{2\pi}} {{\mpi^2}\over{\fpi^4}}( \lambda_1 + 3\lambda_2 )
         + {1\over{4608\pi^3}} {{\mpi^2}\over{\fpi^4}}
 ( 112\beta^2 - 8\alpha\beta - 7\alpha^2 )
\nonumber\\
&-& {3\over{2\pi}} {{\mpi^4}\over{\fpi^6}}\, (\lambda_3 - \lambda_4 ) +
{1\over{432\pi^3}} {{\mpi^4}\over{\fpi^6}}\, \lambda_1
({17\over 4}\alpha - 83\beta ) +
{1\over{108\pi^3}} {{\mpi^4}\over{\fpi^6}}\, \lambda_2
({59\over 8}\alpha - 61\beta )
\nonumber\\
&-& {1\over{165888\pi^5}} {{\mpi^4}\over{\fpi^6}}\,
(71 - {23\over 6}\pi^2 )\alpha^3 -
{1\over{36864\pi^5}} {{\mpi^4}\over{\fpi^6}}\,
(91 - {169\over 9}\pi^2 )\alpha^2\beta
\nonumber\\
&+&{1\over{82944\pi^5}} {{\mpi^4}\over{\fpi^6}}\,
(1487 - {445\over 2}\pi^2 )\alpha\beta^2 -
{1\over{82944\pi^5}} {{\mpi^4}\over{\fpi^6}}\,
({6745\over 2} - {1237\over 3}\pi^2 )\beta^3
\\
&&\nonumber\\
&&\nonumber\\
%( --------------------------------------------------- )
a_2^2 &=&  {1\over{30\pi}}{1\over{\fpi^4}} (\lambda_1 + \lambda_2 )
          + {1\over{7200\pi^3}}{1\over{\fpi^4}}
  ({1\over{8}}\alpha^2 +{13\over{6}}\alpha\beta - {13\over{3}}\beta^2 )
\nonumber\\
&-& {1\over{5\pi}} {{\mpi^2}\over{\fpi^6}}\, (\lambda_3 + \lambda_4 ) -
{1\over{32400\pi^3}} {{\mpi^2}\over{\fpi^6}}\, \lambda_1
({151\over 2}\alpha - 191\beta ) -
{1\over {8100\pi^3}} {{\mpi^2}\over{\fpi^6}}\, \lambda_2
({41\over 2}\alpha - 61\beta )
\nonumber\\
&+& {1\over{9331200\pi^5}} {{\mpi^2}\over{\fpi^6}}\,
({1223\over 16} - 7\pi^2 )\alpha^3 +
{7\over{1382400\pi^5}} {{\mpi^2}\over{\fpi^6}}\,
({113\over 18} - \pi^2 )\alpha^2\beta
\nonumber\\
&+& {1\over{518400\pi^5}} {{\mpi^2}\over{\fpi^6}}\,
( 119 - {257\over 24}\pi^2 )\alpha\beta^2
+ {1\over{18662400\pi^5}} {{\mpi^2}\over{\fpi^6}}\,
({3289\over 2} - 271\pi^2 )\beta^3
\ena

The contributions coming from various orders of the chiral
expansion are easy to distinguish.
The lowest order contributions, corresponding to the amplitude
 \rf{Ap2}, give terms linear in $\alpha$ and $\beta$, proportional to
$\fpi^{-2}$. At this order, only the S- and P-wave scattering lengths and the
S-wave slope parameters are non-vanishing. The one-loop precision brings in two
types of contributions, both proportional to $\fpi^{-4}$: Those which are
linear in $\lambda_1$ and $\lambda_2$ come from $O({\rm p}^4)$ and $O({\rm
p}^5)$ tree graphs, and those which are quadratic in $\alpha , \beta$
correspond to genuine one-loop graphs. Finally, the contributions at order
two-loop, proportional to $\fpi^{-6}$, come from $O({\rm p}^6)$ and $O({\rm
p}^7)$ tree-graphs (terms linear in $\lambda_3$ and $\lambda_4$), from one-loop
graphs with a $\lambda_1$ or $\lambda_2$ vertex, whereas the genuine two-loop
graphs generate the contributions which are cubic in $\alpha ,\beta$.

\newpage

\centerline{\bf Figure Captions}
\bigskip

\begin{itemize}
\item{\bf 1-} Argand diagram of the amplitude $f^0_0$ (i.e.
$\sqrt{1-4M^2_{\pi}/s}\,{\rm Im} f_0^0$ versus
$\sqrt{1-4M^2_{\pi}/s}\,{\rm Re} f_0^0$). The result at one loop is
represented by the dashed line and the result at two loops by the
solid line. Both cases correspond to the same parameter choice
$\alpha=2$, $\beta=1.08$ and $\lambda_i$ as in (4.9) and
(4.10).

\medskip

\item{\bf 2-} Phase shifts computed at increasing chiral order
(see formulas (4.11)-(4.13) and the discussion following them
). The tree, one-loop and two-loop
orders are represented by the dotted, dashed and dash-dotted lines
respectively.
Shown for comparison are the data of Ref.  \cite{rosselet}
and  \cite{EM} for $\delta_0^0$ and Ref.  \cite{Hoog} for
$\delta_0^2$ and $\delta_2^2$. For $\delta_1^1$ and $\delta_2^0$ the
Breit-Wigner contributions of the $\rho$ and the $f_2$ resonances
are shown as crosses.

\medskip

\item{\bf 3-} Illustration of the sensitivity of the phase shifts
to the values of the parameters $\alpha$ and $\beta$. The dotted line
is the result of the standard $\chi$PT $(\alpha=1.04,\ \beta=1.08)$,
the dashed line corresponds to $\alpha=2,\ \beta=1.08$ and the
solid line to $\alpha=3,\beta=1.12$.

\medskip

\item{\bf 4-} The phase shift difference
$\delta_0^0-\delta_1^1$ in the range of energies accessible
in the $K_{l4}$ decays is shown: a) for increasing chiral orders as in
Fig. 2 and, b) for several values of $\alpha$ and $\beta$,
as in Fig. 3.

\medskip

\item{\bf 5-} Influence of the uncertainties affecting the
evaluation of $\lambda_1,...,\lambda_4$ on the prediction for
$\delta_0^0-\delta_1^1$ . The band delimited by solid lines
corresponds to varying the $\lambda_i$'s inside the ranges
indicated in (4.9) and (4.10)
with $\alpha$, $\beta$
fixed to the standard $\chi$PT values, whereas the band delimited by
dashed lines corresponds to $\alpha=2,\ \beta=1.08$.

\medskip

\item{\bf 6-} Plot of the correlation between $\alpha$ and $\beta$ as
inferred from the  Morgan-Shaw universal band given by Eq. \rf{univc}.
The hatched region covers the set of points consistent with Eqs.
\rf{alpha(r)} and \rf{beta(r)}.  If the $O({\rm p^2})$ terms are
ignored, this set shrinks to the doubly-hatched sub-region shown.
See the discussion in Section 5.

\medskip

\item{\bf 7-} Predictions for the scattering lengths
as  functions of
$\alpha$. The inner bands are obtained by varying the $\lambda_i$'s
inside their error bars with $\beta$ fixed at the center of the
Morgan-Shaw band. Including the variations in  $\beta$ allowed by the latter
gives the outer bands.

\medskip

\item{\bf 8-} In a), we show the parameter
$\alpha(r)$ at increasing chiral order (see
Eqs. (5.36) and the following discussion).
The doubly-hatched
area represents uncertainties in the contributions from $\widehat{\rho}_1,~
\widehat{\rho}_2$, while the hatched band includes
the estimated uncertainty of $\alpha^{(2)}$, as described in the text.
In b) we show the
contribution $\alpha^{(2)}_{\rm loop}(r)$ for different values of $\mu$.
Each band corresponds to variations of the scale $\Lambda$ in the range
$M_{\eta}\le\Lambda\le 1$ GeV.

\medskip

\item{\bf 9-} In a), we show the parameter
$\beta(r)$ at increasing chiral order (see
Eq. \rf{beta(r)} and the discussion following Eq. \rf{alpha(r)}).
The hatched area represents
the estimated uncertainty in contribution from
$\beta^{(2)}$.
In b) we show the
contribution $\beta^{(2)}_{\rm loop}(r)$ for different values of $\mu$.
Each band corresponds to variations of the scale $\Lambda$ in the range
$M_{\eta}\le\Lambda\le 1$ GeV.
\medskip

\item{\bf 10-} In a) we show the Gell-Mann--Oakes--Renner ratio $x_{GOR}$ as a
function of the linear combination $(\alpha + 2\beta)/3$ (see Eq.
(5.48)). The hatched area
represents the expected range of $O({\rm p^2})$ contributions.  The
vertical dotted
lines show the range in $(\alpha + 2\beta)/3$ corresponding to the fit
values Eq. (4.15).
In b) we show the
contribution $x^{\rm loop}_{GOR}$ for different values of $\mu$.
Each band corresponds to variations of the scale $\Lambda$ in the range
$M_{\eta}\le\Lambda\le 1$ GeV.

\end{itemize}

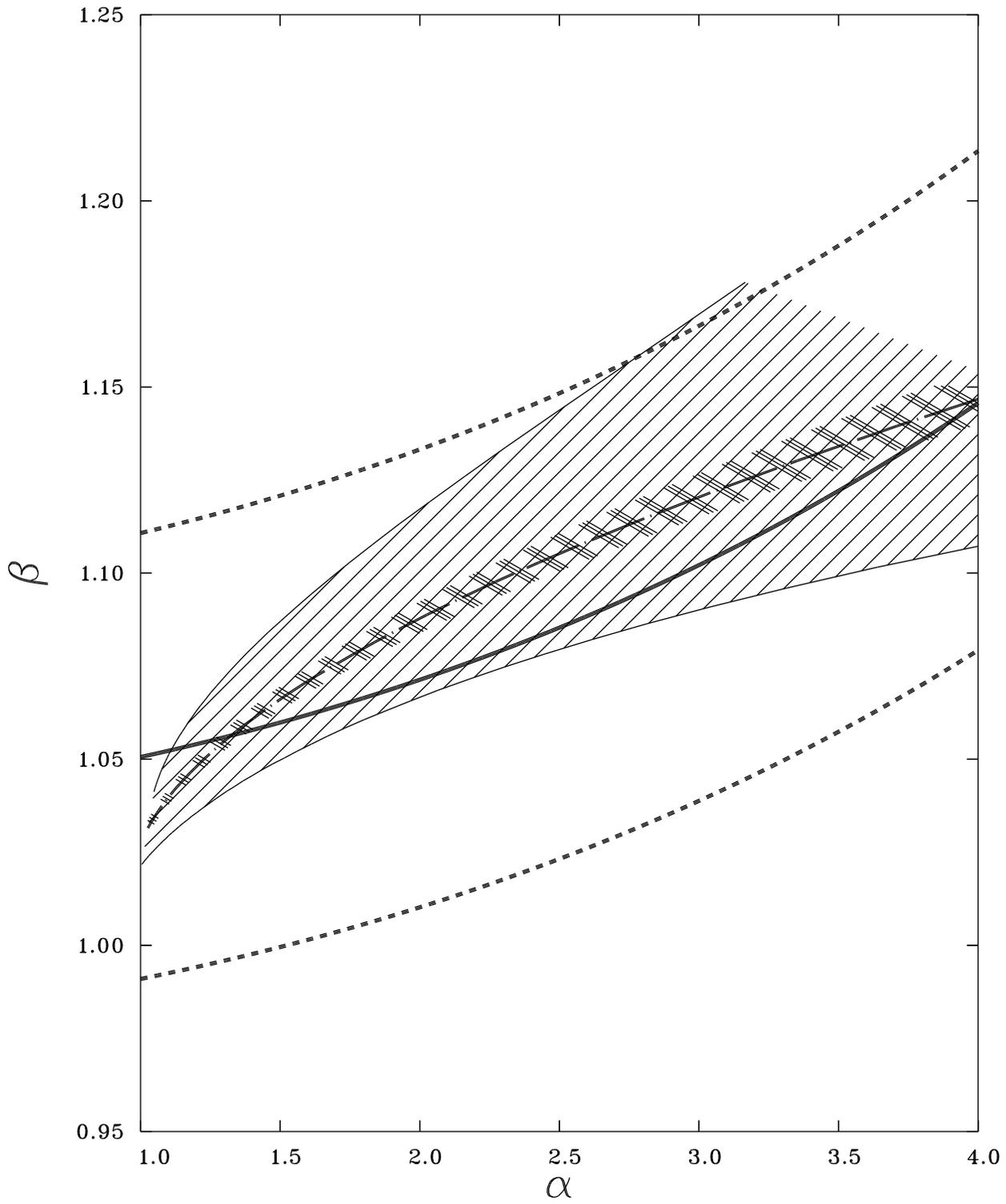
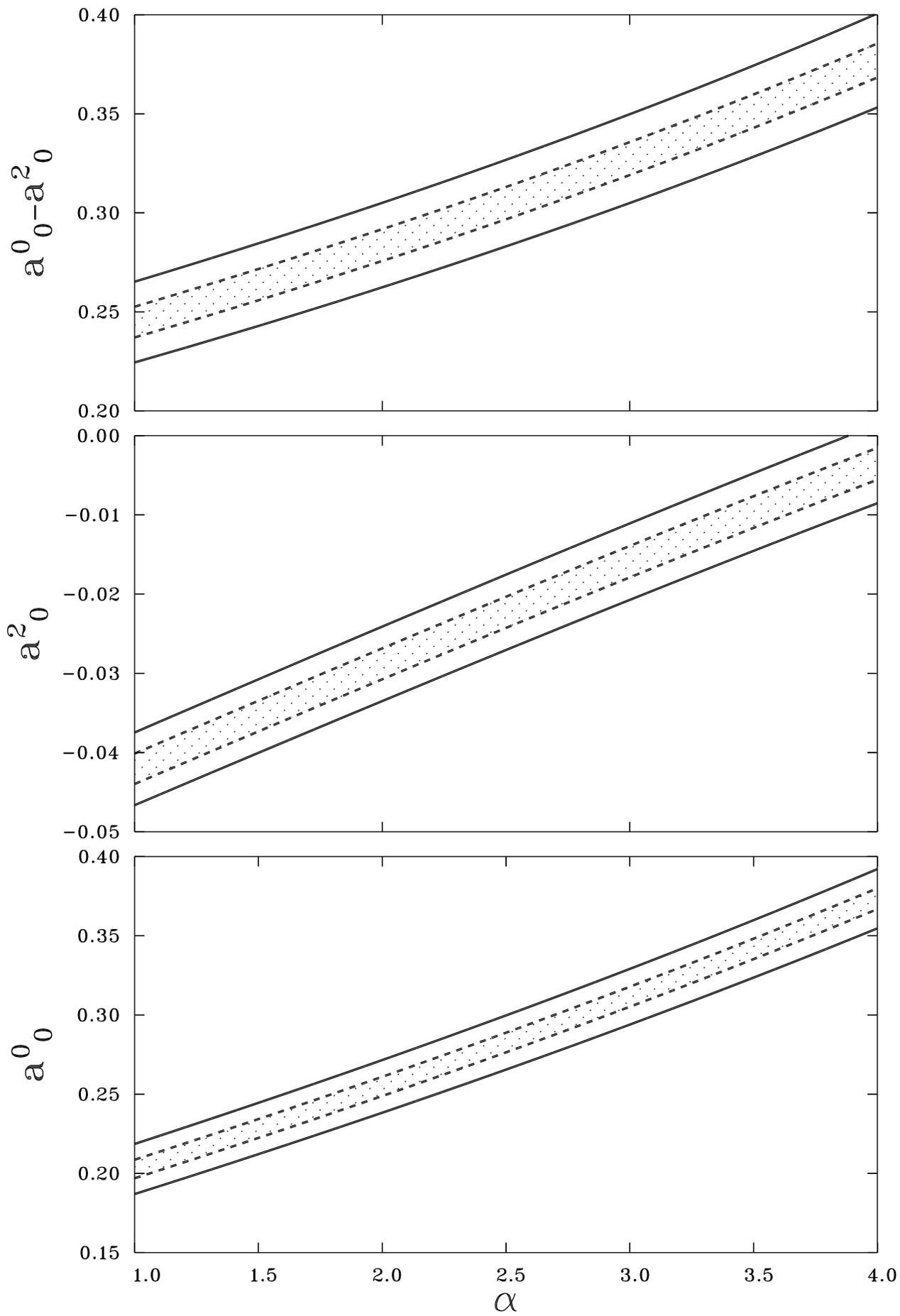
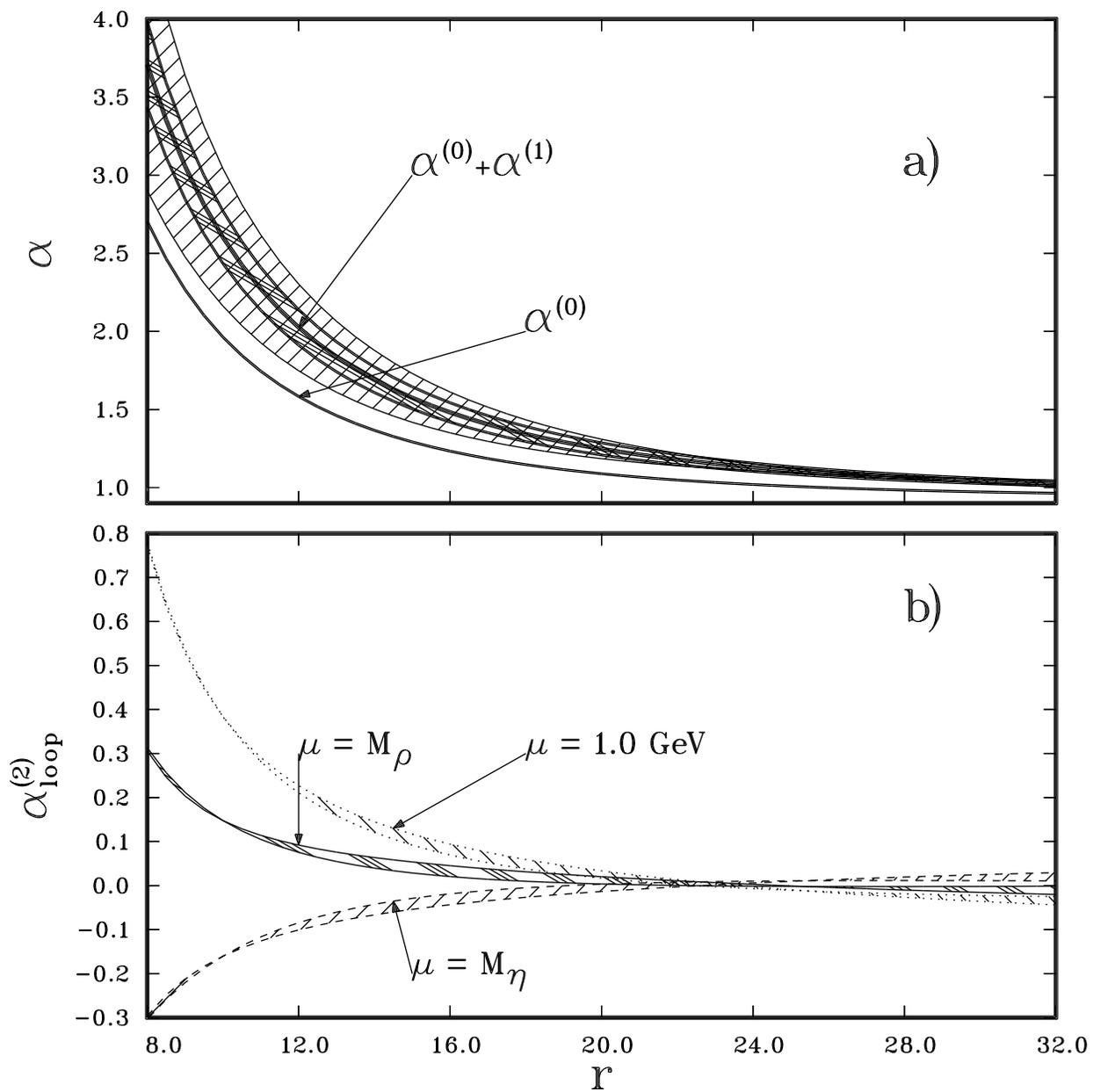
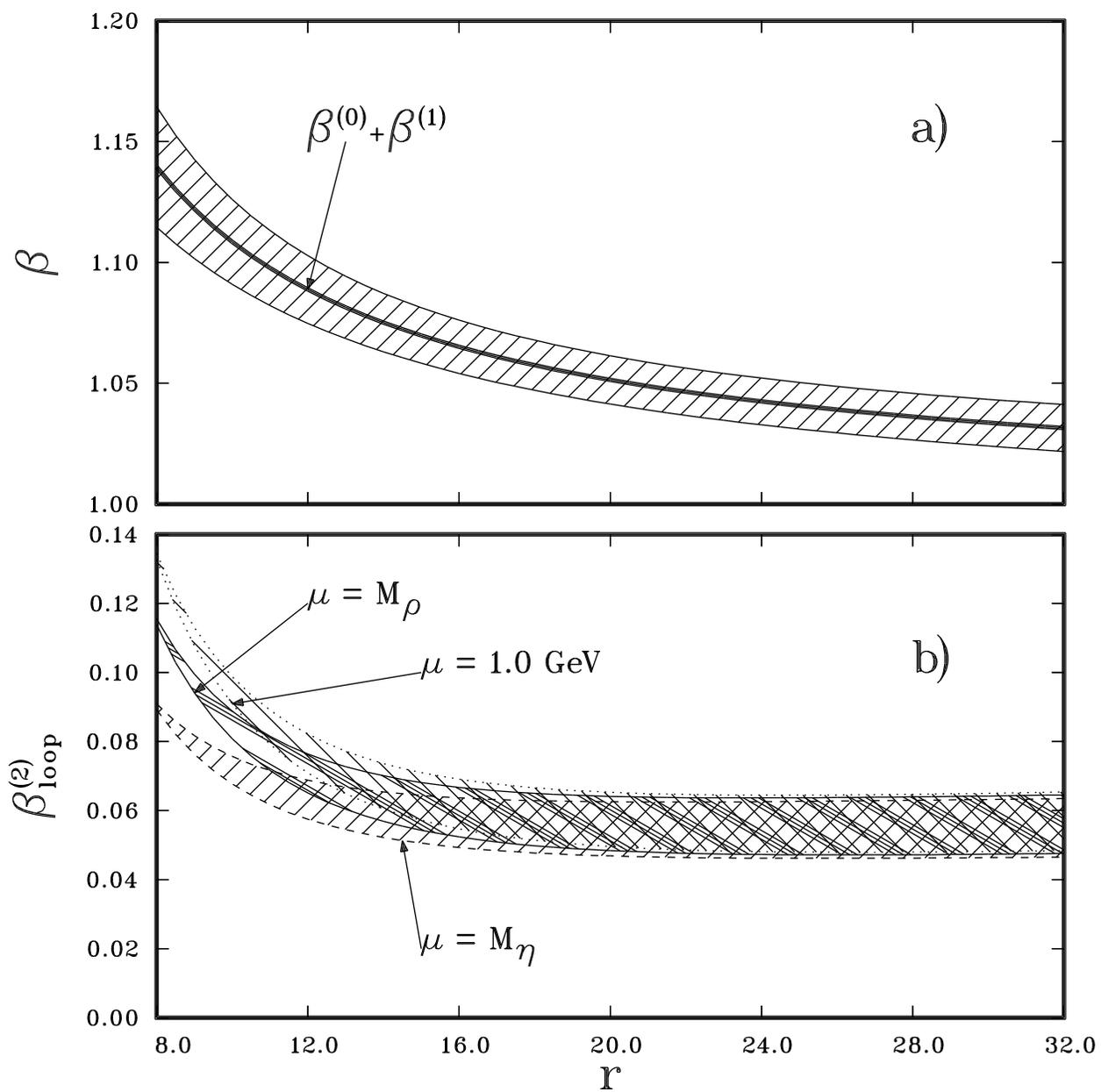
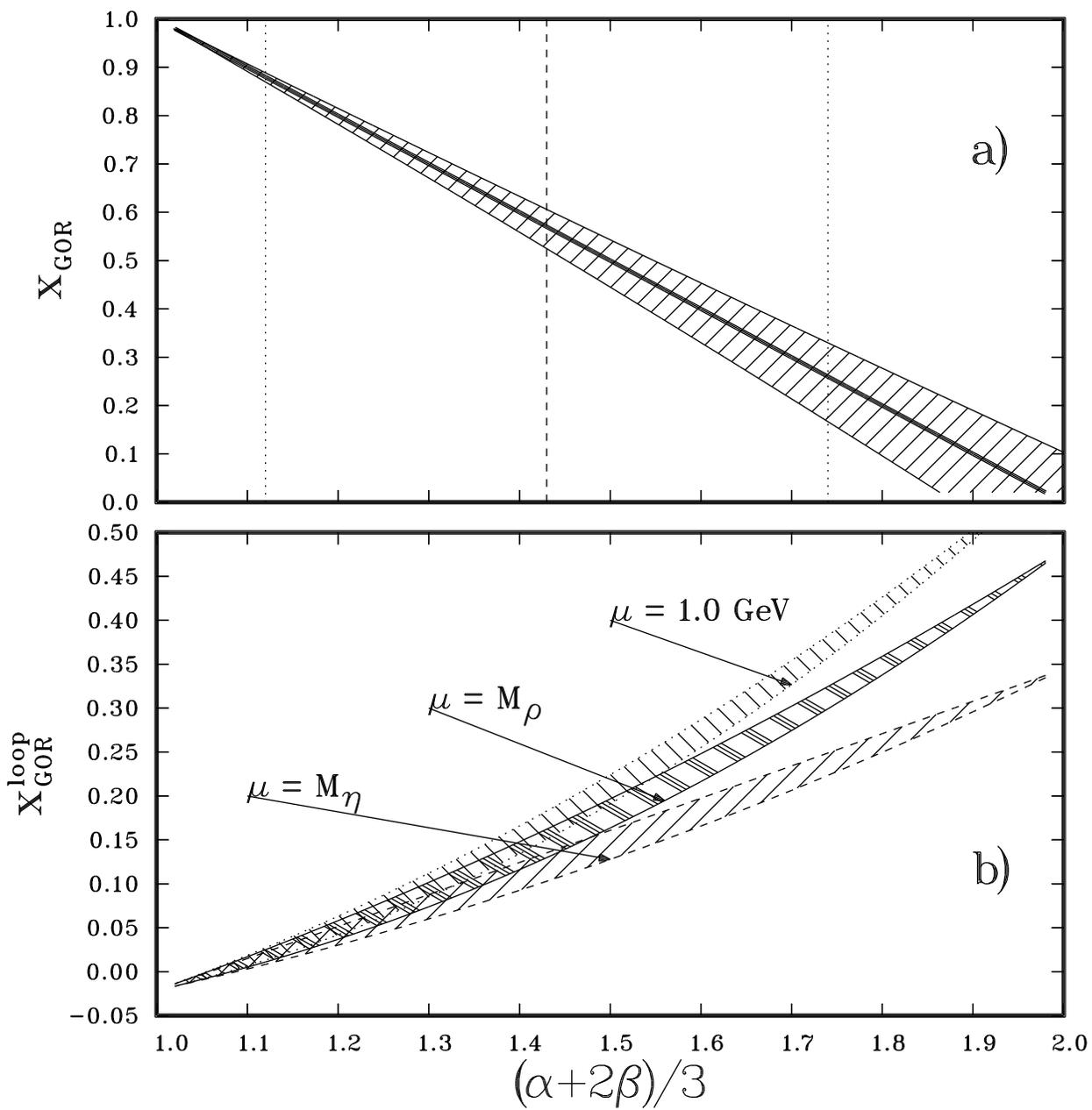
\end{document}